\documentclass[12pt,a4paper]{article}
\usepackage[utf8]{inputenc}
\usepackage{amsmath}
\usepackage{amsfonts}
\usepackage{amssymb}
\usepackage{graphicx}
\usepackage{rotating}
\usepackage{pdflscape}
\usepackage{xspace}
\usepackage{xcolor}
\usepackage{cite}
\usepackage{mciteplus}
\usepackage{upgreek}
\usepackage{hyperref}
\usepackage[all]{hypcap}
\usepackage{ifthen}
\usepackage{authblk}

\usepackage{lineno}

\newboolean{uprightparticles}
\setboolean{uprightparticles}{false}

\newboolean{articletitles}
\setboolean{articletitles}{true}

\def\lhcb {\mbox{LHCb}\xspace}

\def\MagUp {\mbox{\em Mag\kern -0.05em Up}\xspace}

\ifthenelse{\boolean{uprightparticles}}%
{

 \def\Ppi         {\ensuremath{\uppi}\xspace}                 
                  
 \def\Prho        {\ensuremath{\uprho}\xspace}

 \def\Ppsi        {\ensuremath{\uppsi}\xspace}

 \def\PDelta      {\ensuremath{\Delta}\xspace}                 
 \def\PXi      {\ensuremath{\Xi}\xspace}                 
 \def\PLambda      {\ensuremath{\Lambda}\xspace}                 
 \def\PSigma      {\ensuremath{\Sigma}\xspace}                 
 \def\POmega      {\ensuremath{\Omega}\xspace}                 
 \def\PUpsilon      {\ensuremath{\Upsilon}\xspace}                 
 
 \def\PB      {\ensuremath{\mathrm{B}}\xspace}                 
                  
 \def\PD      {\ensuremath{\mathrm{D}}\xspace}

 \def\PJ      {\ensuremath{\mathrm{J}}\xspace}                 
 \def\PK      {\ensuremath{\mathrm{K}}\xspace}

 \def\Pb      {\ensuremath{\mathrm{b}}\xspace}

 \def\Pi      {\ensuremath{\mathrm{i}}\xspace}

 \def\Ps      {\ensuremath{\mathrm{s}}\xspace}

}
{

 \def\Ppi         {\ensuremath{\pi}\xspace}                 
                  
 \def\Prho        {\ensuremath{\rho}\xspace}

 \def\Ppsi        {\ensuremath{\psi}\xspace}                 
                  
 \mathchardef\PDelta="7101
 \mathchardef\PXi="7104
 \mathchardef\PLambda="7103
 \mathchardef\PSigma="7106
 \mathchardef\POmega="710A
 \mathchardef\PUpsilon="7107
                  
 \def\PB      {\ensuremath{B}\xspace}                 
                  
 \def\PD      {\ensuremath{D}\xspace}

 \def\PJ      {\ensuremath{J}\xspace}                 
 \def\PK      {\ensuremath{K}\xspace}

 \def\Pb      {\ensuremath{b}\xspace}

 \def\Pi      {\ensuremath{i}\xspace}

 \def\Ps      {\ensuremath{s}\xspace}

}

\makeatletter
\ifcase \@ptsize \relax
  \newcommand{\miniscule}{\@setfontsize\miniscule{4}{5}}
\or
  \newcommand{\miniscule}{\@setfontsize\miniscule{5}{6}}
\or
  \newcommand{\miniscule}{\@setfontsize\miniscule{5}{6}}
\fi
\makeatother

\DeclareRobustCommand{\optbar}[1]{\shortstack{{\miniscule (\rule[.5ex]{1.25em}{.18mm})}
  \\ [-.7ex] $#1$}}

\def\squark    {{\ensuremath{\Ps}}\xspace}

\def\bquark    {{\ensuremath{\Pb}}\xspace}

\def\pion   {{\ensuremath{\Ppi}}\xspace}

\def\pip    {{\ensuremath{\pion^+}}\xspace}
\def\pim    {{\ensuremath{\pion^-}}\xspace}

\def\rhomeson {{\ensuremath{\Prho}}\xspace}
\def\rhoz     {{\ensuremath{\rhomeson^0}}\xspace}

\def\kaon    {{\ensuremath{\PK}}\xspace}
  \def\Kbar    {{\kern 0.2em\overline{\kern -0.2em \PK}{}}\xspace}

\def\KorKbar    {\kern 0.18em\optbar{\kern -0.18em K}{}\xspace}

\def\Kp      {{\ensuremath{\kaon^+}}\xspace}
\def\Km      {{\ensuremath{\kaon^-}}\xspace}

\def\Kstarz  {{\ensuremath{\kaon^{*0}}}\xspace}

  \def\Dbar    {{\kern 0.2em\overline{\kern -0.2em \PD}{}}\xspace}
\def\D       {{\ensuremath{\PD}}\xspace}

\def\DorDbar    {\kern 0.18em\optbar{\kern -0.18em D}{}\xspace}
\def\DtwoorDtwobar {\kern -0.25em\optbar{\kern 0.25em D_2^*}{}\xspace}
\def\Dz      {{\ensuremath{\D^0}}\xspace}

\def\Ds      {{\ensuremath{\D^+_\squark}}\xspace}

\def\B       {{\ensuremath{\PB}}\xspace}
\def\Bbar    {{\ensuremath{\kern 0.18em\overline{\kern -0.18em \PB}{}}}\xspace}

\def\BorBbar    {\kern 0.18em\optbar{\kern -0.18em B}{}\xspace}

\def\BzorBzbar  {\kern 0.18em\optbar{\kern -0.18em B}{}^0\xspace}
\def\Bu      {{\ensuremath{\B^+}}\xspace}

\def\Bp      {{\ensuremath{\Bu}}\xspace}

\def\jpsi     {{\ensuremath{{\PJ\mskip -3mu/\mskip -2mu\Ppsi\mskip 2mu}}}\xspace}

  \def\Y#1S{\ensuremath{\PUpsilon{(#1S)}}\xspace}

\def\Lz          {{\ensuremath{\PLambda}}\xspace}
\def\Lbar        {{\ensuremath{\kern 0.1em\overline{\kern -0.1em\PLambda}}}\xspace}
\def\LorLbar    {\kern 0.18em\optbar{\kern -0.18em \PLambda}{}\xspace}

\def\Lb      {{\ensuremath{\Lz^0_\bquark}}\xspace}

\def\to                 {\ensuremath{\rightarrow}\xspace}

\def\CP                {{\ensuremath{C\!P}}\xspace}

\def\AT#1     {\ensuremath{A_{\mathrm{T}}^{#1}}\xspace}           

\def\C#1      {\ensuremath{\mathcal{C}_{#1}}\xspace}                       
\def\Cp#1     {\ensuremath{\mathcal{C}_{#1}^{'}}\xspace}                    
\def\Ceff#1   {\ensuremath{\mathcal{C}_{#1}^{\mathrm{(eff)}}}\xspace}        
\def\Cpeff#1  {\ensuremath{\mathcal{C}_{#1}^{'\mathrm{(eff)}}}\xspace}       
\def\Ope#1    {\ensuremath{\mathcal{O}_{#1}}\xspace}                       
\def\Opep#1   {\ensuremath{\mathcal{O}_{#1}^{'}}\xspace}                    

\newcommand{\tev}{\ifthenelse{\boolean{inbibliography}}{\ensuremath{~T\kern -0.05em eV}\xspace}{\ensuremath{\mathrm{\,Te\kern -0.1em V}}}\xspace}
\newcommand{\gev}{\ensuremath{\mathrm{\,Ge\kern -0.1em V}}\xspace}
\newcommand{\mev}{\ensuremath{\mathrm{\,Me\kern -0.1em V}}\xspace}
\newcommand{\kev}{\ensuremath{\mathrm{\,ke\kern -0.1em V}}\xspace}
\newcommand{\ev}{\ensuremath{\mathrm{\,e\kern -0.1em V}}\xspace}
\newcommand{\gevc}{\ensuremath{{\mathrm{\,Ge\kern -0.1em V\!/}c}}\xspace}
\newcommand{\mevc}{\ensuremath{{\mathrm{\,Me\kern -0.1em V\!/}c}}\xspace}
\newcommand{\gevcc}{\ensuremath{{\mathrm{\,Ge\kern -0.1em V\!/}c^2}}\xspace}
\newcommand{\gevgevcccc}{\ensuremath{{\mathrm{\,Ge\kern -0.1em V^2\!/}c^4}}\xspace}
\newcommand{\mevcc}{\ensuremath{{\mathrm{\,Me\kern -0.1em V\!/}c^2}}\xspace}

\def\gsim{{~\raise.15em\hbox{$>$}\kern-.85em
          \lower.35em\hbox{$\sim$}~}\xspace}
\def\lsim{{~\raise.15em\hbox{$<$}\kern-.85em
          \lower.35em\hbox{$\sim$}~}\xspace}

\def\pt         {\mbox{$p_{\rm T}$}\xspace}

\def\tell1  {TELL1\xspace}
\def\ukl1   {UKL1\xspace}

\newcommand{\eg}{\mbox{\itshape e.g.}\xspace}
\newcommand{\ie}{\mbox{\itshape i.e.}\xspace}

\newcommand{\Xvec}{\ensuremath{\mathbf{X}}\xspace}
\newcommand{\xvec}{\ensuremath{\mathbf{x}}\xspace}
\newcommand{\yvec}{\ensuremath{\mathbf{y}}\xspace}

\newcommand{\mvec}{\ensuremath{\mathbf{m}}\xspace}

\newcommand{\ampl}{\ensuremath{\mathcal{A}}\xspace}
\newcommand{\thetaveca}{\ensuremath{\mathbf{\Theta}_{\mathcal{A}}}\xspace}
\newcommand{\thetavec}{\ensuremath{\mathbf{\Theta}}\xspace}
\newcommand{\loglh}{\ensuremath{\ln\mathcal{L}}\xspace}

\title{Efficient description of experimental effects in amplitude analyses}
\author{Abhijit Mathad$^{1,2}$, Daniel O'Hanlon$^{3}$\thanks{Now at H.H. Wills Physics Laboratory, University of Bristol, Bristol, United Kingdom},\\ Anton Poluektov$^{2,4}$\thanks{Corresponding author. E-mail address: \texttt{Anton.Poluektov@cern.ch}}, Raul Rabadan$^{4}$}
\affil{\small
${}^{1}$ Physik-Institut, Universit\"{a}t Z\"{u}rich, Z\"{u}rich, Switzerland\\
${}^{2}$ Department of Physics, University of Warwick, Coventry, United Kingdom\\
${}^{3}$ INFN Sezione di Bologna, Bologna, Italy\\
${}^{4}$ Aix Marseille Univ, CNRS/IN2P3, CPPM, Marseille, France
}

\begin{document}


\maketitle

\begin{abstract}
  \noindent

  Amplitude analysis is a powerful technique to study hadron decays. A significant complication in these analyses is
  the treatment of instrumental effects, such as background and selection efficiency variations, in the multidimensional
  kinematic phase space. This paper reviews conventional methods to estimate efficiency and background
  distributions and outlines the methods of density estimation using Gaussian processes and artificial neural networks.
  Such techniques see widespread use elsewhere, but have not gained popularity in use for amplitude analyses.
  Finally, novel applications of these models are proposed, to estimate background density in the signal region from the sidebands in multiple dimensions, and a more general method for model-assisted density estimation using artificial neural networks.
\end{abstract}

\clearpage

\tableofcontents

\clearpage

\section{Introduction}

\label{sec:introduction}


Amplitude analysis of hadron decays is a powerful technique employed in many flavour physics
studies, such as measurements of \CP violation, searches for effects beyond the Standard Model,
spectroscopic studies of excited hadrons, and searches for previously unobserved hadronic states.
In this kind of analysis, multidimensional kinematic distributions of the decay products of a parent particle are studied to
reveal the dynamical structure of the decay amplitude~\cite{Dalitz:1953cp}.
In addition to the decay dynamics, the kinematic distributions are in general affected by non-uniform
acceptance, or detection efficiency, and background density, which need to be accounted for in the fit.

In this paper we briefly review the conventional approaches,
recall a few already known but rarely used methods, and, finally, propose new techniques to
model non-uniform acceptance and background distributions.
The proposed techniques not only offer more accurate descriptions of these distributions,
but also provide improved avenues to control the systematic uncertainties arising from conventional approaches. Furthermore, the ability to make efficient use of expensive detailed simulation will be of key importance in the future, when data rates are expected to grow faster than the availability of computing resources~\cite{Bozzi:2657832}

A simple, yet typical example of an amplitude analysis is the study of the two-dimensional distribution
of a three-body decay of a (pseudo)scalar meson into three (pseudo)scalar mesons: Dalitz plot analysis~\cite{Dalitz:1953cp, Back:2017zqt}.
This is the simplest case where the
decay has internal degrees of freedom, yet the amplitude is a function of only two kinematic variables.
In more complicated cases, such as decays involving non-spin-zero states or decays with greater than three particles in the final state,
one is required to analyse kinematic distributions in more than two dimensions.
Here we focus on the simple two-dimensional case, however  the approaches presented
here can be easily generalised to the cases with higher dimensionality.

We deliberately avoid any quantitative direct comparisons of the performance for the illustrated techniques:
the optimal technique for each individual analysis depends on many factors,
such as the size of the data sample; dimensionality of the kinematic phase space;
requirements on statistical and systematic uncertainties;
complexity of the amplitude model; signal selection procedure; or structure of the background contributions. As such, it is important to investigate a variety of complementary approaches.

The structure of the paper is as follows: the formalism of multidimensional maximum-likelihood
fits is recalled in Section~\ref{sec:aman_formalism} and non-parametric methods to deal with background and
acceptance are presented. The samples used to illustrate the background and
acceptance parameterisation techniques are described in Section~\ref{sec:toymc}.
Conventional techniques to parametrise the acceptance distribution are illustrated in Section~\ref{sec:acc-comp}.
Further, two rarely used but yet efficient approaches are presented: a technique using Gaussian processes (Section~\ref{sec:gp}) and
density estimation with artificial neural networks (Section~\ref{sec:annde}).
Finally, two novel approaches are proposed: the technique for inter- or extrapolation of background density
from the sidebands using one of the methods above (Section~\ref{sec:interpolation}),
and a model-assisted parameterisation of background or acceptance density using neural networks (Section~\ref{sec:madenn}).

\section{Formalism of amplitude analyses}
\label{sec:aman_formalism}

A typical amplitude analysis in flavour physics deals with the distribution of kinematic observables $\xvec$ that
characterise the multibody decay. The goal is to determine the unknown parameters $\thetaveca$ that characterise the amplitude
of the decay $\ampl(\xvec|\thetaveca)$. Given a set of decay candidates characterised by a vector of observables $\xvec_i$ ($1\leq i \leq N$) obtained in an experiment,
an unbinned maximum-likelihood fit is performed to infer the model parameters $\thetaveca$. The negative logarithm of the likelihood, $-2\loglh$, minimised in the fit, is of the form
\begin{equation}
\label{eq:nll}
  -2\loglh = -2\sum_{i=1}^N \ln F(\xvec_i|\thetaveca),
\end{equation}
where $N$ is the size of sample being fit, $F(\xvec|\thetaveca)$ is the normalised probability density of the decay that depends on the model $\ampl(\xvec|\thetaveca)$, with a normalisation term $I$, the acceptance $\epsilon(\xvec)$ and the normalised probability density function for the background events $B(\xvec)$,
\begin{equation}
  F(\xvec|\thetaveca) = \frac{f_{\rm sig}\epsilon(\xvec)|\ampl(\xvec|\thetaveca)|^2}{I} + f_{\rm bkg}B(\xvec).
  \label{eq:decay_density}
\end{equation}
$f_{\rm sig}$ and $f_{\rm bkg}$ are the relative fractions of signal and background contributions, respectively.
Another instrumental effect that needs to be taken into account in the fit, particularly if the amplitude contains narrow resonant states, is the finite resolution of the kinematic observables $\xvec$. This effect is beyond the scope of this paper and is not considered. 

The contribution of background events is typically obtained by analysing the distribution of selection variables, $\mvec$.
In the simplest case, $\mvec$ is a single variable that is taken to be the combined invariant mass
of the final state particles, which typically peaks at the mass of the parent particle for the signal events, and is distributed
more uniformly for the background. However, other parameters of the event can also be included in the background selection.
Alternatively, instead of treating the background contribution explicitly as shown in Eq.~(\ref{eq:decay_density}), one can also assign
to each candidate $i$ in the data sample $\xvec_i$ a weight $w_i$, such that the background contribution is statistically subtracted.
This procedure will be discussed in more detail in Section~\ref{sec:background}.

While the amplitude $\ampl(\xvec|\thetaveca)$ is driven by the model of the decay dynamics and is the primary object under study,
the experimental effects of background and non-uniform acceptance exist as ``nuisance'' objects that, nevertheless, have to be
modelled accurately for correct interpretation of the results.
Their description, especially in the case of the multidimensional kinematic phase space of the decay, often
presents a major difficulty in an analysis. Below we review several conventional techniques employed in amplitude analyses to deal with effects of background and non-uniform acceptance.

\subsection{Treatment of non-uniform acceptance}

Non-uniform acceptance is usually handled either explicitly, using a parametric or non-parametric model of the decay
density as shown in Eq.~(\ref{eq:decay_density}), or in an implicit way, by including its effect in the normalisation term of the likelihood. In the latter case, the scattered data from simulation can be directly used, and no functional representation of the acceptance is required.

To demonstrate the implicit approach, let us consider Eq.~(\ref{eq:nll}) that is being minimised in
the unbinned fit (the background contribution has been omitted for simplicity):
\begin{equation}
\begin{split}
-2\ln\mathcal{L} = & -2\sum\limits_{i=1}^{N}\ln \left(\frac{|\ampl(\xvec_i|\thetaveca)|^2\epsilon(\xvec_i)}{I}\right)\\
                 = & -2\sum\limits_{i=1}^{N}\ln |\ampl(\xvec_i|\thetaveca)|^2 -2\sum\limits_{i=1}^{N}\ln\epsilon(\xvec_i) + 2N\ln\left(\frac{V}{M}\right)\\
 		   & + 2N\ln\left(\sum\limits_{j=1}^{M}|\ampl(\yvec_j|\thetaveca)|^2 \epsilon(\yvec_j)\right).
\end{split}
\end{equation}
Here the normalisation term $I$ is calculated by taking the mean of the values of the density
function on a uniformly distributed sample $\yvec_j$ ($1\leq j\leq M$) in a volume $V$. The constant terms that do not
depend on the parameters of the model $\thetaveca$ can be omitted, which leads to the following expression:
\begin{equation}
    -2\ln\mathcal{L} =
    -2\sum\limits_{i=1}^{N}\ln |\ampl(\xvec_i|\thetaveca)|^2 +
    2N\ln\left(\sum\limits_{j=1}^{M}|\ampl(\yvec_j|\thetaveca)|^2\epsilon(\yvec_j)\right).
\end{equation}
The second term in the above equation can be seen as the sum of $|\ampl|^2$ calculated over the sample $\yvec_j$ distributed
uniformly over the decay phase space, where each event $j$ enters with the weight $\epsilon(\yvec_j)$.
The weight of each event can also be interpreted as the probability of that event passing the detector acceptance.
Such an interpretation hints at a way to prepare the normalisation sample:
one has to generate the decays uniformly in the decay phase space, and then simulate the reconstruction
and selection of the events. The retained events will serve as a normalisation sample for the likelihood and
no further corrections to the acceptance are required. In a real analysis, however, additional weighting of the
normalisation sample may be needed to account for the imperfections in the simulated sample.

With this approach, no explicit parameterisation of the acceptance distribution is needed, and therefore it is often
used in amplitude analyses with more degrees of freedom than the two-dimensional Dalitz plot, such as $\Lb\to\jpsi p\Km$~\cite{LHCb-PAPER-2015-029}, $\Bp\to\jpsi\phi\Kp$~\cite{LHCb-PAPER-2016-018}, or $\Dz\to\Km\pip\pip\pim$~\cite{LHCb-PAPER-2017-040} decays where the amplitudes are described in a five- or 
six-dimensional phase space. This method is, however, statistically sub-optimal, since it does not
exploit the fact that the acceptance distribution can be assumed to be at least locally smooth. As a result, these analyses usually require simulation data sizes several times larger than the real data samples.

Explicit modelling of the acceptance function is more typically used in two-di\-men\-sional Dalitz-plot
analyses. The techniques often used are two-dimensional polynomials
with the parameters obtained from fitting a simulated data sample~\cite{LHCb-PAPER-2014-070}, two-dimensional histograms smoothed
with cubic splines~\cite{LHCb-PAPER-2014-036, Aaij:2019jaq}, and kernel density estimation~\cite{Poluektov:2014rxa, LHCb-PAPER-2016-061, Aaij:2020ypa}.
Analyses in more than two dimensions do sometimes also use an explicit acceptance parameterisation, albeit
with assumptions on the factorisation of the acceptance in some variables~\cite{LHCb-PAPER-2015-029} to reduce the dimensionality.

\subsection{Treatment of background contributions}

\label{sec:background}

As in the case of acceptance, backgrounds can also be treated in the amplitude analyses either in an explicit or implicit fashion. The implicit inclusion of the background into the likelihood fit
can be performed using the {\it sPlot} technique~\cite{Pivk:2004ty, Xie:2009rka}, where each event is
assigned a weight that depends on the value of the selection variables $\mvec$.
These weights are positive in the signal-dominated regions of the selection variables and negative in the background-dominated regions, and as such the contribution of the background events can be statistically subtracted from the likelihood.

This procedure does not require explicit parameterisation of the background density, however, it suffers some
drawbacks. Firstly, it assumes that the kinematic observables $\xvec$ are uncorrelated with the
selection variables $\mvec$, which as will be demonstrated in Section~\ref{sec:interpolation}, is an assumption that in general
is not well motivated. The presence of correlations will thus introduce bias in the fit results, especially if the background level
is large. Secondly, since the procedure does not make any assumptions on the functional
form of the background density, it results in larger statistical uncertainties on the results than when a reasonable functional form is assumed.

\section{Illustrative simulated samples}
\label{sec:toymc}

For the purposes of illustration, the $\Ds\to\Kp\pim\pip$ decay is considered in this paper.
As this is a three-body decay with scalar particles in
both the initial and final states, its dynamics are fully characterised by two kinematic variables.
This decay is also convenient as all the final states of the decay are charged tracks, which makes the selection
easier to be implemented in a simplified Monte-Carlo simulation framework. There are no
identical particles in the final state, avoiding any need to deal with Bose symmetrisation of the kinematic phase space.
Finally, being a singly Cabibbo-suppressed decay, it is interesting from a physics point of view, since it can exhibit significant \CP violation.

Here we choose to parametrise the phase space in terms of the
two square Dalitz-plot observables $\xvec=(m', \theta')$~\cite{Aubert:2005sk}, defined as
\begin{equation}
  \begin{split}
    m' = & \frac{1}{\pi}\arccos\left(2\frac{m(\Kp\pim)-m_{K\pi}^{\rm min}}{m_{K\pi}^{\rm max}-m_{K\pi}^{\rm min}} - 1\right),  \\
    \theta' = & \frac{1}{\pi}\theta(\Kp\pim),
  \end{split}
\end{equation}
where $m(\Kp\pim)$ is the invariant mass of the $\Kp\pim$ combination, $m_{K\pi}^{\rm min} = M_{K}+M_{\pi}$ and
$m_{K\pi}^{\rm max} = M_B - M_{\pi}$ are the minimum and maximum values of $m(\Kp\pim)$ variations,
$M_{K}$ and $M_{\pi}$ are the masses of $K$ and $\pi$ mesons, respectively~\cite{Patrignani:2016xqp},
and $\theta(\Kp\pim)$ is the helicity angle of the $\Kp\pim$ combination (the angle between the $\Kp\pim$ resonance and the lone hadron, in the resonance rest frame). While the square Dalitz plot was proposed for
analyses of $B$ mesons in order to give more weight to the interference regions between different two-body combinations that predominantly occur in the corners of the conventional Dalitz plot, in our case it is
used purely to avoid complications related to the curved boundaries of the conventional Dalitz plot.

Although the three-body Dalitz-plot analysis is the simplest case for the techniques considered here, they scale well with
dimensionality of the kinematic phase space, and can be applied for more complicated amplitude analyses. Similarly, the exact
definition of the kinematic phase space (conventional or square Dalitz plots, or various representations for four-body
kinematics) is not a limitation for the techniques under study.

\subsection{Efficiency distribution for $\Ds\to\Kp\pim\pip$}

\label{sec:toymc_acc}

The sample of $\Ds\to\Kp\pim\pip$ decays is simulated using a simplified Monte-Carlo technique, where only kinematic properties of the
initial and final state particles are considered. The production of $\Ds$ mesons and reconstruction of decay products is inspired by the
conditions of \lhcb experiment~\cite{Alves:2008zz}, however the numerical values of the parameters used in the simulation
are largely arbitrary and differ from those at \lhcb.

For the initial $\Ds$ mesons, the transverse momentum $p_T$ (the component of momentum perpendicular to the $\hat{z}$--axis, which in
the case of a proton-proton collider corresponds to the direction of the beams) is generated according to an exponential distribution
with a mean of $1$\gev.\footnote{Natural units with $c=\hbar=1$ are used in this paper}
The angle $\theta_{T}$, which is the angle between the direction of $\Ds$ momentum and the $\hat{z}$--axis, is generated such
that the pseudorapidity, $\eta = -\ln\tan(\theta_{T}/2)$, is distributed uniformly in the range from 2 to 5 (which roughly corresponds to the fiducial volume of LHCb). Assuming that the $\Ds$ decay is spherically symmetric, the momenta of the final state products are
generated such that they are uniform in the square Dalitz-plot variables $m'$ and $\theta'$.
These momenta in the \Ds rest frame are then boosted to the laboratory frame.

To simulate the selection of \Ds candidates in the experiment, only the candidates that satisfy certain kinematic criteria are retained:
the total momentum $p$ and the transverse momentum $\pt$ of each track are required to exceed
$3.0$ and $0.4$\gev, respectively; the \pt of at least one of the final state tracks is required to be greater than $1.0$\gev; the $\pt$
of the \Ds candidate is required to be greater than $2.0$\gev; and the sum of transverse momenta of the three tracks is required to
exceed $3.0$\gev.

\begin{figure}
  \centering
  \includegraphics[width=0.495\textwidth]{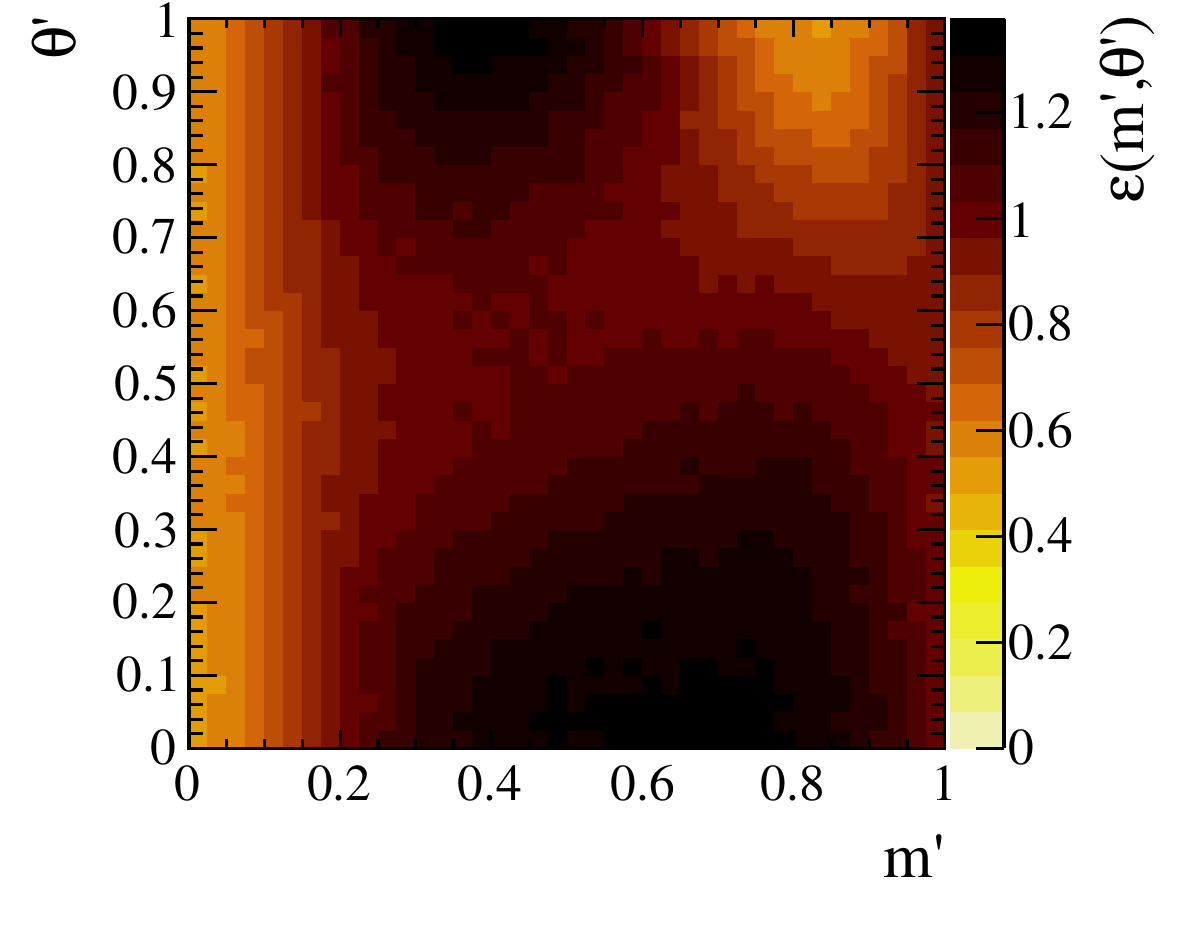}
  \caption{
  Relative efficiency variation over the square Dalitz-plot variables for $\Ds\to\Kp\pim\pip$ sample obtained from high-statistics Monte-Carlo simulation. The distribution is normalised such that the average efficiency equals $1$.
	 }
  \label{fig:eff_sim}
\end{figure}

The square Dalitz-plot distribution for the retained candidates is shown in Fig.~\ref{fig:eff_sim}.
Since we are only interested in relative variations of the efficiency,
it is normalised such that its average over the full kinematic phase space equals $1$. The same applies to
other generated and estimated two-dimensional efficiency and background distributions presented in this paper.
The plot in Fig.~\ref{fig:eff_sim} uses a high-statistics sample of $4\times 10^6$ events after selection, where a smaller sample of $10^5$ selected candidates is used in the examples presented elsewhere in this paper to estimate this distribution.

\subsection{Combinatorial background density for $\Ds\to\Kp\pim\pip$}
\label{sec:toymc_bkg}

The simulated combinatorial background contribution to the $\Ds\to\Kp\pim\pip$ decays contain purely random
combinations of three tracks, as well as the combinations of $\rhomeson(770)^0\to \pip\pim$ or $\kaon^*(892)^0\to K^+\pi^-$ with a $\Kp$ or a $\pip$,
respectively. The kaon and pion tracks, as well as the $\Kstarz$ and $\rhoz$ resonances, are generated uniformly in
pseudorapidity, $\eta$, and with an exponential $\pt$
distribution (with a mean \pt of $0.3$\gev, $0.6$\gev and $2.0$\gev for pions, kaons and resonances, respectively).
The fractions of $\rhoz$ and $\Kstarz$ in the combinations before applying selection cuts are set to be $20$\% and $10$\%, respectively.
The invariant masses of the $\rhoz$ and $\Kstarz$ resonances are generated according to the relativistic Breit-Wigner distribution, with masses and
widths equal to their world-average values~\cite{Patrignani:2016xqp}, and their decay products are generated assuming that the resonances
are unpolarised (\ie they are isotropic in the resonance rest frame and are uncorrelated with the third track).

In a second stage, the three charged tracks are combined to form the $\Ds$ candidate. Only the tracks with \pt greater than
$0.4$\gev and total momentum greater than $3$\gev are used to make the candidates. Finally, a kinematic fit is performed which adjusts the
momenta of the final state tracks in such a way that the invariant mass of the three-body combination is coincident with the world-average \Ds mass $M_{\Ds}=1.97\gev$~\cite{Patrignani:2016xqp}.
The three-dimensional distribution of the invariant mass, $m_D \equiv m(\Kp\pim\pip)$, of the three particles before the kinematic fit,
and the square Dalitz-plot variables $m'$ and $\theta'$ after the kinematic fit, are used to extract the density of the background
events in the signal region.

\begin{figure}



  \includegraphics[width=0.33\textwidth]{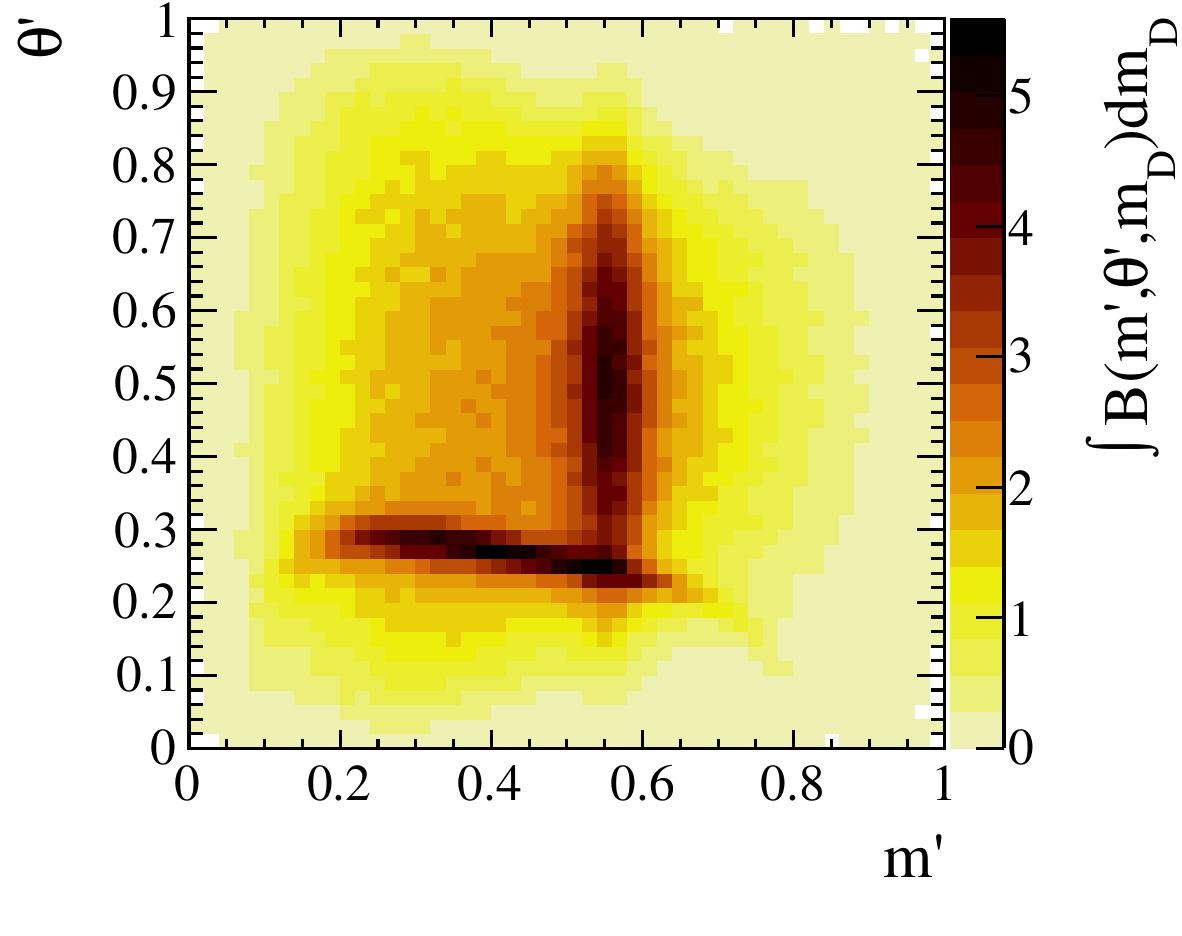}
  \put(-55, 100){\colorbox{white}{\small (a)}}
  \includegraphics[width=0.33\textwidth]{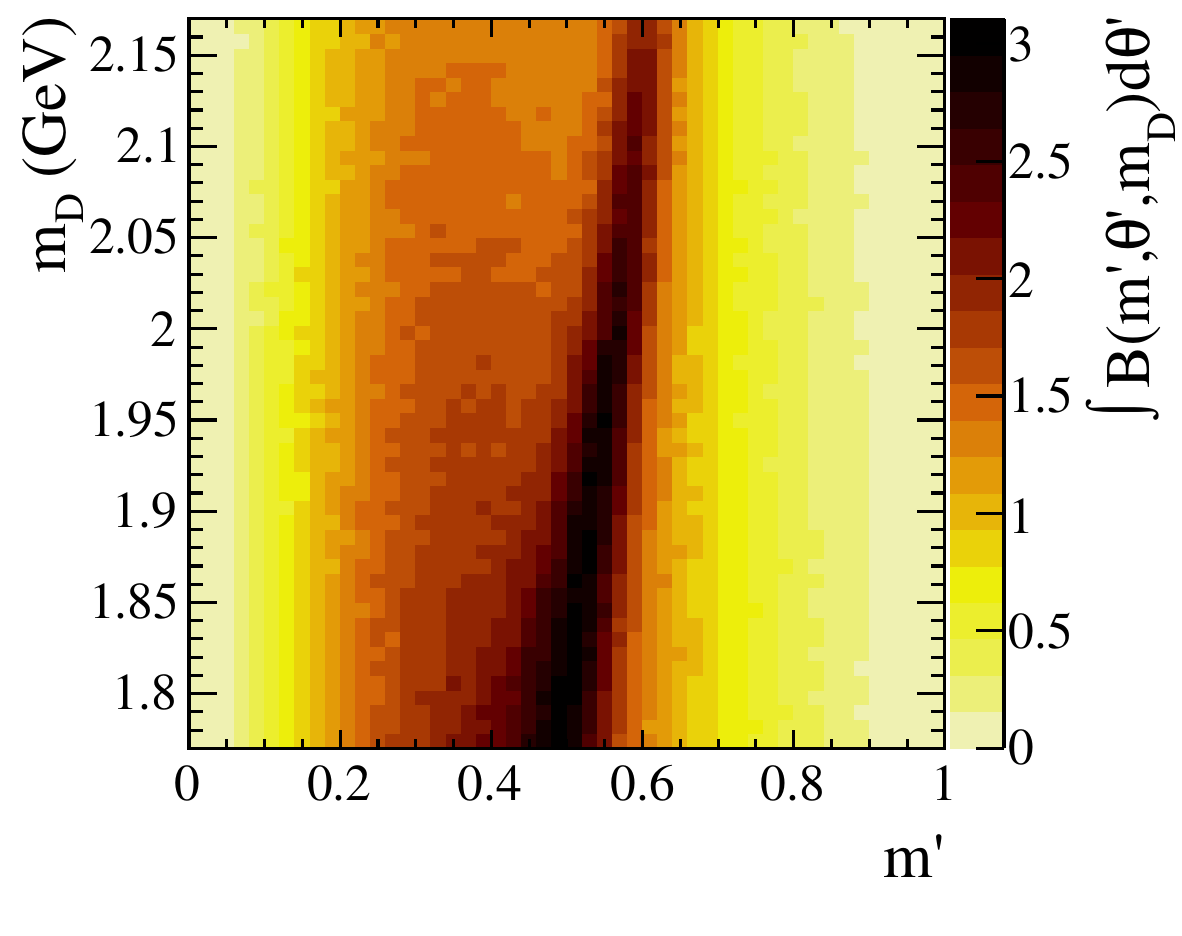}
  \put(-55, 100){\colorbox{white}{\small (b)}}
  \includegraphics[width=0.33\textwidth]{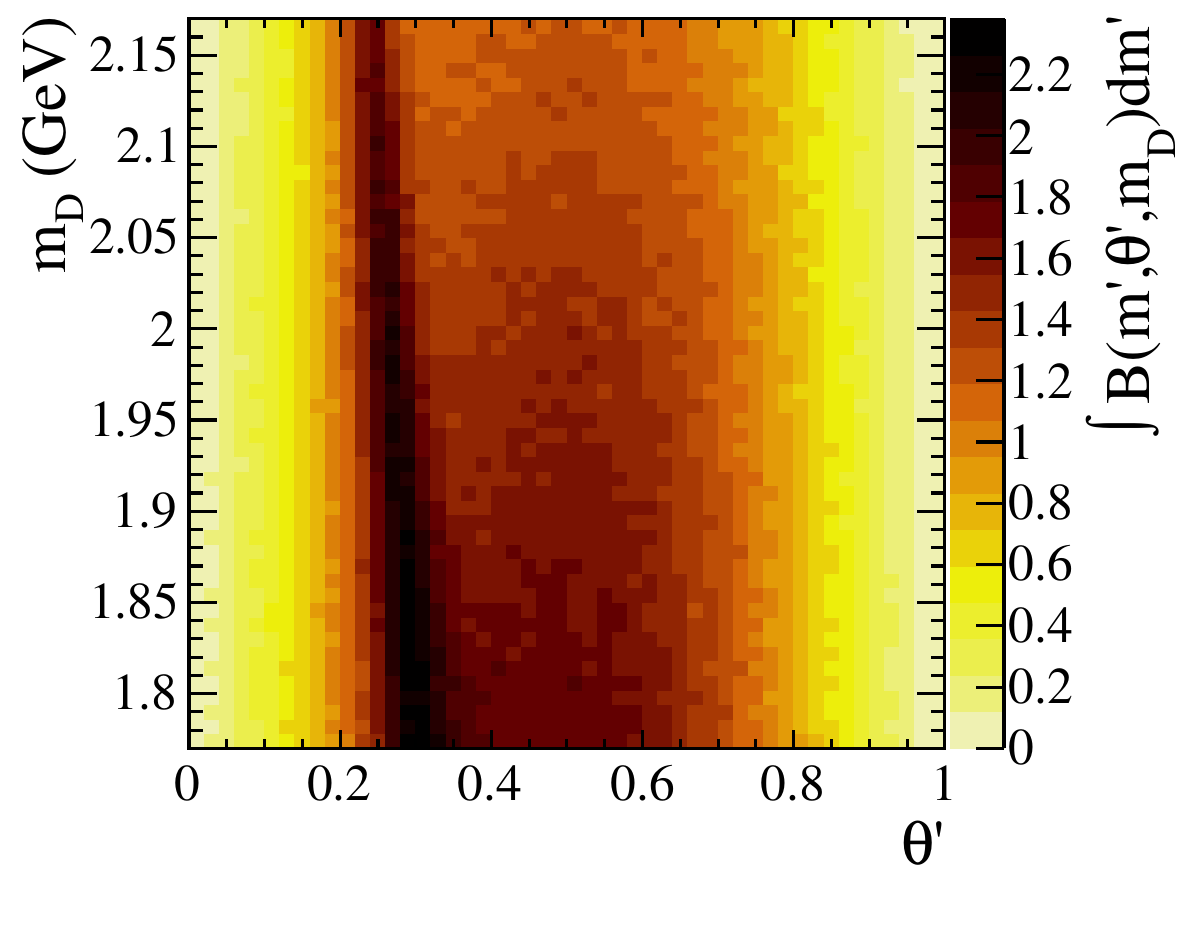}
  \put(-55, 100){\colorbox{white}{\small (c)}}

  \includegraphics[width=0.33\textwidth]{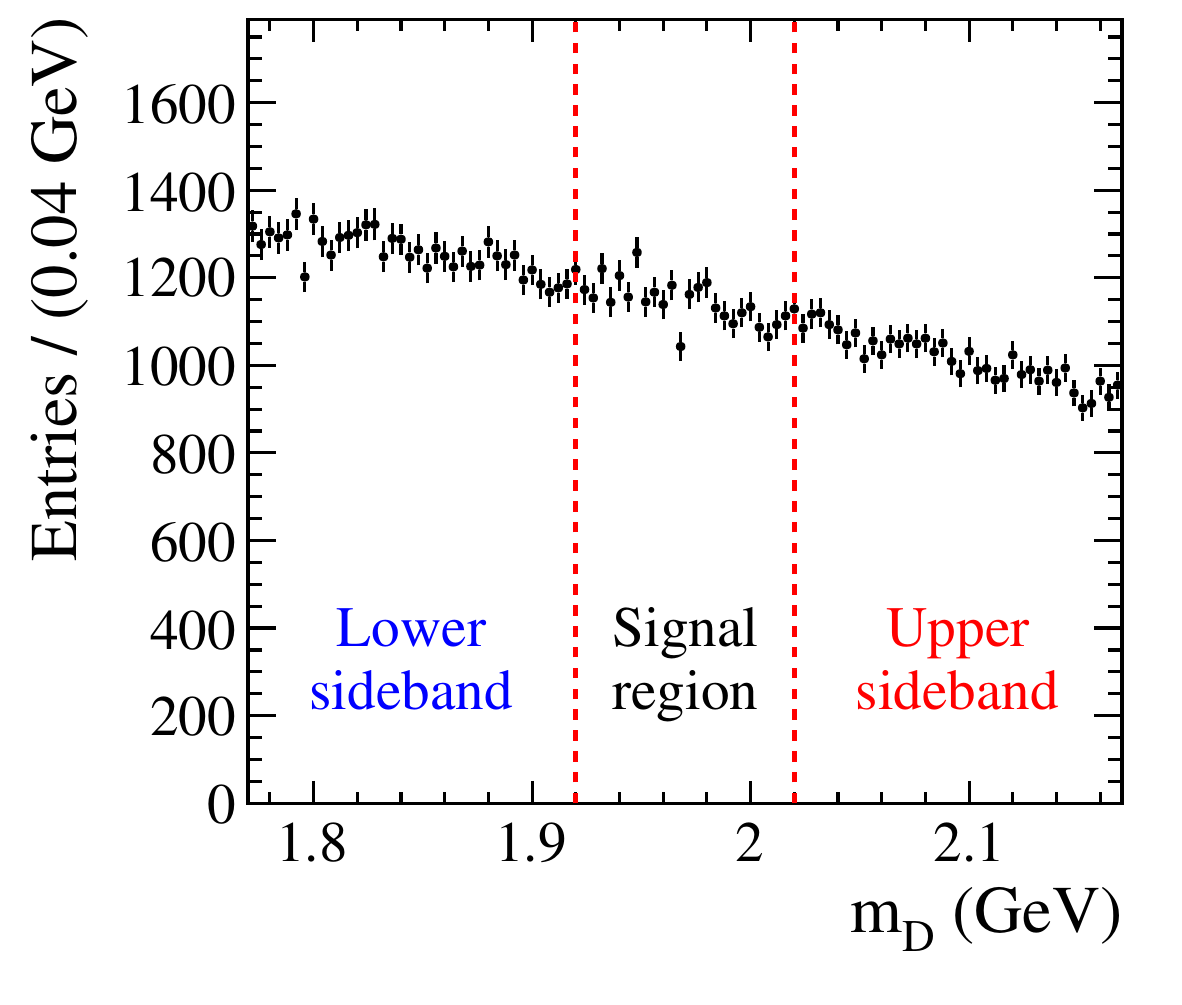}
  \put(-30, 105){\colorbox{white}{\small (d)}}
  \includegraphics[width=0.33\textwidth]{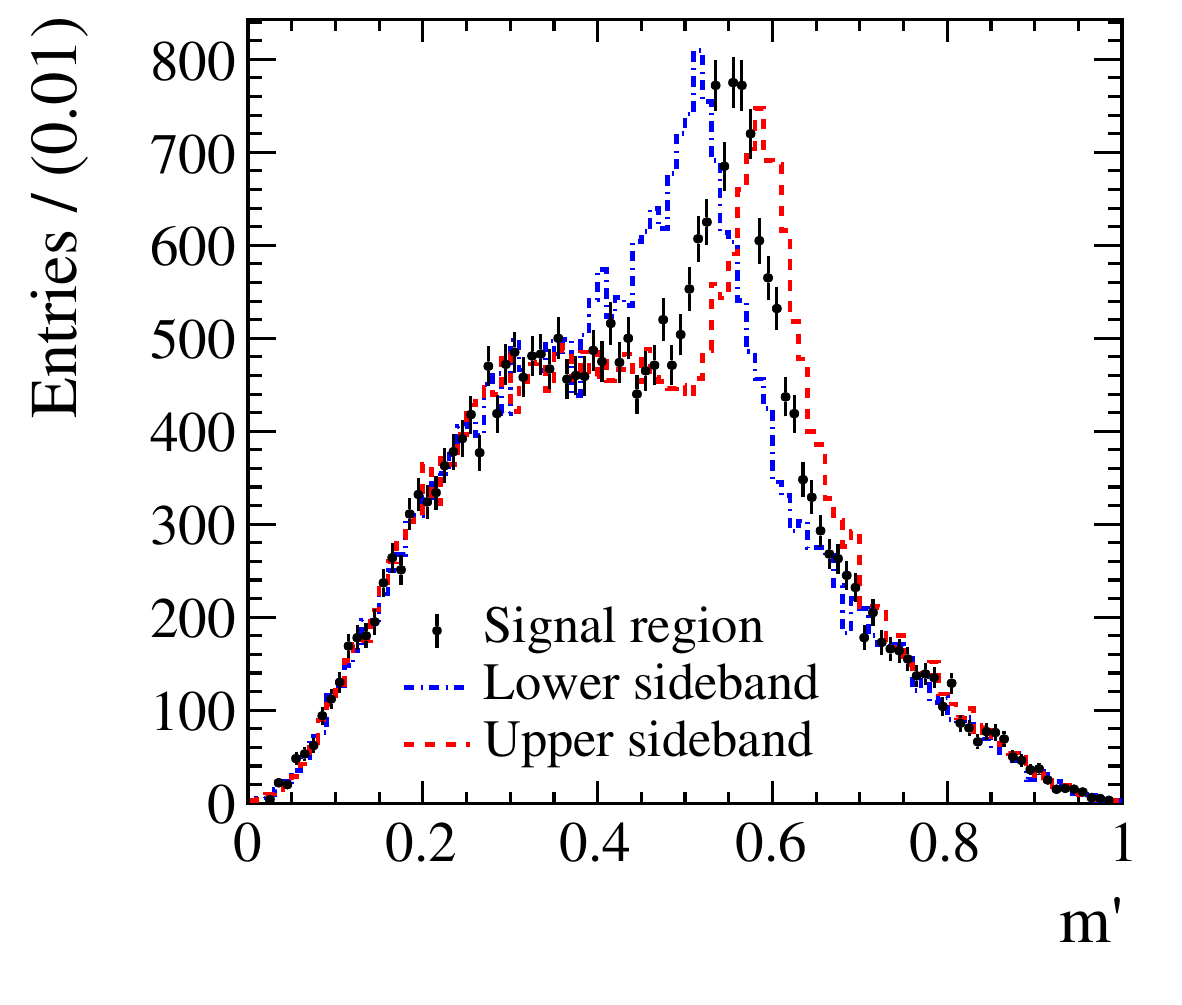}
  \put(-30, 105){\colorbox{white}{\small (e)}}
  \includegraphics[width=0.33\textwidth]{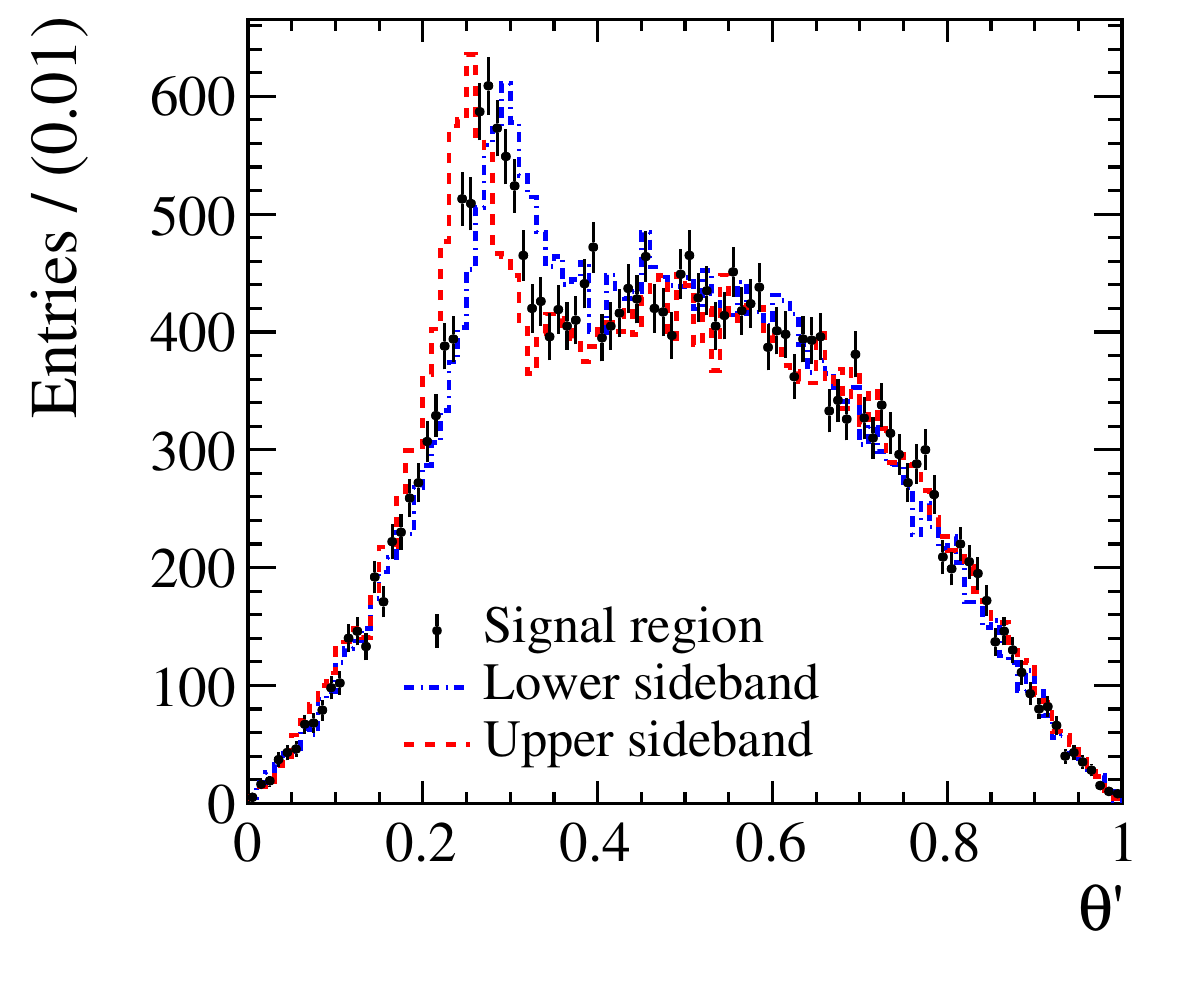}
  \put(-30, 105){\colorbox{white}{\small (f)}}

  \caption{Simulated combinatorial background to $\Ds\to\Kp\pim\pip$ decay. Two-dimensional
           (a) $m'$ vs. $\theta'$
           (b) $m'$ vs. $m_D$, and
           (c) $\theta'$ vs. $m_D$ projections, one-dimensional projections onto
           (d) $m_D$ variable with the definition of signal and sideband regions,
           and projections of signal and sideband regions onto
           (e) $m'$ and (f) $\theta'$ variables.
           Two-dimensional distributions are normalised such that the average density is equal to one.
          }
  \label{fig:bkg_sidebands}
\end{figure}

The distributions of each variable with the definition of signal and sideband regions are shown in Fig.~\ref{fig:bkg_sidebands}.
To clearly show the features of the background density, the two-dimensional projections in Fig.~\ref{fig:bkg_sidebands}(a,b,c) are
obtained with the high-statistics Monte-Carlo sample of $4\times 10^6$ events satisfying the selection requirements.
The examples of background density estimation in this paper, as well as the one-dimensional projections in Fig.~\ref{fig:bkg_sidebands}(d,e,f)
use the smaller sample of $10^5$ candidates. The following examples assume that
the background density is estimated only using the sideband regions of the $m_D$ distribution, defined as
$1.77 < m_D < M_{\Ds}-0.2\gev$ (lower sideband) and $2.17 > m_D > M_{\Ds}+0.2\gev$ (upper sideband), where the aim is to
estimate the density in the signal region $|m_D - M_{\Ds}|<0.2\gev$ (see Fig.~\ref{fig:bkg_sidebands}(d)).

As seen in Fig.~\ref{fig:bkg_sidebands}(e,f), background densities in
the signal and sideband regions are clearly different, with the positions of the peaking structures due to
$\Kstarz$ and $\rhoz$ resonances shifted in the sidebands with respect to the signal region. In this particular case,
it is purely explained by the kinematic fit procedure applied to the background sample. In a real analysis, this can also
be caused by dependence of background production properties on the selection variable, $m_D$.

\section{Conventional acceptance parameterisations}
\label{sec:acc-comp}

In this section, we present the estimation of the acceptance variation over the amplitude fit variables using conventional methods that involve use of Legendre polynomials and cubic splines.
These methods use histograms to estimate the local density of events before and after a selection requirements, and then interpolate the efficiency value between the bin centres of the histogram.

Numerous analyses use multidimensional orthogonal Legendre polynomials\cite{LHCb-PAPER-2014-070, LHCB-PAPER-2015-051, LHCb-PAPER-2016-025, LHCb-PAPER-2018-029} to parameterise the variation of the acceptance.
In $d$ dimensions these consist of the product of $d$ $n_d$-order polynomials, where $n_d$ can be different for each dimension, that describe the form of the acceptance.
As such, the number of free parameters in this method are $\mathcal{O}(n_1 \times n_2 \times ... \times n_{d})$, exhibiting the power-law growth in the number of dimensions.
The coefficients of these polynomials are then extracted in a maximum-likelihood fit.
As the optimal order of each of these polynomials is \emph{a priori} unknown, often cross-validation and/or regularisation is used to prevent overfitting.

Another method commonly used is interpolation between the histogram bin centres via cubic splines~\cite{LHCb-PAPER-2014-036, LHCB-PAPER-2017-033, LHCB-PAPER-2018-034}.
Here, the scale of the variation over the phase-space is determined by the initial histogram bin size, and therefore to avoid overfitting the size and location of bins are optimised before the spline function is calculated.
Unlike in the case for the Legendre polynomials, no free parameters exist for the spline functions (except for the values at the bin centres).
Therefore these are particularly susceptible to over-fitting, as they simply ``connect'' the points.

\begin{figure}
  \centering
  \includegraphics[width=0.34\textwidth]{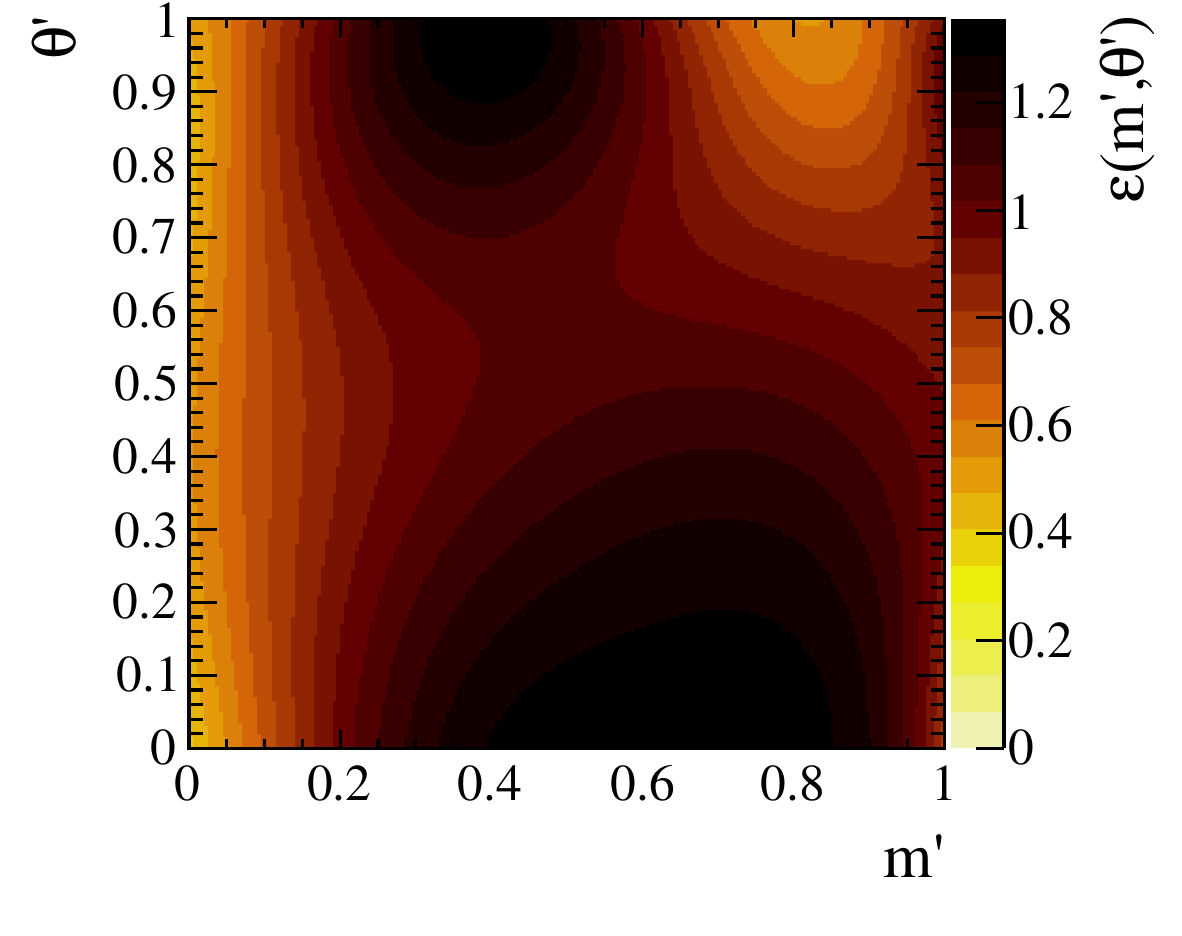}
  \put(-54, 35){\colorbox{white}{\small (a)}}
  \includegraphics[width=0.32\textwidth]{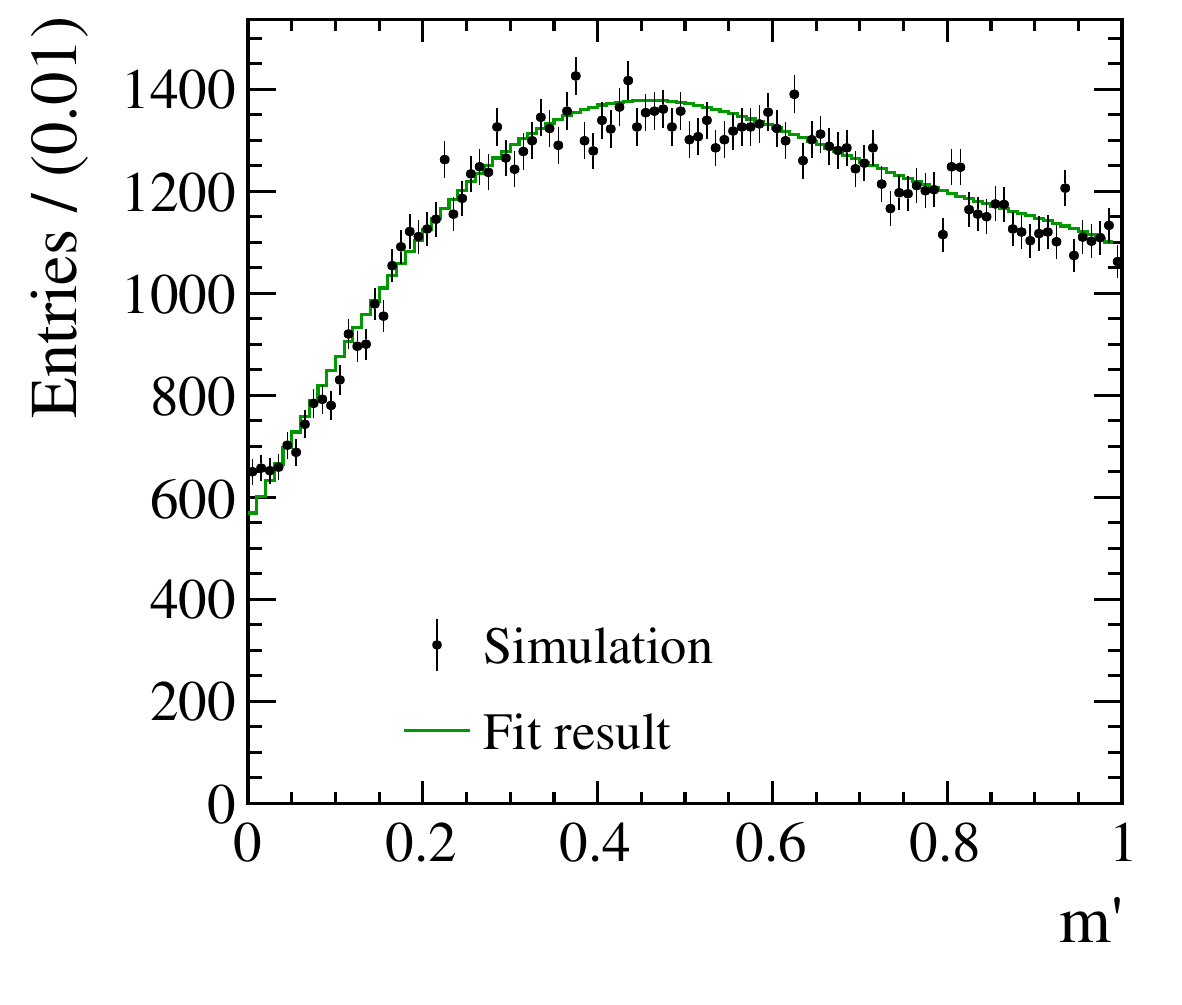}
  \put(-32, 35){\colorbox{white}{\small (b)}}
  \includegraphics[width=0.32\textwidth]{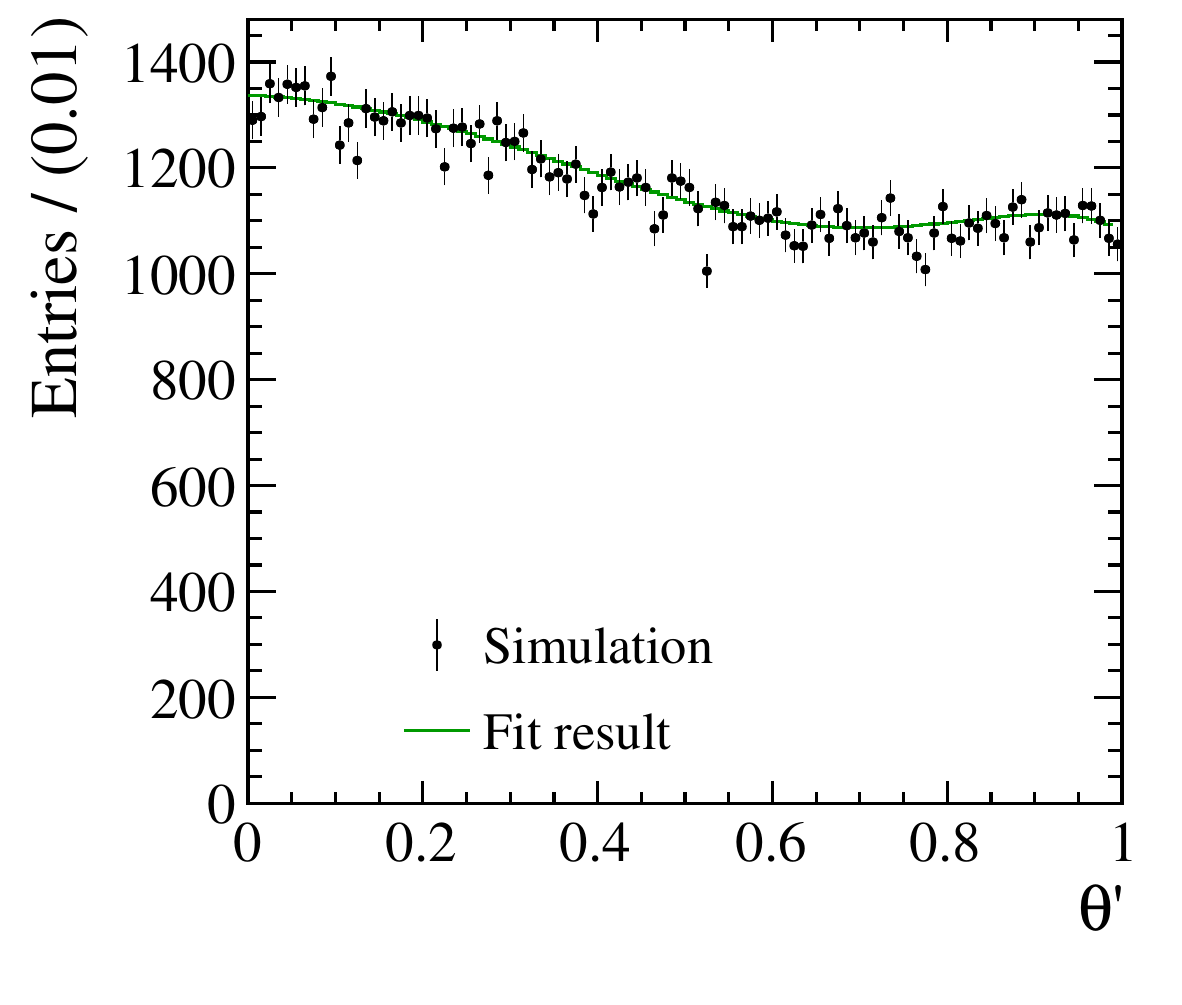}
  \put(-32, 35){\colorbox{white}{\small (c)}}
  \\

  \includegraphics[width=0.34\textwidth]{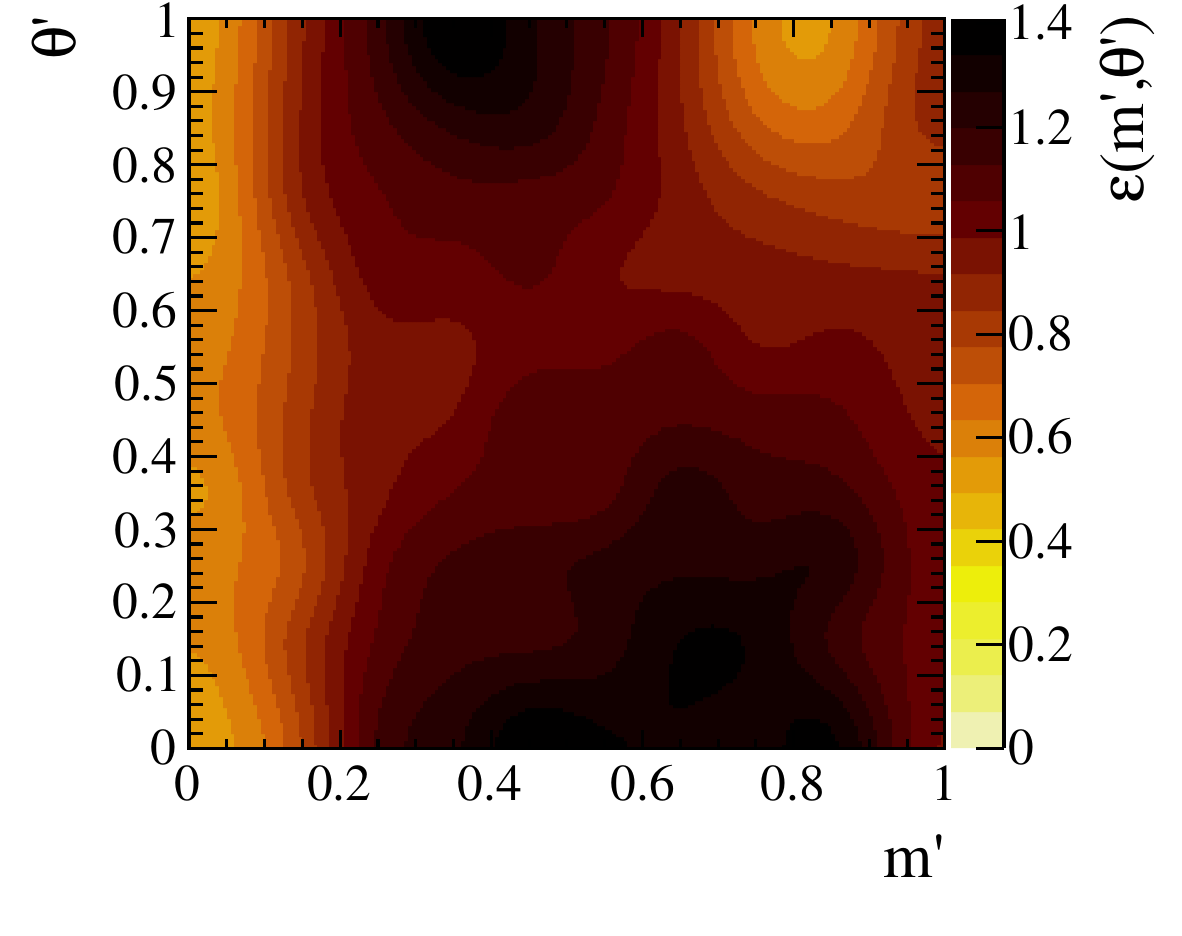}
  \put(-54, 35){\colorbox{white}{\small (d)}}
  \includegraphics[width=0.32\textwidth]{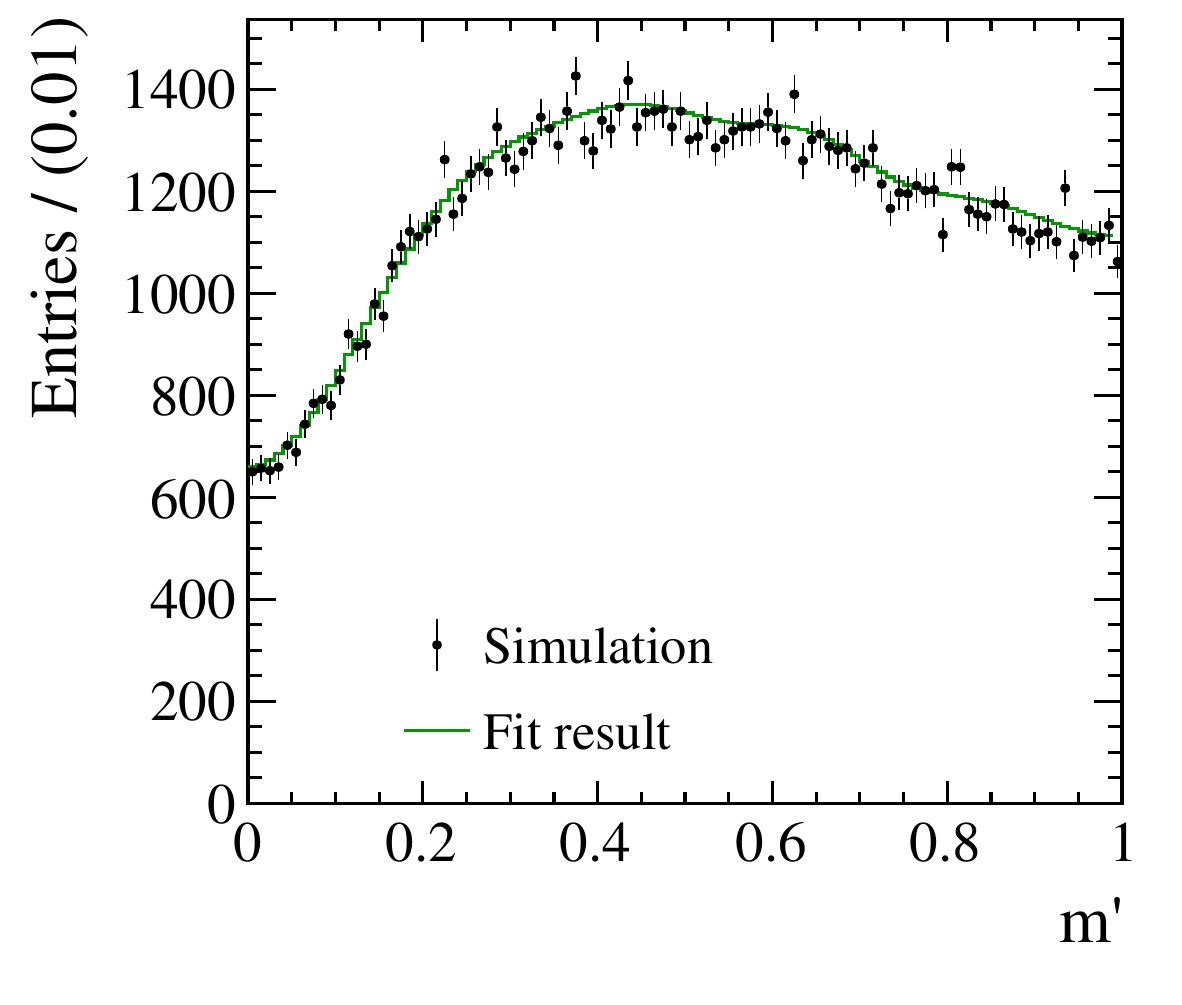}
  \put(-32, 35){\colorbox{white}{\small (e)}}
  \includegraphics[width=0.32\textwidth]{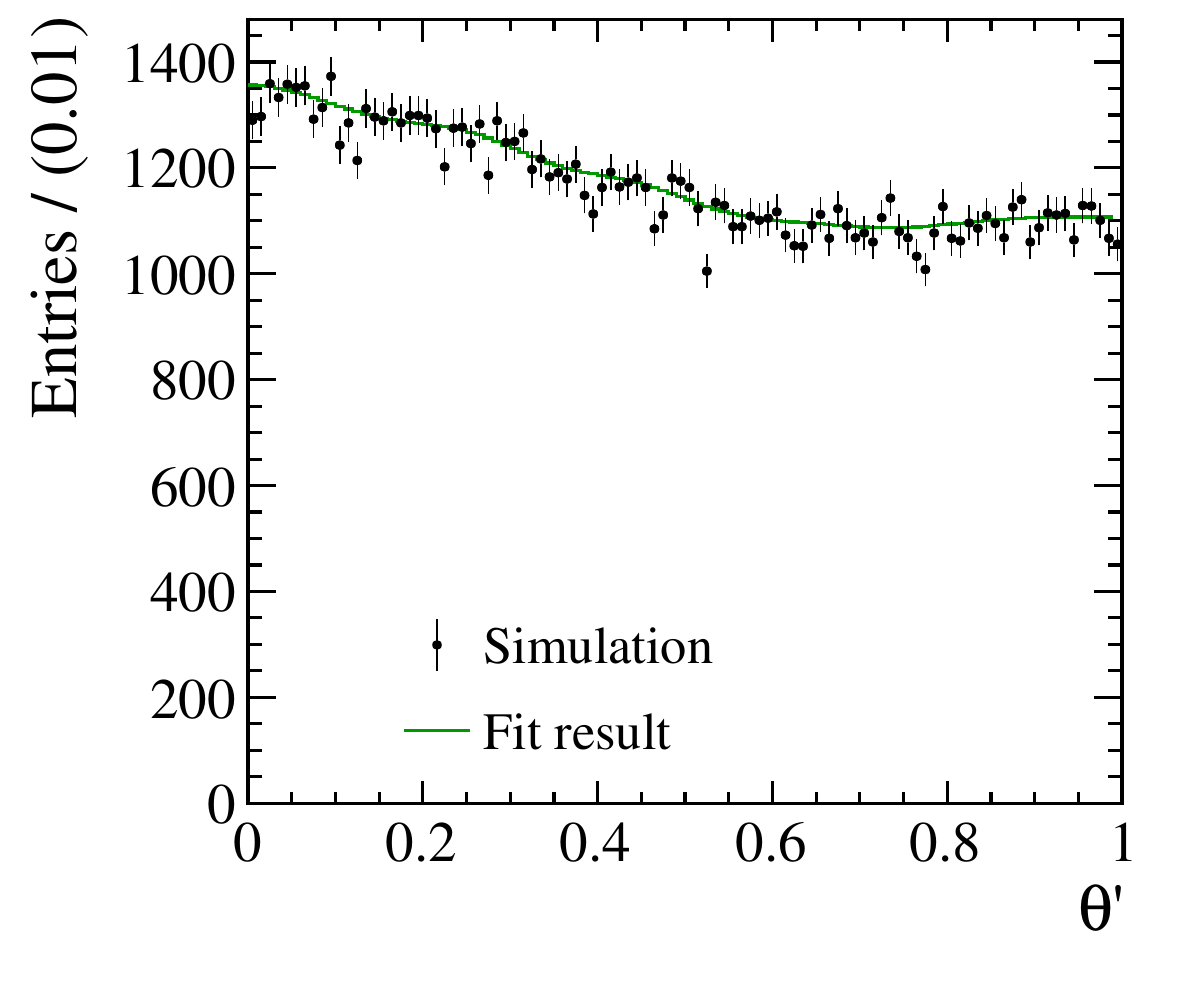}
  \put(-32, 35){\colorbox{white}{\small (f)}}
  \caption{Estimate of the acceptance variation using (top) Legendre polynomial and (bottom) cubic spline model over (left to right) two-dimensional $m'$ vs. $\theta'$ variables, $m'$ and $\theta'$ projections. }
  \label{fig:eff_comparison}
\end{figure}

In Figure~\ref{fig:eff_comparison}, we show the results of these two conventional approaches~\cite{numpy, scipy}.
For the Legendre polynomial fit, additive $L_1$ and $L_2$ regularisation terms were included into the likelihood
to reduce overfitting and improve likelihood fit stability~\cite{zou2005regularization}:
\begin{equation}
  -2\log\mathcal{L}_{\rm reg} = -2\log\mathcal{L} + L_1 + L_2
\end{equation}
with
\begin{equation}
  L_1 = \lambda_1\sum_{i,j}|c_{ij}|
\end{equation}
and
\begin{equation}
  L_2 = \lambda_2\sum_{i,j}c^2_{ij},
\end{equation}
where $c_{ij}$ are the coefficients of the polynomials and $\lambda_1$ and $\lambda_2$ are regularisation parameters.
A cross-validation procedure was performed to determine the optimal degree of the Legendre polynomials
(set to be equal in each dimension for simplicity), as well as the magnitude of each of the $L_1$ and $L_2$ terms.
This resulted in a polynomial degree of $n = 8$ in each dimension, an $L_1$ regularisation parameter $\lambda_{1} = 0.01$, and an $L_2$
regularisation parameter $\lambda_{2} = 0.1$. The reduced $\chi^2$ for this fit is calculated using an independent test set, and the number of effective degrees of freedom is calculated using bootstrap resampling, and yields a value of $\chi^2/{\text{nDoF}} = 2517/2429 = 1.04$.

For the fit with cubic splines, a $10\times10$ binning in $(m', \theta')$ is used, and is chosen \emph{ad hoc} such that the bin size is similar to the smallest structure size in these variables ($\sim 0.1$), but not so fine as to exhibit dramatic overfitting if these are compared to the underlying distribution. Here the reduced $\chi^2$ is calculated using the number of bins in the $\chi^2$ test ($50 \times 50$) minus the number of points fitted by the spline ($100$), and a value of $\chi^2/{\text{nDoF}} = 2460/2400 = 1.03$ is obtained.

\section{Gaussian processes}

\label{sec:gp}

A Gaussian process~\cite{rasmussen2006gaussian} is a statistical model which associates each point in the input space with a normally distributed random variable. The joint distribution of these random variables is also a normal distribution, yielding a closed-form expression for the model at any arbitrary point the space. Fortunately, to avoid having to fit for the parameters of an infinite number of normal distributions, Gaussian processes are completely determined by parameterising the covariance matrix using a covariance function. As such, it is possible to obtain model estimates in a large number of dimensions with relatively few parameters, which can then be used to extrapolate the behaviour of the model with reliable estimates of the uncertainty. Furthermore, these parameters can be robustly extracted directly from the data, which gives Gaussian processes an advantage over other ``non-parametric'' models, such as kernel density estimates or piece-wise spline interpolation. A pedagogical introduction to Gaussian processes can be found in Ref.~\cite{roberts2013gaussian}, and other applications in high-energy physics can be found in Refs.~\cite{Frate:2017mai, Bozson:2018asz}

Given some input data vector $\Xvec=(\xvec_1, \ldots, \xvec_N)$, of length $N$ (where the elements $\xvec_i$ can themselves be vectors of arbitrary dimension), the output, $y$, of a Gaussian process is defined as
\begin{equation}
\label{eq:GP}
y \sim \mathcal{N}\left(0, \Sigma(\Xvec; \thetavec)\right),
\end{equation}
where $\mathcal{N}$ is a multivariate normal distribution with zero mean, and the covariance matrix, $\Sigma(\Xvec; \thetavec)$, given by
\begin{equation}
    \Sigma(\Xvec; \thetavec) = \begin{bmatrix}
           k(\xvec_0, \xvec_0; \thetavec) & k(\xvec_1, \xvec_0; \thetavec) & \hdots &k(\xvec_N, \xvec_0; \thetavec) \\[0.3em]
           k(\xvec_0, \xvec_1; \thetavec) & k(\xvec_1, \xvec_1; \thetavec) & \hdots &k(\xvec_N, \xvec_1; \thetavec) \\[0.3em]
           \vdots & \vdots & \ddots & \vdots \\[0.3em]
           k(\xvec_0, \xvec_N; \thetavec) & k(\xvec_1, \xvec_N; \thetavec) & \hdots & k(\xvec_N, \xvec_N; \thetavec)
         \end{bmatrix}.
\end{equation}
Here $k(\xvec_i, \xvec_j; \thetavec)$ is a covariance function, with hyperparameter vector $\thetavec$, that is defined between any two input points, $\xvec_i, \xvec_j$, that are elements of $\Xvec$.
For a new input point $\xvec_*$, the conditional probability of predicting an output that is equal to the true unknown value $y_*$, given the previously observed outputs, $Y=(y_0, \ldots, y_N)$, follows a normal distribution,
\begin{equation}
P(y_*|Y) \sim \mathcal{N}(\Sigma_*\Sigma^{-1}Y^T, \Sigma_{**} - \Sigma_*\Sigma^{-1}\Sigma_*^T),
\end{equation}
where $\Sigma_* = [k(\xvec_*, \xvec_0; \thetavec), k(\xvec_*, \xvec_1; \thetavec), \ldots, k(\xvec_*, \xvec_N; \thetavec)]$, and $\Sigma_{**} = k(\xvec_*, \xvec_*; \thetavec)$.
The best estimate of the true value, $y_*$, is equal to the mean of the above probability distribution,
\begin{equation}
\hat{y}_* = \Sigma_*\Sigma^{-1}Y^T,
\end{equation}
and the uncertainty is the square-root of the variance,
\begin{equation}
{\rm Var}(y_*) = \Sigma_{**} - \Sigma_*\Sigma^{-1}\Sigma_*^T.
\end{equation}
The negative logarithm of the likelihood for this construction is given by
\begin{equation}
-2\log p(Y|\mathbf{\Xvec},\thetavec) = Y \Sigma^{-1} Y^T + \ln |\Sigma| + N \ln 2 \pi.
\label{eq:nll_gp}
\end{equation}
The vector of hyperparameters $\thetavec$ of the covariance function can then be inferred by minimising the negative logarithm of the likelihood (Eq.~\ref{eq:nll_gp}),
or obtained via marginalisation using suitable priors and Markov chain Monte-Carlo.

One such covariance function is the Mat\'{e}rn function,
\begin{equation}
k_{\nu}(d; \thetavec = \{ \sigma, \nu\} ) = \sigma^2 \frac{2^{1-\nu}}{\Gamma(\nu)} \left( \sqrt{2\nu} d \right)^{\nu} K_{\nu} \left( \sqrt{2\nu} d \right)
\end{equation}
where $d = ||(\xvec_j - \xvec_i)/\boldsymbol{\rho}||$ is the scaled distance between two points in the input space, $\Gamma$ is the gamma function, $K_{\nu}$ is the modified Bessel function of the second kind, $\boldsymbol{\rho}$ is a vector of length scales, $\nu$ is a non-negative parameter, and $\sigma$ controls the absolute magnitude of the covariance. The Mat\'{e}rn function is defined in terms of the distance between two points, rather than the location of each point, and therefore describes a stationary distribution. For half-integer values of $\nu$, this can be expressed as a product of an exponential function and a polynomial of order $p = \nu - 1/2$. The parameters $\nu$ can be thought of as controlling the smoothness of the function, and when $\nu \rightarrow \infty$, the Mat\'{e}rn function converges to the squared-exponential covariance function
\begin{equation}
\lim_{\nu \rightarrow \infty} k_v(d) = \sigma^2 \exp \left( -\frac{d^2}{2} \right).
\end{equation}

Here we use the Mat\'{e}rn function with $\nu = \frac{5}{2}$ throughout, as this provides a good balance between replicating and smoothing the observed structures, however the parameter values, and the best model choice in general, depends on the data in question. These are ideally selected using cross-validation, or a similar procedure, and can be considered as a source of systematic uncertainty on the final distribution.

In reality, one would also want to describe distributions that are non-stationary (\ie, where the mean in Eq.~(\ref{eq:GP}) is non-zero). However, due to the linearity of the model, this represents a simple subtraction of the mean function from the observed data, and therefore there is no loss in generality due to this description. Specific assumptions on the mean distribution will be discussed for the applications described in Sections~\ref{sec:gpAcceptance} and~\ref{sec:interpolation}.

\subsection{Acceptance parameterisation}

\label{sec:gpAcceptance}

The dataset described in Section~\ref{sec:toymc} parameterises the acceptance in terms of the square Dalitz-plot variables, $m'$ and $\theta'$. As the Gaussian process does not estimate the density directly (although there are modifications that would permit this~\cite{murray2009gaussian}), the density in $(m', \theta')$-space is estimated first by a uniformly binned histogram, with $50$ bins in each axis. The location of the centres of these bins are then the input points, $\Xvec$, defined above, where $\xvec_i = [m'_i, \theta'_i]$. As these were generated uniformly in $(m', \theta')$-space, the acceptance probability in each bin is simply the reciprocal of the bin content, which is the output, $Y$, of the Gaussian process. Therefore for each input $\xvec_i = [m'_i, \theta'_i]$ there is an associated output $y_i$. As the acceptance is positive definite everywhere, the resulting Gaussian process does not represent a stationary process, and as such, a mean function constant in $(m', \theta')$ is added to the Gaussian process to account for this scaling.

An advantage of the Gaussian process is that it is relatively robust to statistical fluctuations due to low sample sizes, as the uncertainty at each point is estimated directly. As such, the aforementioned binning can be considerably finer than in other methods, provided that the assumption of normally distributed uncertainties holds (processes where the likelihood is replaced with a Poisson distribution can also be used, however this is less computationally tractable than in the Gaussian case, as Poisson distributions are not closed under linear combination).

\begin{figure}


  \includegraphics[width=0.34\textwidth]{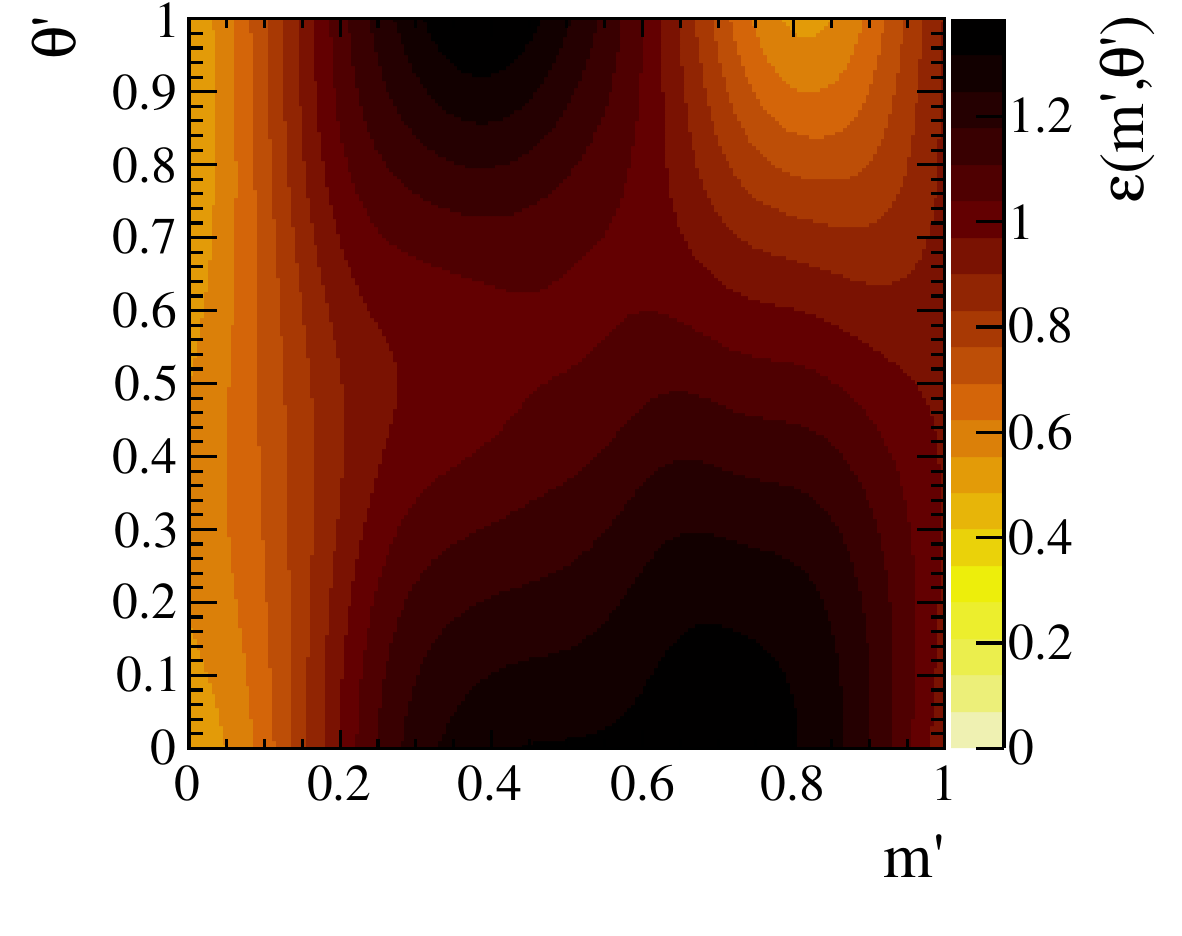}
  \put(-54, 35){\colorbox{white}{\small (a)}}
  \includegraphics[width=0.32\textwidth]{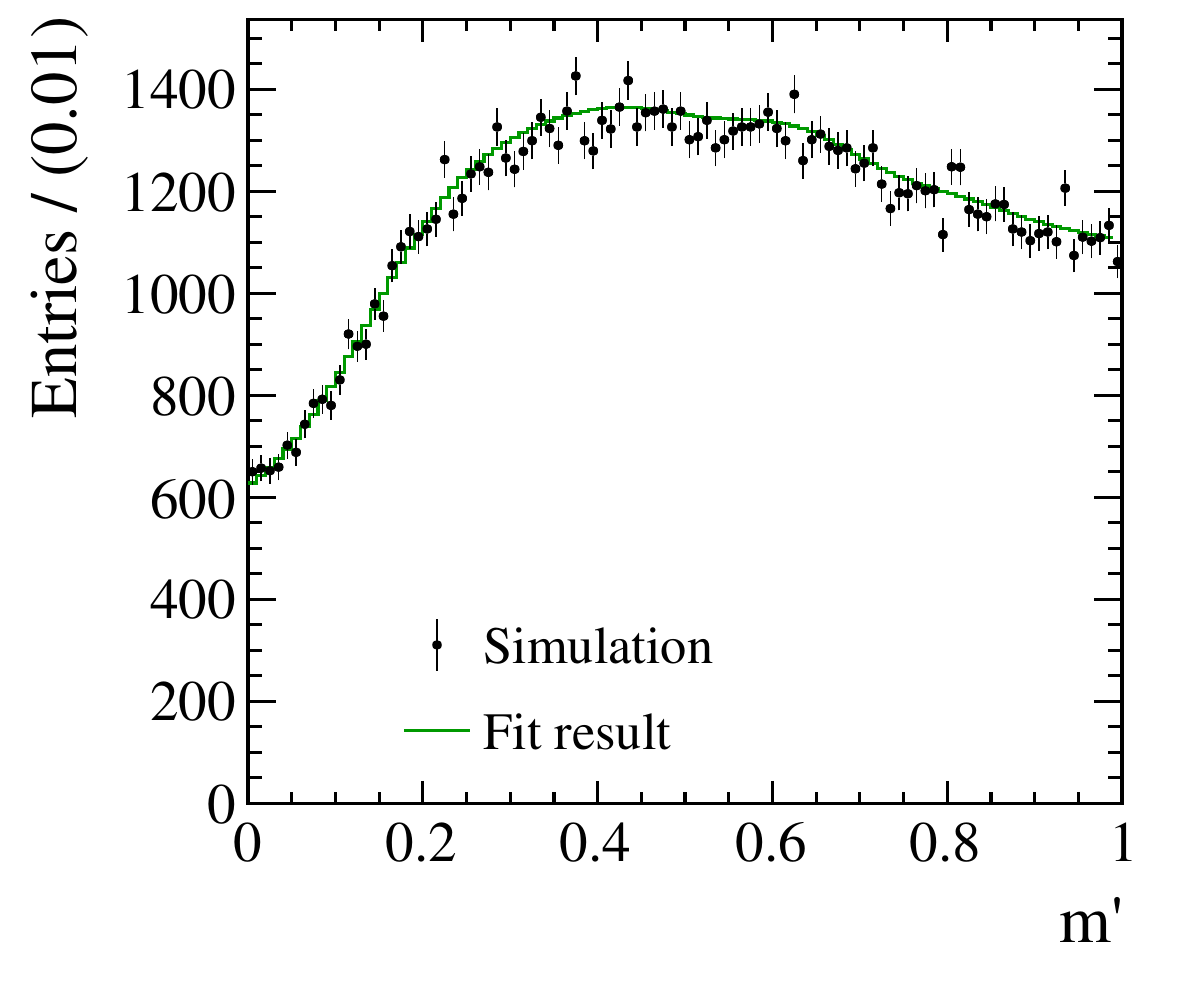}
  \put(-32, 35){\colorbox{white}{\small (b)}}
  \includegraphics[width=0.32\textwidth]{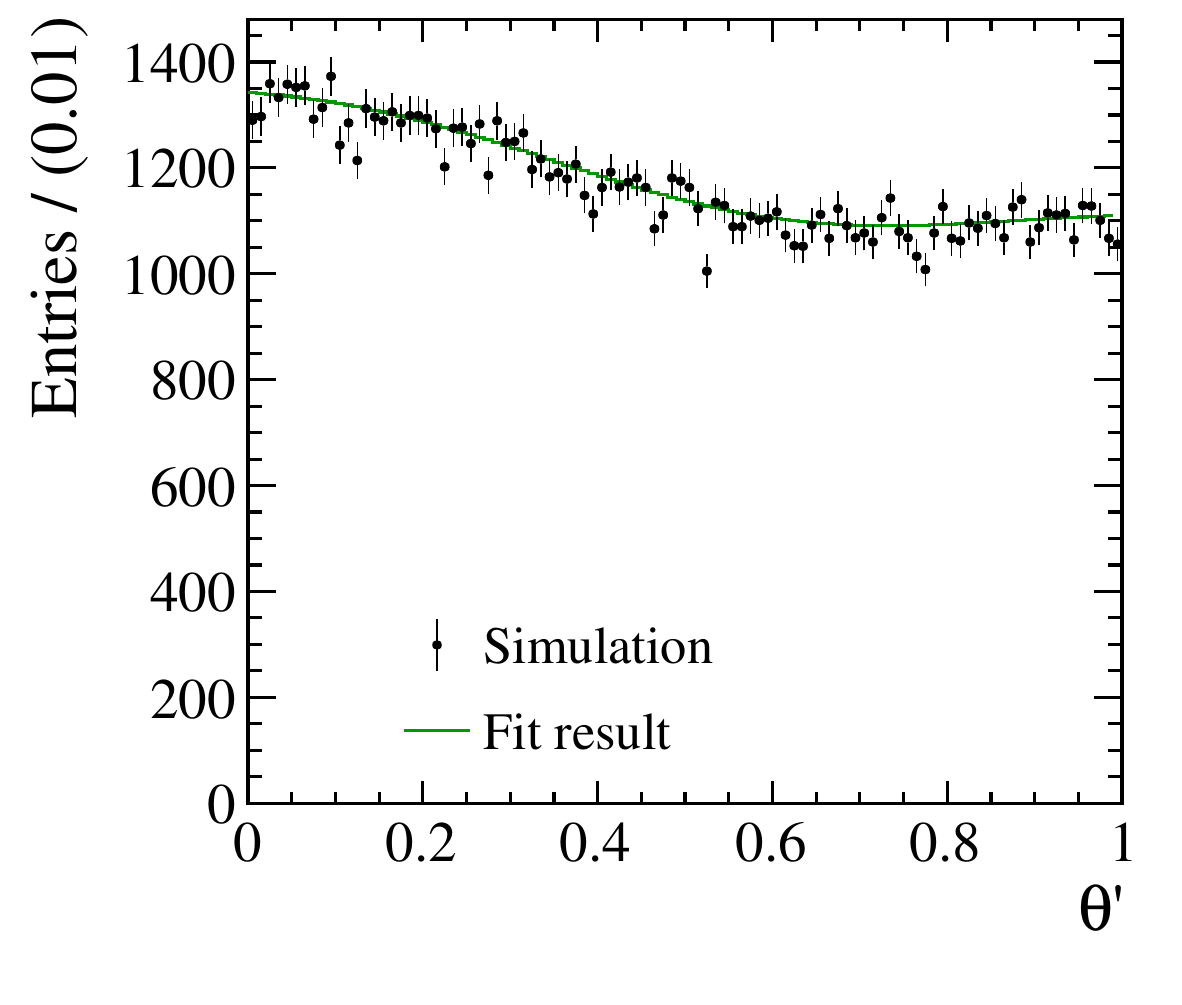}
  \put(-32, 35){\colorbox{white}{\small (c)}}

  \caption{Result of the density estimation of the simulated sample of $\Ds\to\Kp\pim\pip$ decays using Gaussian process method
          (a) in two square Dalitz-plot variables $m'$ and $\theta'$, and projections of the two-dimensional distribution
          onto (b) $m'$ and (c) $\theta'$ variables. }
  \label{fig:gp_eff_fit_2d}
\end{figure}

Here, Gaussian processes are implemented using the GPy package~\cite{gpy} (and the functionality described in this section and Section~\ref{gp_bkg} can be obtained with the package in Ref.~\cite{gp_package}), and the resulting acceptance model as a function of $m'$ and $\theta'$ can be seen in Figure~\ref{fig:gp_eff_fit_2d}. The parameters of this model are the overall scale of the Mat\'{e}rn function, $\sigma_{\rm GP}^2$, the overall scale of the constant mean function $\sigma_{\rm mean}^2$, the characteristic length scale over which points covary for each dimension, $\rho_{m'}$ and $\rho_{\theta'}$, and a term describing the additive Gaussian noise at each point, $\epsilon$. This fit was performed using a maximum-likelihood approach, and the resulting parameter values can be found in Table~\ref{tab:effGP}.

\begin{table}
  \caption{Parameters of the Gaussian process fit to the simulated $\Ds\to\Kp\pim\pip$ decays, with a Mat\'{e}rn kernel and a constant mean function.}
  \label{tab:effGP}
  \begin{center}
  \begin{tabular}{|l|c|}
    \hline
    Model parameter        & Value \\
    \hline
    $\sigma_{\rm GP}^2$    & $648$ events$^2$\\
    $\sigma_{\rm mean}^2$  & $547$ events$^2$ \\
    $\rho_{m'}$            & $0.4$ \\
    $\rho_{\theta'}$       & $1.0$ \\
    $\epsilon$             & $44.8$ events \\
    \hline
  \end{tabular}
  \end{center}
\end{table}

The $\chi^2$ per number of degrees of freedom of this acceptance model with respect to the data is evaluated using an independent dataset of simulated $\Ds\to\Km\pip\pim$ decays. Here, the effective number of degrees of freedom is calculated approximately as the number of bins in the $\chi^2$ test ($50 \times 50$), minus the number of model parameters ($5$), and a value of  $\chi^2/{\text{nDoF}} = 2471/2495 = 0.99$ is obtained. This indicates that the model reproduces the underlying distribution very well, where the smallness of the $\chi^2$ value compared to the values reported in Section~\ref{sec:acc-comp} is mostly due to the fact that this reproduction can be achieved with a small number of parameters.

As the only parameters that scale with dimensionality are the characteristic length scale of the Mat\'{e}rn function, the increase in the number of parameters is linear in the dimensionality. Furthermore, the time complexity of Gaussian processes is also linear in the dimensionality, making these very efficient in high dimensions compared to other parameterisations. Unfortunately however, the complexity is cubic in the input data size, due to the dependence on a matrix inversion, and therefore these do not scale well with large data sizes. Nevertheless, methods exist to mitigate this, such as the use of binned data in the strategies described here, or by selection of a small number of ``pseudo'' inputs~\cite{snelson2006sparse}.

\section{Density estimation with neural networks}

\label{sec:annde}

Multivariate techniques such as artificial neural networks (ANNs) or boosted decision trees provide
an alternative approach to parametrise multidimensional probability density from scattered
data~\cite{LIKAS2001167, Viaud:2016lwk}. 
The approach involving ANNs exploits a property of neural networks, 
where a ``feed forward'' network (when layers of neurons are arranged in a non-cyclical structure), 
with smooth activation functions can approximate any continuous function. 
Here, the parameters of the ANN are treated as free parameters in a maximum-likelihood fit to the unbinned data, 
performed by treating the negative logarithm of the likelihood as a custom loss function. 
Since this technique does not require binning the input data, and in general ANNs have successfully 
shown their ability for multivariate generalisation, we expect that density estimation approach using ANNs 
can become useful for multidimensional amplitude analyses. 
This approach is demonstrated below for the parameterisation
of the two-dimensional acceptance of the $\Ds\to\Kp\pim\pip$ decay, described in Section~\ref{sec:toymc}.


The outputs of the $n^{\rm th}$ neuron of the $l^{\rm th}$ layer in the ANN is given by
\begin{equation}
  a_{n,l+1} = f\left( \sum_m w_{nm,l} a_{m,l} + b_{n,l}\right),
\end{equation}
where $w_{nm,l}$ is the matrix of weights, $b_{n,l}$ is the vector of biases for $l^{\rm th}$ layer,
and $f(x)$ is a non-linear activation function. For the estimated density to be smooth,
it is convenient to use a smooth differentiable activation function such as a sigmoid function,
\begin{equation}
  f(x) = \frac{1}{1+e^{-x}}.
\end{equation}
In this structure, the first layer of neurons is the input layer, and accepts kinematic variables $\xvec$ as inputs, while the output neuron return a single scalar, the density estimate $P(\xvec)$

Density estimation can be performed by treating the weights and biases $\thetavec\equiv\{w_{nm,l}, b_{n,l}\}$ of the ANN
as free parameters, and minimising the negative logarithm of the likelihood,
\begin{equation}
  -2\ln\mathcal{L} = -2\sum_{i=1}^{N}\ln P(\xvec_i|\thetavec) + 2N \ln\left( \sum_{j=1}^{M} P(\yvec_i|\thetavec) \right),
  \label{eq:nn_nll}
\end{equation}
where $\xvec_i$ ($i=1\ldots N$) are data points, and $\yvec_i$ ($j=1\ldots M$) is a
uniformly distributed sample used for normalisation. 
The function~(\ref{eq:nn_nll}) is used as the loss function to train the ANN given the training sample $\xvec_i$.

As in many applications of machine learning techniques, special care needs to be taken to avoid overfitting, where the model configuration or parameters become too specialised to the training dataset, and therefore fail to generalise properly. In the case of density estimation with ANN, overfitting manifests itself as isolated peaks of the PDF around training data points. Regularisation techniques, where the likelihood is explicitly penalised to promote smoothness or sparsity, are thus essential to control overfitting.

It was found that regularisation which penalises large neuron weights (and therefore those that result in large gradients of the density function) works well in the typical cases when density is expected to be smooth.
Specifically, an $L_2$ regularisation term of the form
\begin{equation}
  L_2 = \lambda_2 \sum_{n,m,l} w^2_{nm, l}
\end{equation}
is added to the loss function~(\ref{eq:nn_nll}), where $\lambda_2$ is the regularisation parameter
that ultimately controls the smoothness of the PDF.

Additional constraints can be imposed by the choice of the input variables. For example, in the case of a decay process where angular observables are necessary to describe it, the periodicity of the density as a function of angles can be enforced by using two variables, $\sin\phi$ and $\cos\phi$, as the input variables for the ANN instead of a single angular variable $\phi$. Similarly, if the density is expected to be an even function of a kinematic observable (\eg cosine of a helicity angle), the square of this  might be a better choice as an input to the ANN. 

\begin{figure}


  \includegraphics[width=0.34\textwidth]{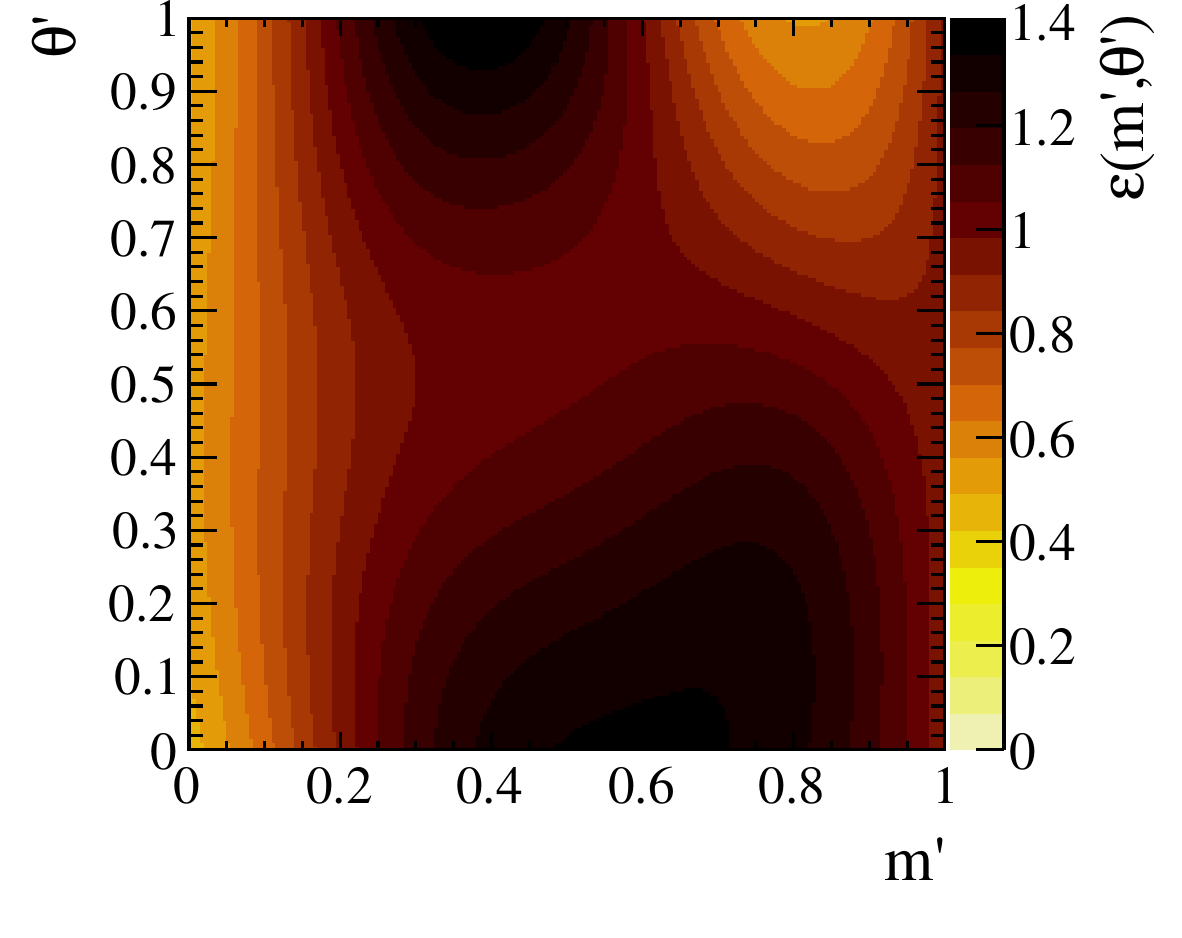}
  \put(-54, 35){\colorbox{white}{\small (a)}}
  \includegraphics[width=0.32\textwidth]{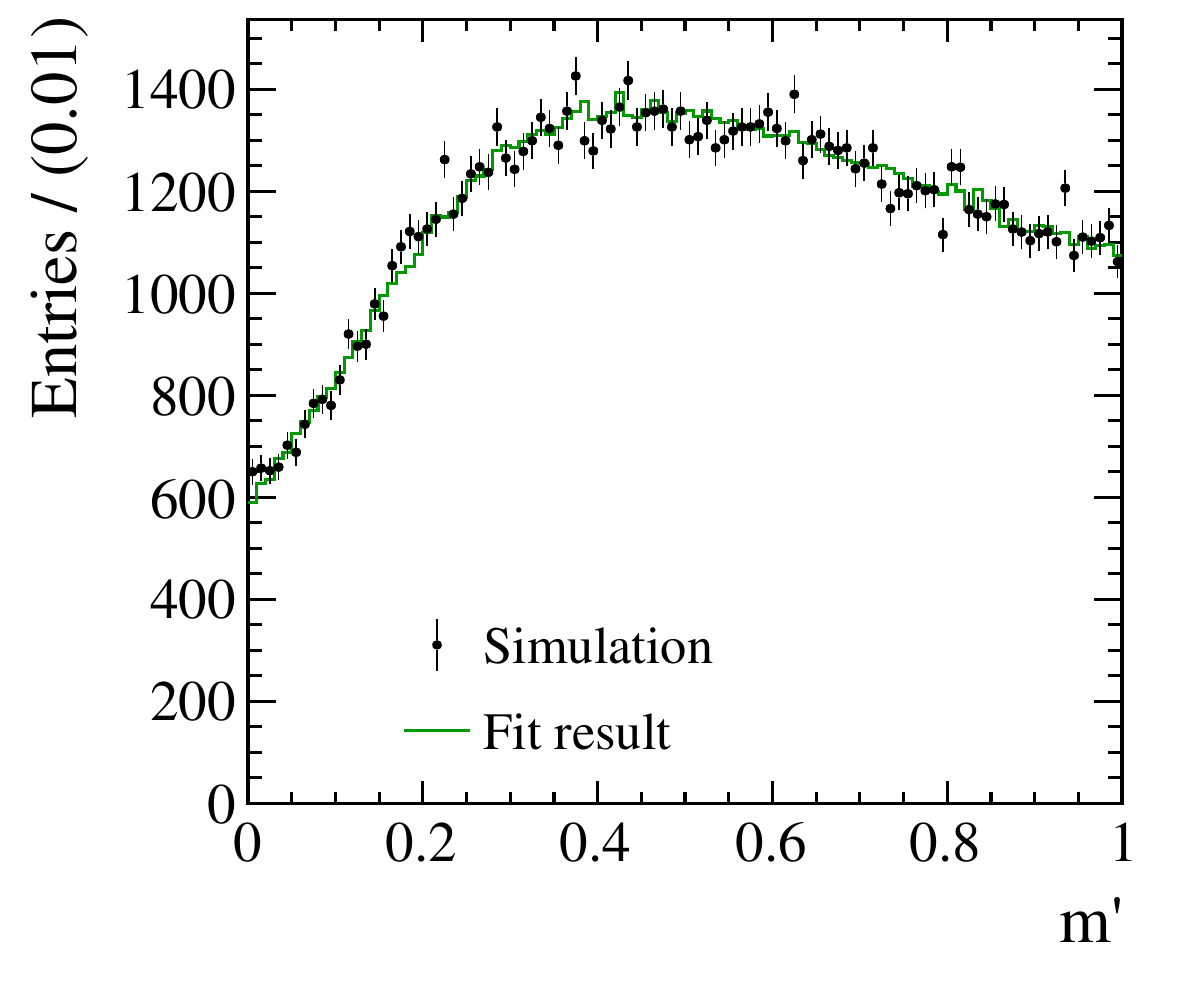}
  \put(-32, 35){\colorbox{white}{\small (b)}}
  \includegraphics[width=0.32\textwidth]{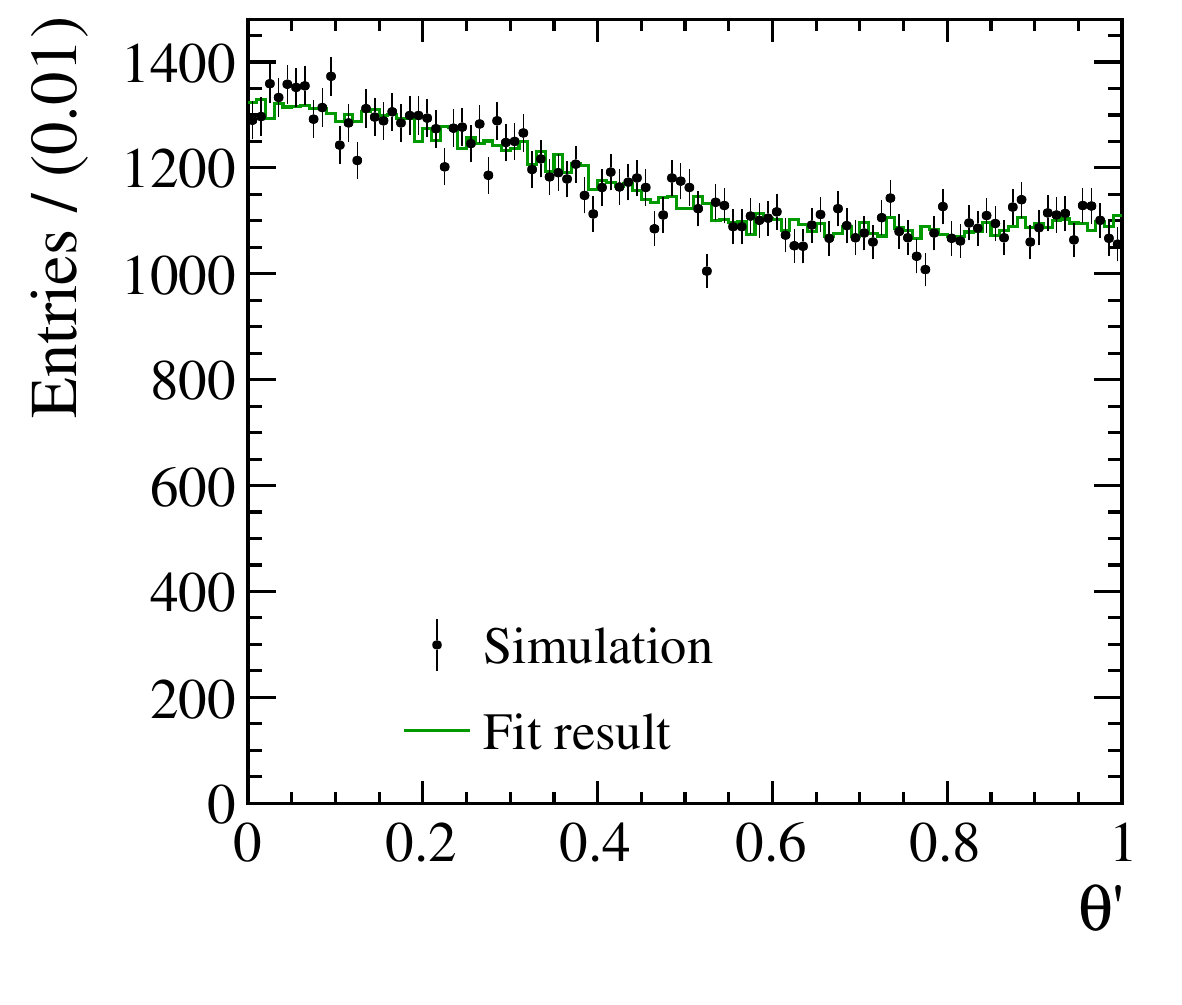}
  \put(-32, 35){\colorbox{white}{\small (c)}}

  \caption{Result of the density estimation of the simulated sample of $\Ds\to\Kp\pim\pip$ decays using an ANN
          (a) in the two-dimensional square Dalitz-plot variables $m'$ and $\theta'$, and projections of this distribution
          onto the (b) $m'$ and (c) $\theta'$ variables. }
  \label{fig:eff_fit_2d}
\end{figure}

The density estimation of the $\Ds\to\Kp\pim\pip$ decay acceptance is performed with an ANN consisting of four hidden
layers with the number of neurons, from the first to fourth layer, equalling 32, 64, 32, and 8. The regularisation parameter
$\lambda_2$ is chosen to be equal to 0.1. Normalisation is performed with $5\times 10^5$ events distributed
uniformly over the space of inputs, the square Dalitz plot. The likelihood minimisation is performed using the TensorFlow
framework~\cite{tensorflow2015-whitepaper} and the Adam optimiser~\cite{Kingma2014AdamAM} with learning rate of $10^{-3}$. The resulting estimate of the density after 30\,000 training epochs (passes through the data) is shown in Fig.~\ref{fig:eff_fit_2d}. The $\chi^2$ value obtained for the fit with these conditions for $50\times 50$ bins equals 2418.3. It is difficult to estimate the effective number of degrees of freedom for a training with regularisation. By varying the regularisation parameter one can obtain a broad range of $\chi^2$ values, both lower (which might indicate overtraining) and higher (indicating a bias in density estimation). Cross-validation can be used to obtain the optimal value for the $\lambda_2$ parameter and control overtraining. 

The code used to produce the results in this paper with ANNs is available at Ref.~\cite{annde}. This package uses the {\tt TensorFlowAnalysis} and {\tt AmpliTF} libraries~\cite{tfa, amplitf} for the implementation of the functions related to flavour physics analyses within {\tt TensorFlow} framework.

\section{Extrapolation of background density from sidebands}

\label{sec:interpolation}


As highlighted in Section~\ref{sec:background}, the conventional methods that either use sideband distributions or the {\it sPlot} technique
to determine the combinatorial background contribution can in general introduce systematic biases, since they ignore correlations between the amplitude fit
variables and the selection variables (such as the combined invariant mass of the final state particles $m_D$).
The bias can become  pronounced if only one of the sidebands can be used to estimate the background (\eg due to presence of specific peaking
backgrounds in the other sideband as often happens in $B$ meson decays). The
{\it sPlot} procedure also introduces additional statistical uncertainty compared to the parametric approach
due to the lack of any assumptions on the behaviour of the background.

To overcome these issues, one can add the selection variables to the background parameterisation.
For example, in the case of Dalitz-plot analysis, a 3D fit can be performed to obtain the probability density function $P(m', \theta', m_D)$,
which can then be used to extrapolate the desired combinatorial PDF $B(m', \theta')$ in the signal region.
We present in this section two such approaches, one using a Gaussian process~\cite{O'Hanlon:2289593}, and another using an ANN~\cite{Mathad:2646802},
to extrapolate the combinatorial background PDF from both the upper and lower sidebands of $m_D$ to the signal region.
To illustrate the performance of both of these approaches, simulated combinatorial background of $\Ds\to\Kp\pim\pip$ decay is used, as described in Section~\ref{sec:toymc}.

\subsection{Gaussian process background fit}
\label{gp_bkg}

In the Gaussian process method, the fit variables ($m', \theta'$) and selection variable ($m_D$) taken from the sideband
sample are first binned to obtain a local estimate of the density, and the parameters of the covariance function are then inferred by fitting the model using the location of the bin centres and their respective yields. As mentioned in Section~\ref{sec:gp}, the model is fairly robust to variations in the choice of the location and size of these bins, providing that they capture sufficient variation in the input variables.

Here, the $\Ds\to\Kp\pim\pip$ lower, $m_D \in [1.77, 1.92]$ \gev, and upper, $m_D \in [2.02, 2.17]$ \gev, $\Ds$ sideband described in Section~\ref{sec:toymc_bkg} are separated into three bins of $0.05$ \gev each, for a total of six bins in $m_D$. In each of these bins, the square Dalitz plot is separated into $20 \times 20$ uniform bins (with bin size $0.05 \times 0.05$), for a total of $6 \times 20 \times 20  = 2400$ inputs to the Gaussian process.

\begin{figure}
  \centering
  \includegraphics[width=0.8\textwidth]{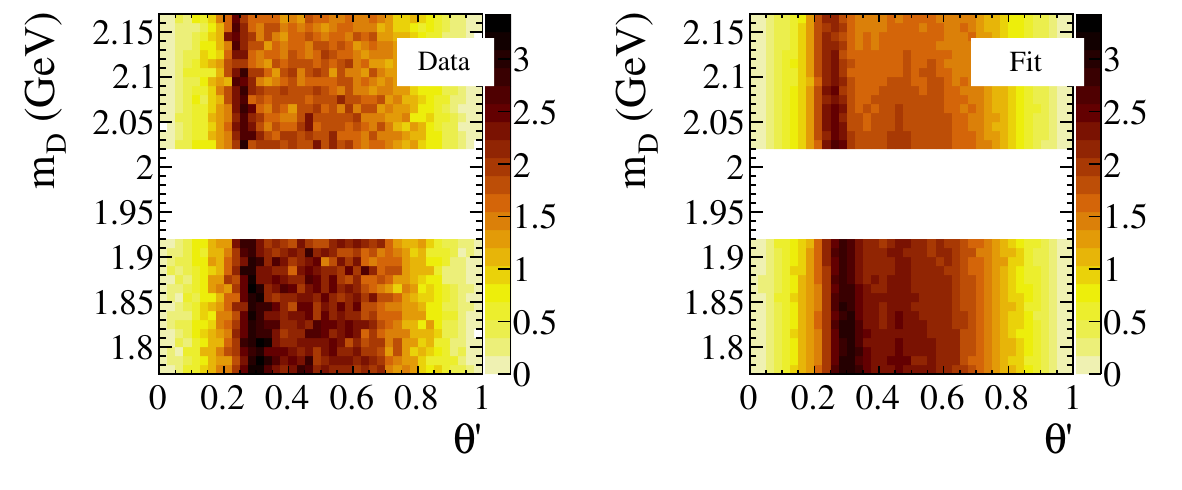}
  \includegraphics[width=0.8\textwidth]{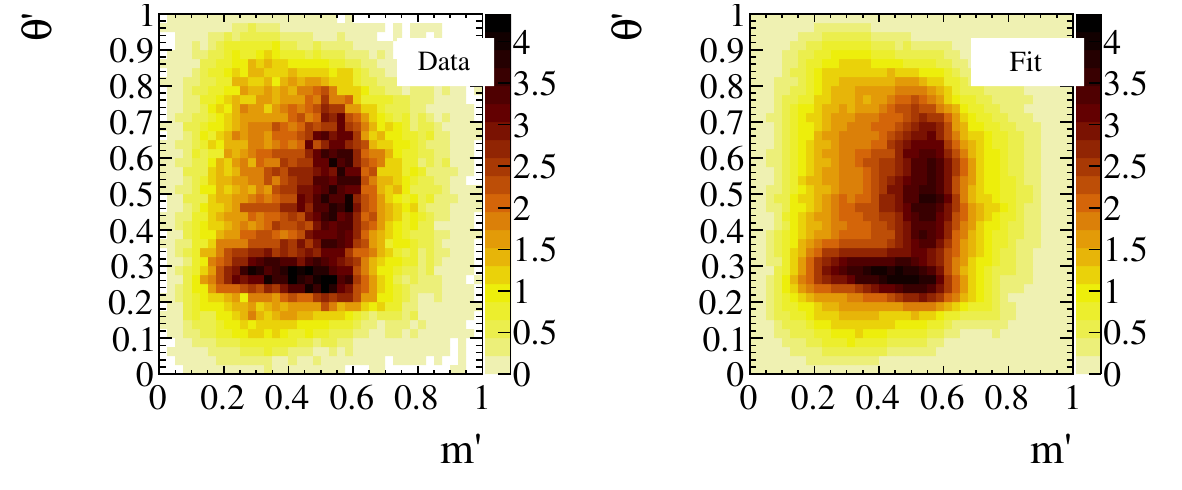}
  \includegraphics[width=0.8\textwidth]{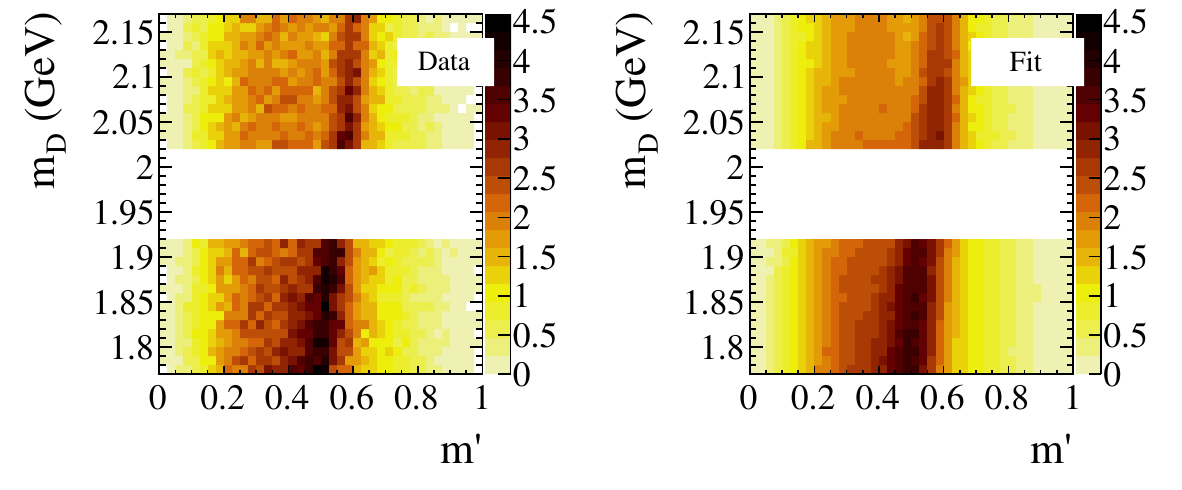}

  \includegraphics[width=0.32\textwidth]{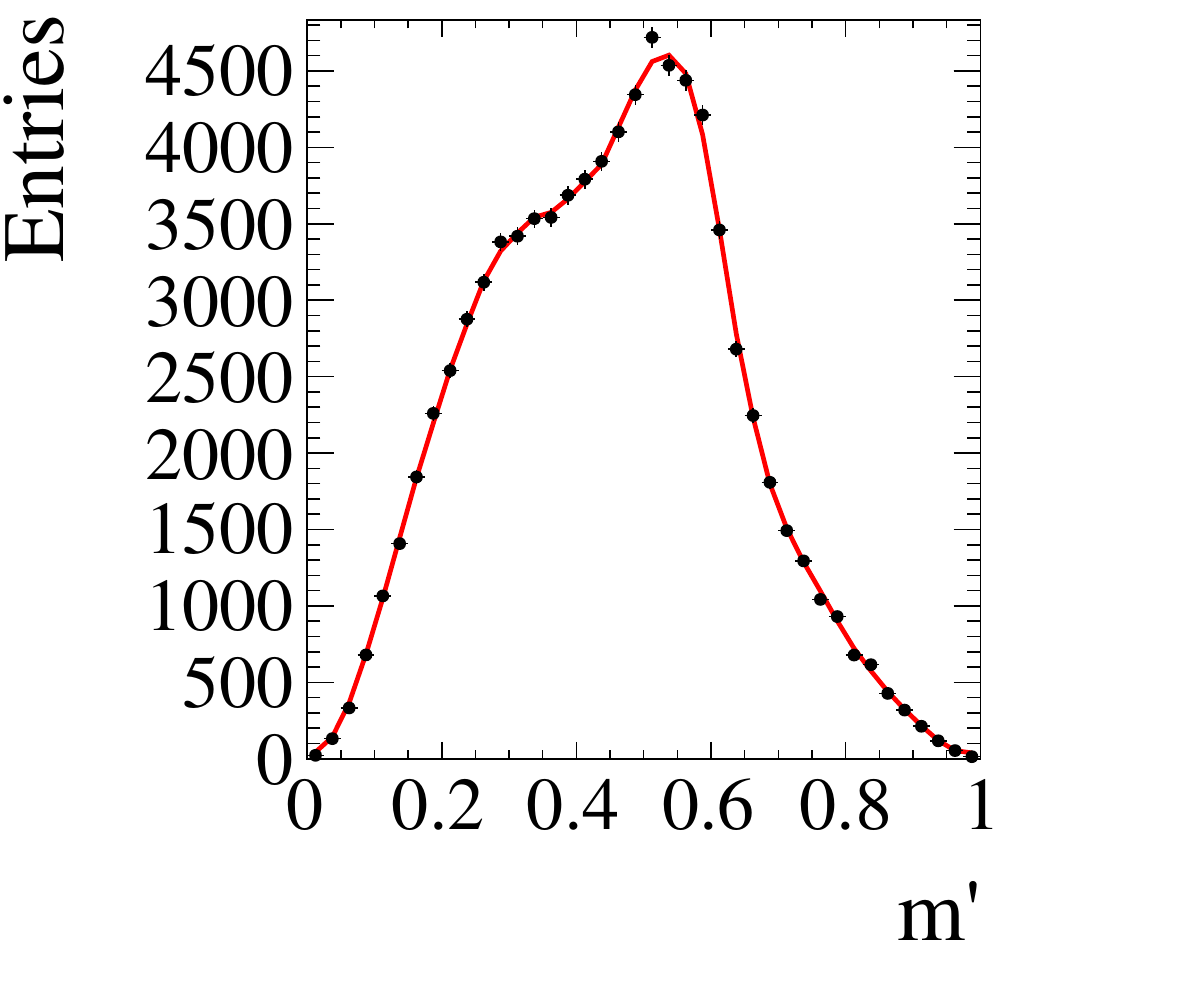}
  \includegraphics[width=0.32\textwidth]{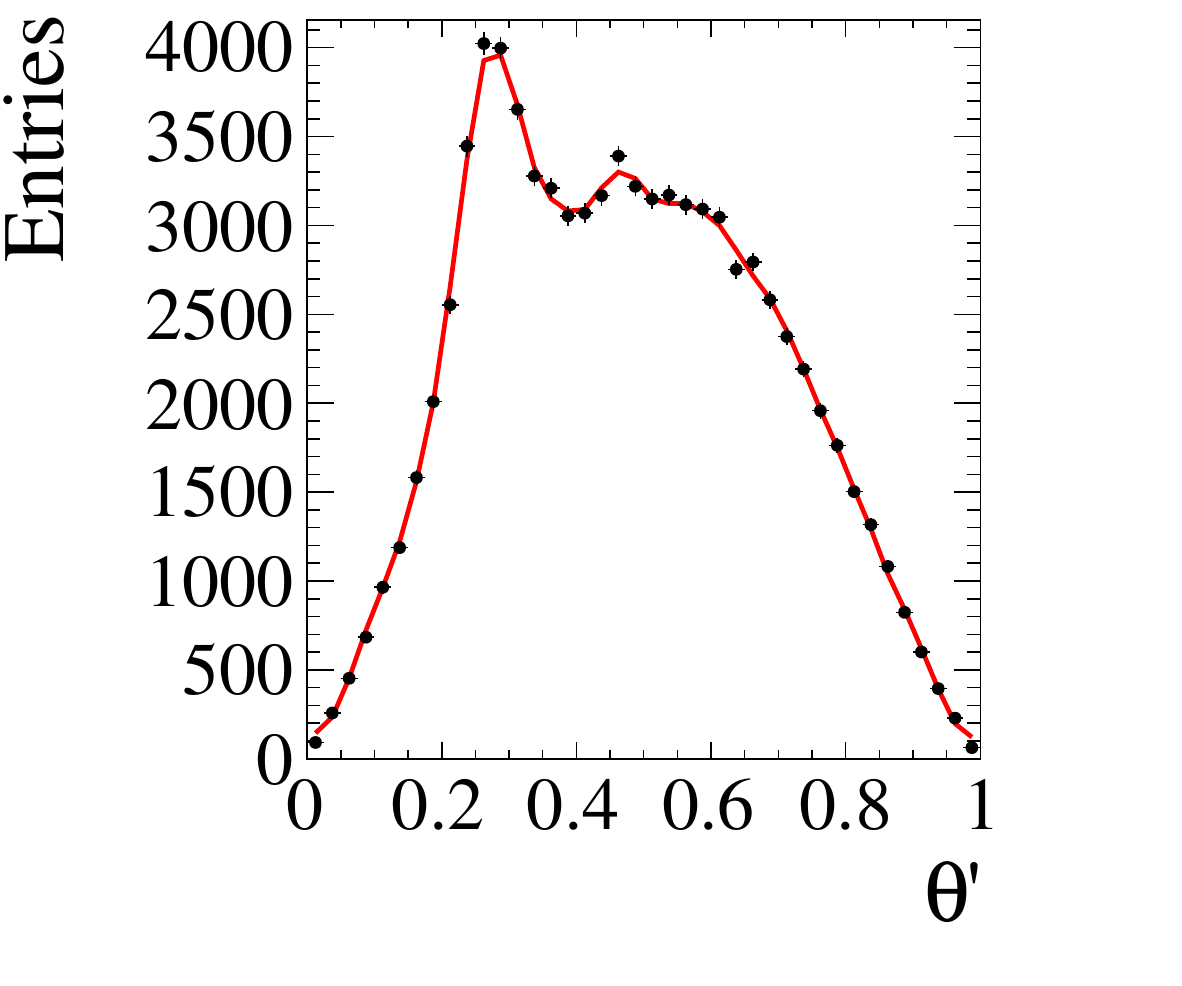}
  \includegraphics[width=0.32\textwidth]{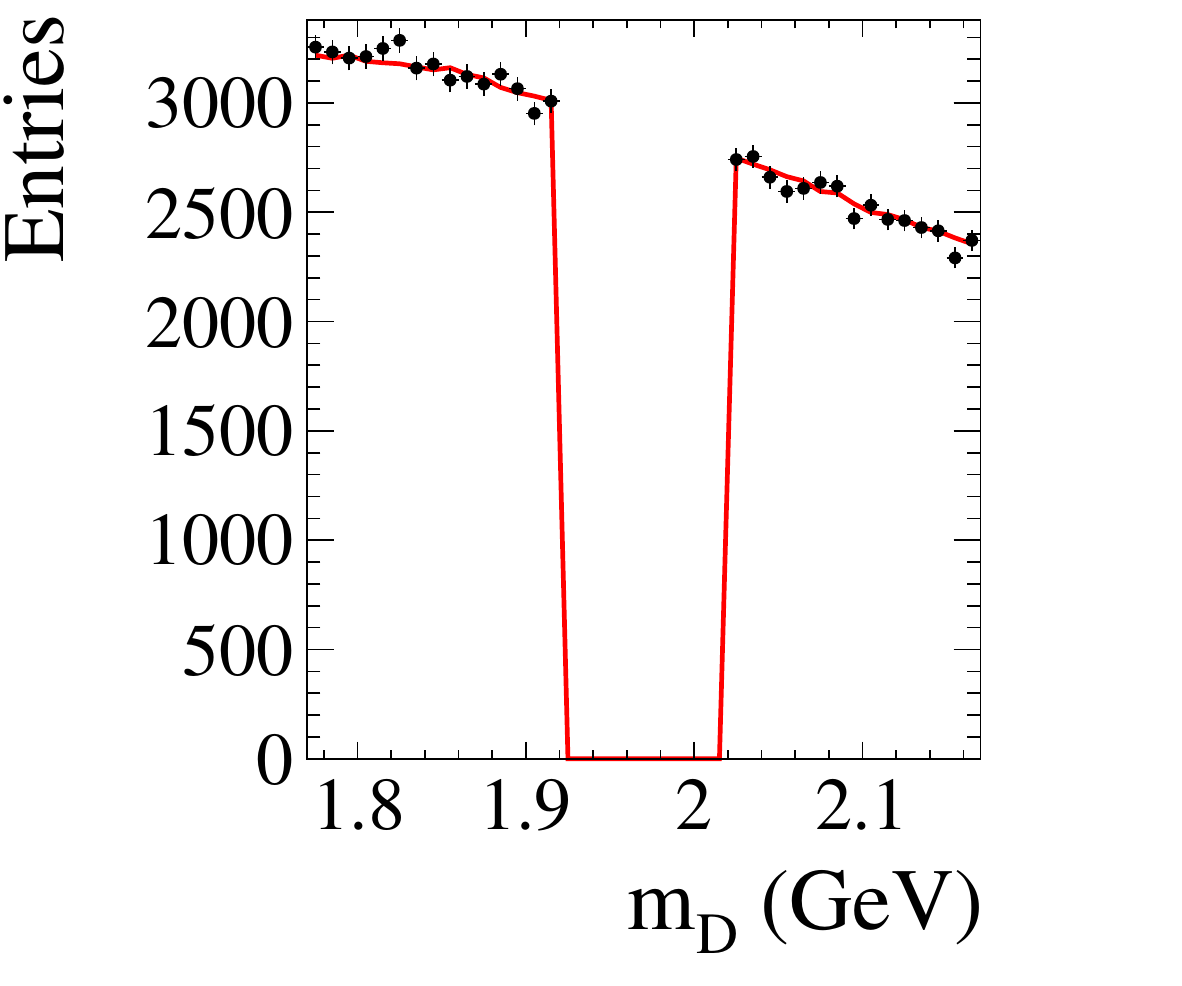}

  \caption{
  Results of the estimation of combinatorial background density in sideband regions using Gaussian process.
  Two-dimensional (first row) $m_D$ vs. $m'$, (second row) $\theta'$ vs. $m'$, and (third row) $m_D$ vs. $\theta'$ projections
  of the (left) simulated background sample and (right) density predictions from the fit model.
  (Bottom row) One-dimensional projections onto (from left to right) $m'$, $\theta'$ and $m_D$.
  }
  \label{fig:gp_bkg_train_3d}
\end{figure}

The Gaussian process uses the Mat\'{e}rn kernel, defined in Section~\ref{sec:gp}, with $\nu = \frac{5}{2}$, along with a constant mean function in $(m', \theta')$, and a linear mean function in $m_D$. Results of the estimation of the background density in the sideband regions of $m_D$ variable
are presented in Fig.~\ref{fig:gp_bkg_train_3d}. Here it can be seen specifically that the Gaussian process model estimates well the variation in the resonance structure in $(m', \theta')$ as a function of $m_D$ due to the kinematical constraints, permitting accurate estimation of this background structure in the unobserved signal region. The corresponding kernel parameters that are extracted from the data can be seen in Table~\ref{tab:bkgGP}.

\begin{figure}


  \includegraphics[width=0.34\textwidth]{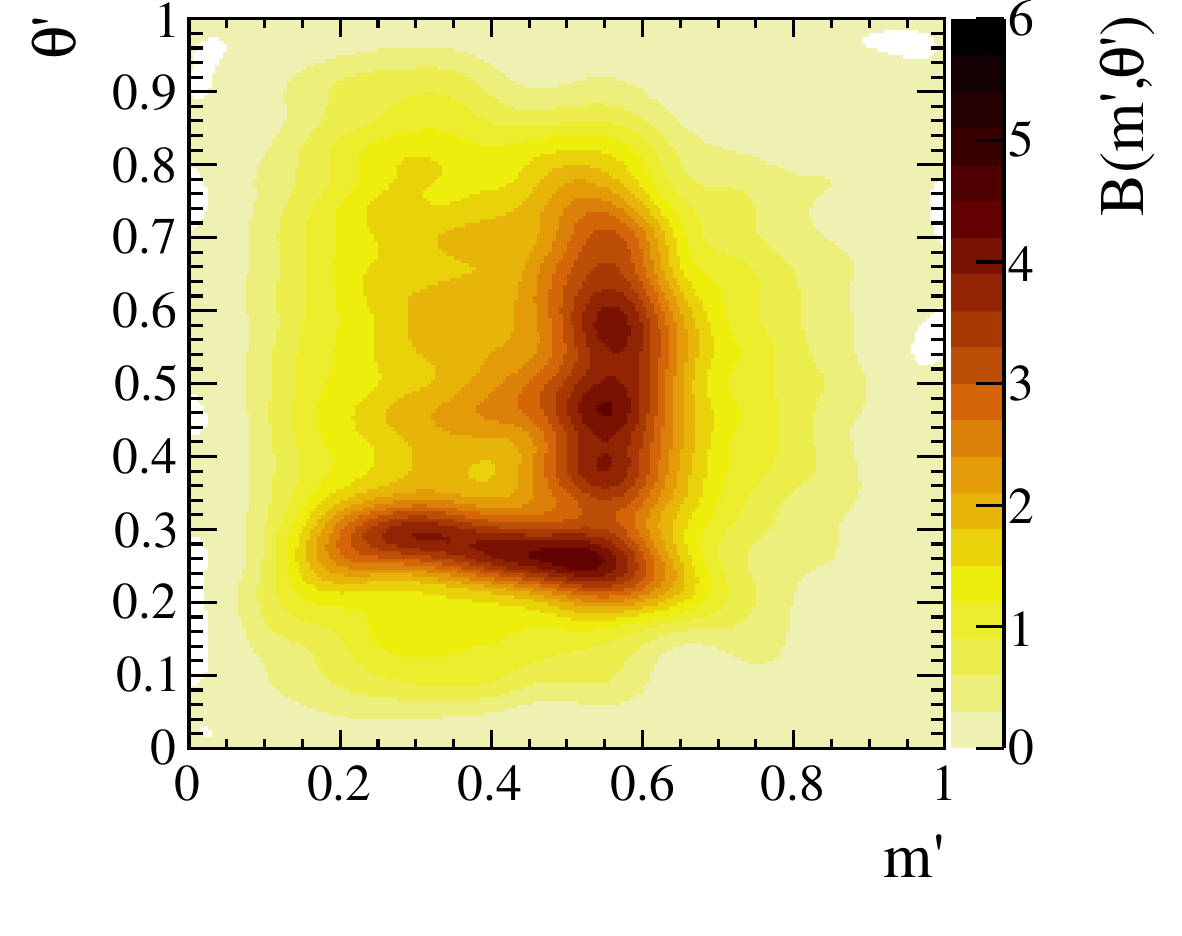}
  \put(-54, 100){\colorbox{white}{\small (a)}}
  \includegraphics[width=0.32\textwidth]{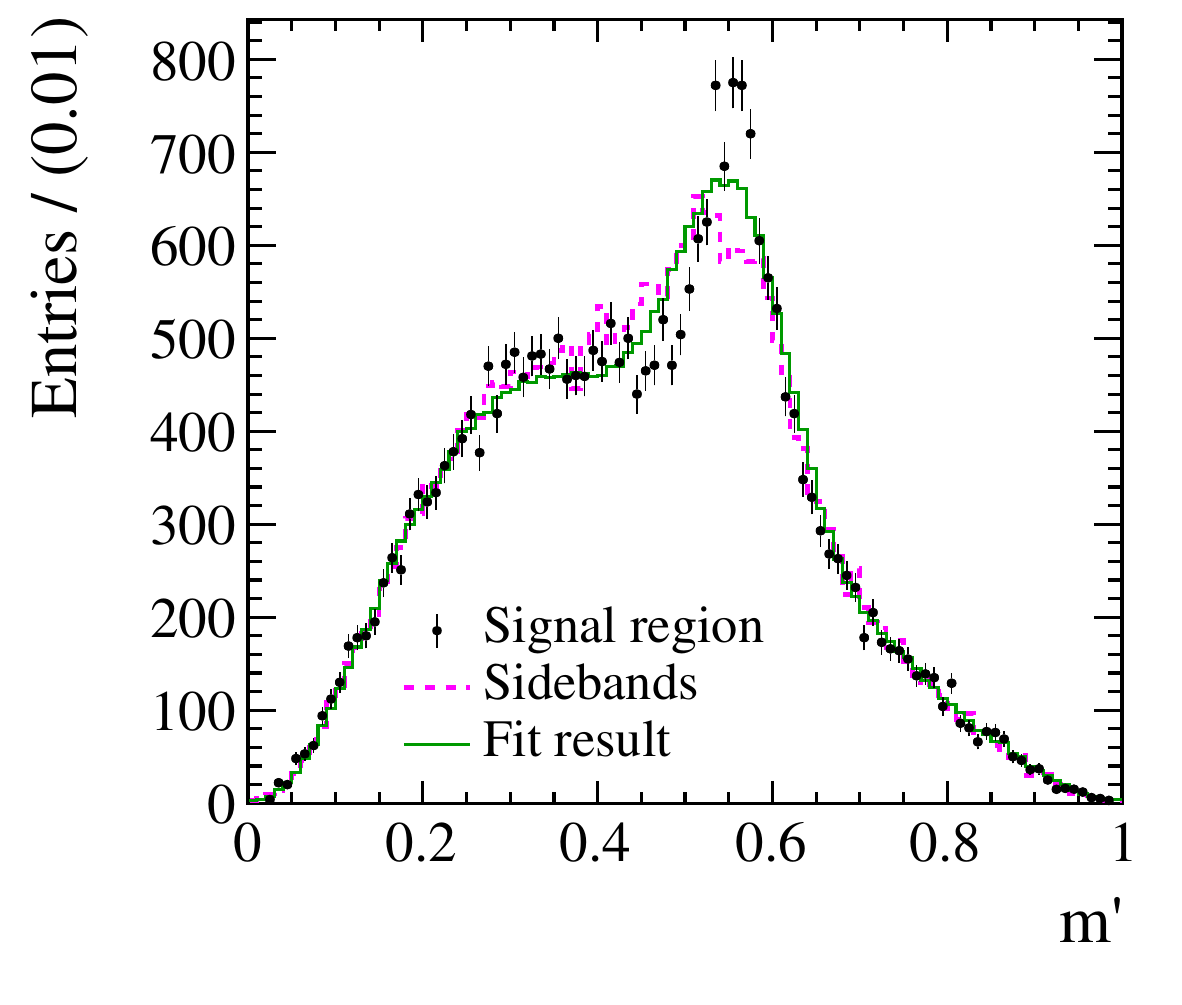}
  \put(-32, 100){\colorbox{white}{\small (b)}}
  \includegraphics[width=0.32\textwidth]{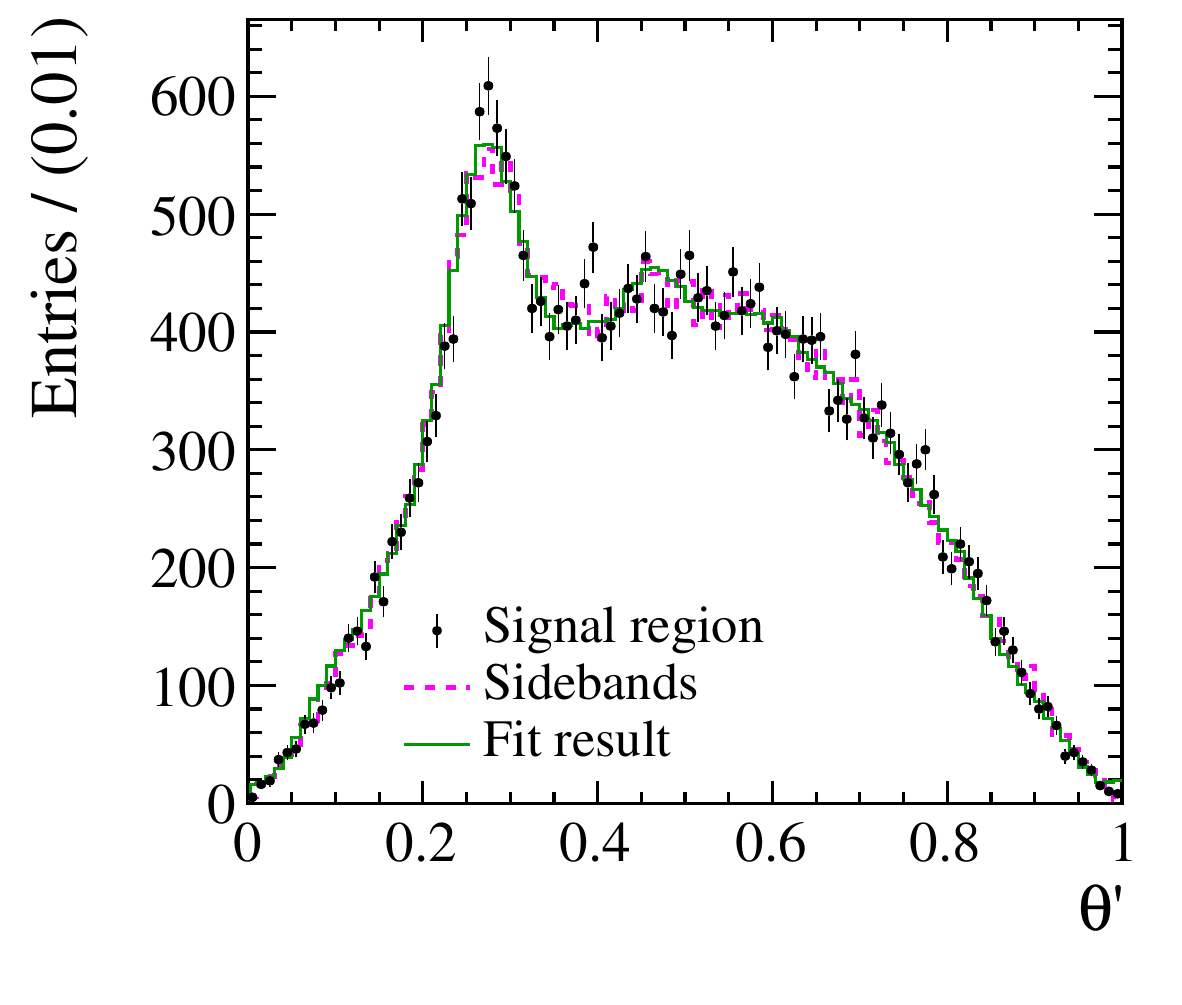}
  \put(-32, 100){\colorbox{white}{\small (c)}}

  \caption{Results of the interpolation of the combinatorial background density in the signal region using Gaussian process:
           (a) two-dimensional density and (b,c) its projections. }
  \label{fig:gp_bkg_fit_3d}
\end{figure}

The distribution of the background in the signal region is obtained by querying the Gaussian process at $m_D = 1.97$ \gev,
the results of which can be seen in Figure~\ref{fig:gp_bkg_fit_3d}.
While the resulting density somewhat smears the narrow structure seen in $m'$ distribution (Fig.~\ref{fig:gp_bkg_fit_3d}(b)),
the bias of the distribution is clearly smaller than that obtained from the simple projections of the distribution in the sidebands.

\begin{table}
  \caption{Parameters of the Gaussian process fit to the simulated $\Ds\to\Km\pip\pim$ background, with a Mat\'{e}rn kernel, constant mean function in $(m', \theta')$, and linear mean function in $m_D$.}
  \label{tab:bkgGP}
  \begin{center}
  \begin{tabular}{|l|c|}
    \hline
    Model parameter        & Value \\
    \hline
    $\sigma_{\rm GP}^2$    & $279$ events$^2$\\
    $\sigma_{{\rm mean\ } m_D}^2$  & $43.0$ events$^2$ \\
    $\sigma_{\rm mean\ DP}^2$  & $33.6$ events$^2$ \\
    $\rho_{m_D}$           & $0.27$ \gev \\
    $\rho_{m'}$            & $0.12$ \\
    $\rho_{\theta'}$       & $0.07$ \\
    $\epsilon$             & $22.8$ events \\
    \hline
  \end{tabular}
  \end{center}
\end{table}

\subsection{Neural network background fit}

If there are no narrow structures in the background, one can consider a background PDF that is positive-definite,
reasonably smooth in $m_D$, and is sufficiently generic in square Dalitz-plot variables, such as
\begin{equation}
  P(m', \theta', m_D) = |P_1(m', \theta') + e^{-\alpha m_D} P_2(m', \theta')|^2
  \label{eq:bkg_density_nn1}
\end{equation}
where $P_{1,2}(m', \theta')$ are the functions modelled with ANN.
One can then perform an unbinned fit of $P(m', \theta', m_D)$ to sideband data, with regularisation to avoid overtraining,
with the weights and biases $\thetavec_{1,2}$ of ANN functions $P_{1,2}$ and $\alpha$ as the free parameters.
The background in the signal region can then be extrapolated using the trained model as $B(m', \theta') = P(m', \theta', m_D = M_\Ds)$.

In the presence of narrow structures in the amplitude that vary as a function of $m_D$
(such as the resonant $\Kstarz$ and $\rhoz$ contributions in the $\Ds\to\Kp\pim\pip$ sample, see Fig.~\ref{fig:bkg_sidebands}), the
approximation shown in Eq.(\ref{eq:bkg_density_nn1}) may not work well. As ANN density estimation can be performed in multiple dimensions,
one can estimate the background density in the selection variable $m_D$ in addition to that in the Dalitz-plot variables, $m'$ and $\theta'$.
Therefore, as an alternative to the PDF of the form~(\ref{eq:bkg_density_nn1}), the full three-dimensional ANN can be used to
parametrise the background as a function of square Dalitz-plot variables and $m_D$, where additional regularisation has to be applied
to ensure continuity as a function of the selection variable $m_D$. The latter can be done by adding an extra penalty term in the likelihood
which penalises configurations where the neurons in the input layer have large weights corresponding to $m_D$ variable. Such regularisation will effectively result in the ANN where the first input layer consists of features that are slowly varying as a function of $m_D$. 

\begin{figure}
  \centering
  \includegraphics[width=0.8\textwidth]{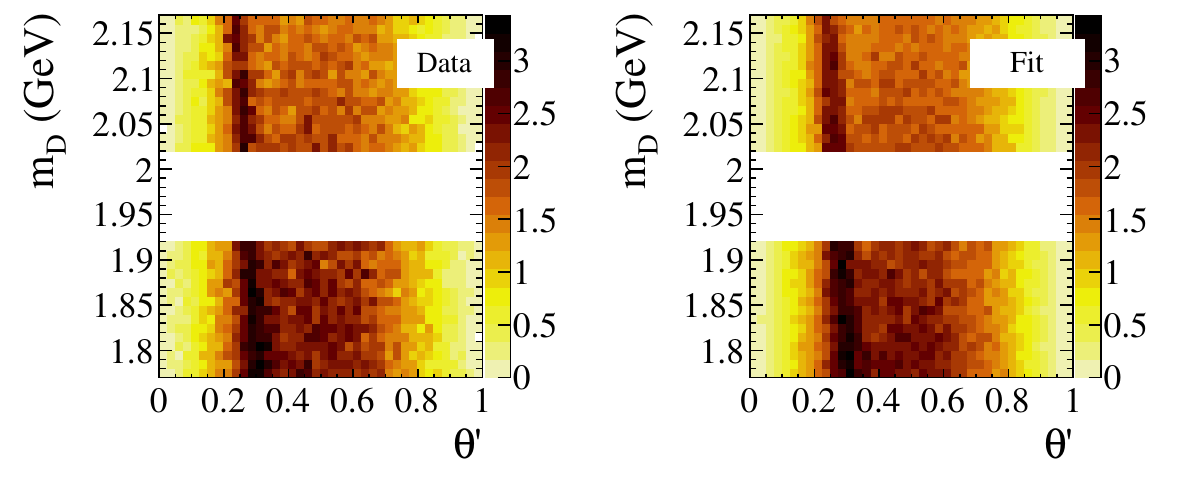}
  \includegraphics[width=0.8\textwidth]{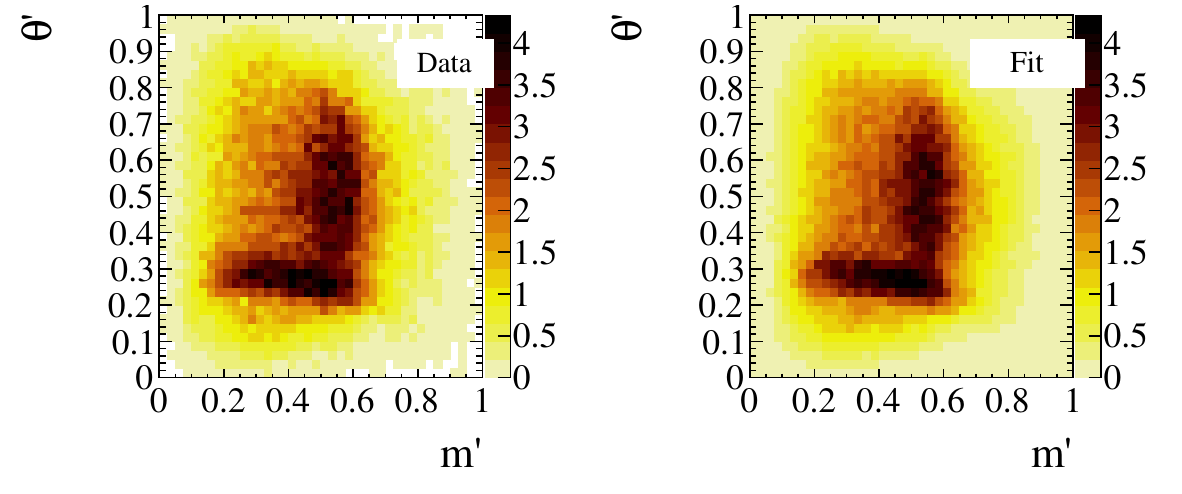}
  \includegraphics[width=0.8\textwidth]{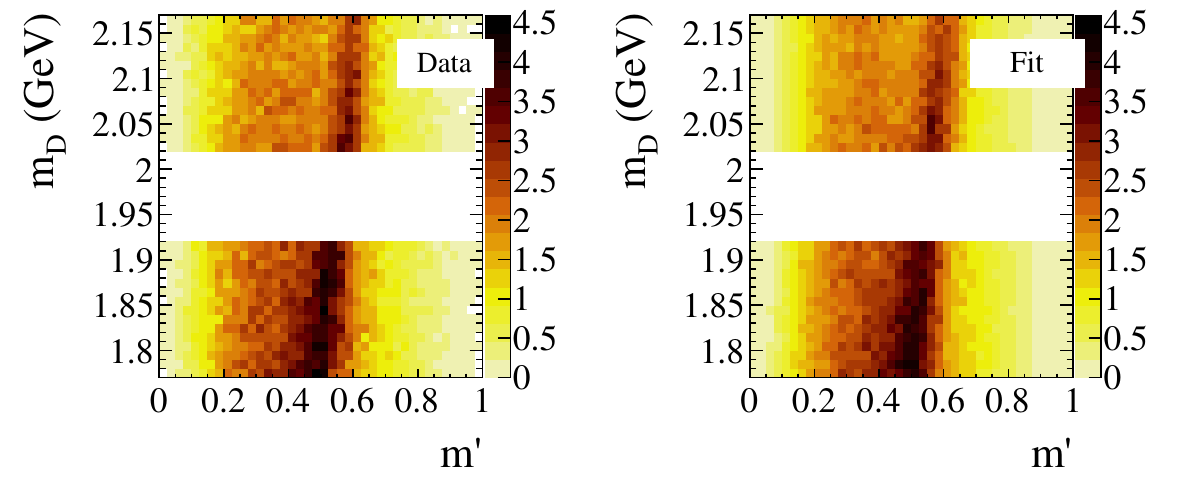}

  \includegraphics[width=0.32\textwidth]{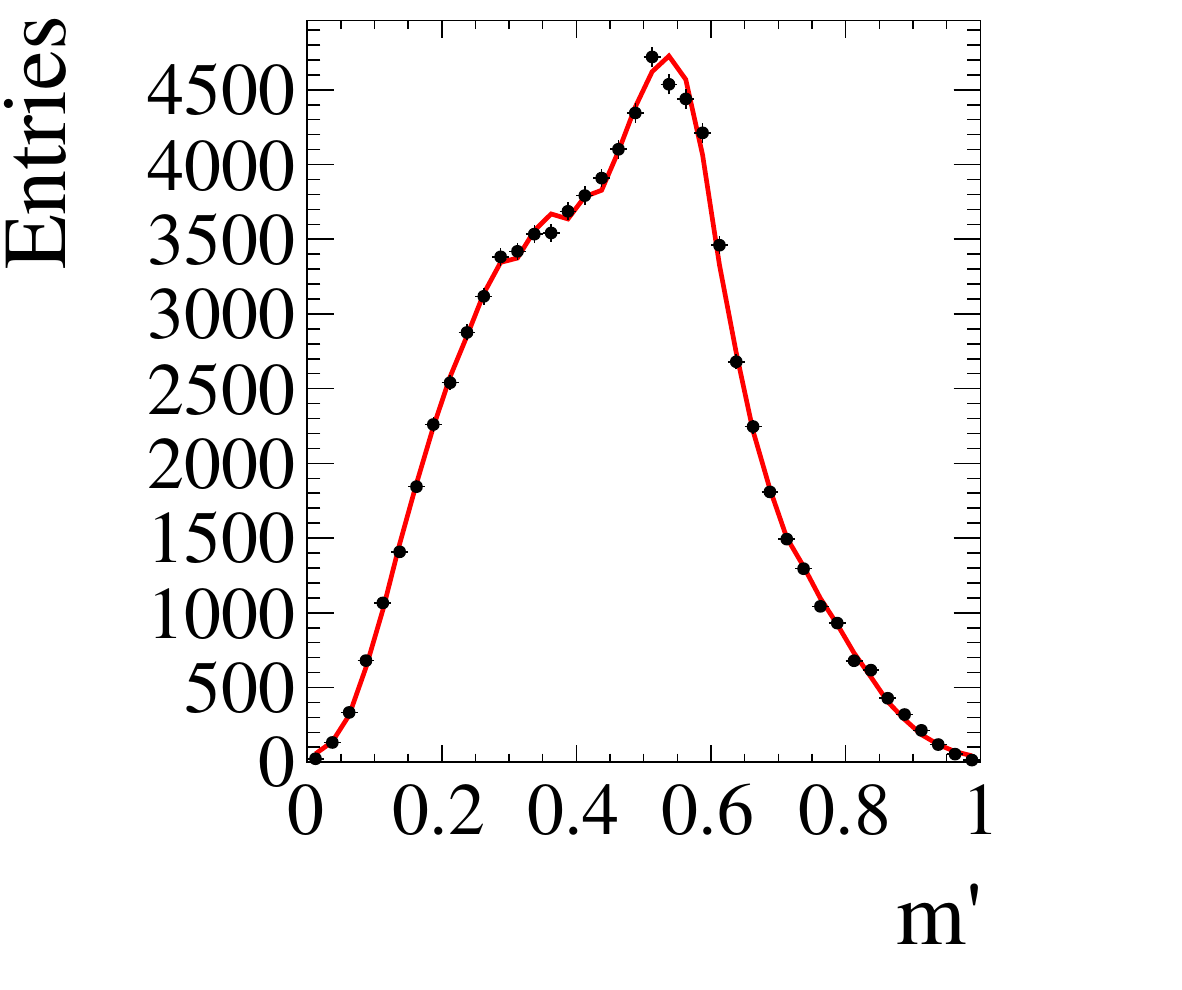}
  \includegraphics[width=0.32\textwidth]{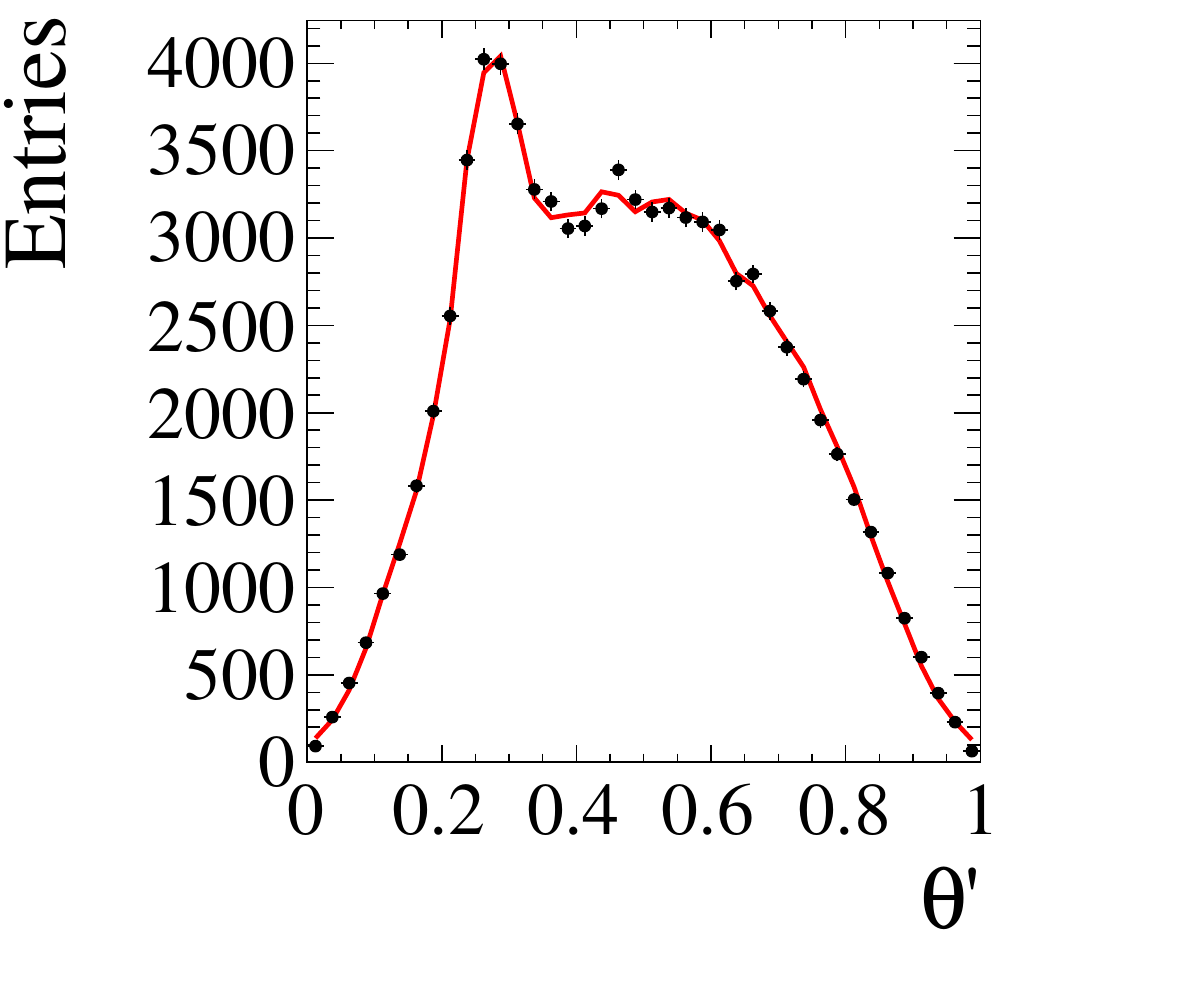}
  \includegraphics[width=0.32\textwidth]{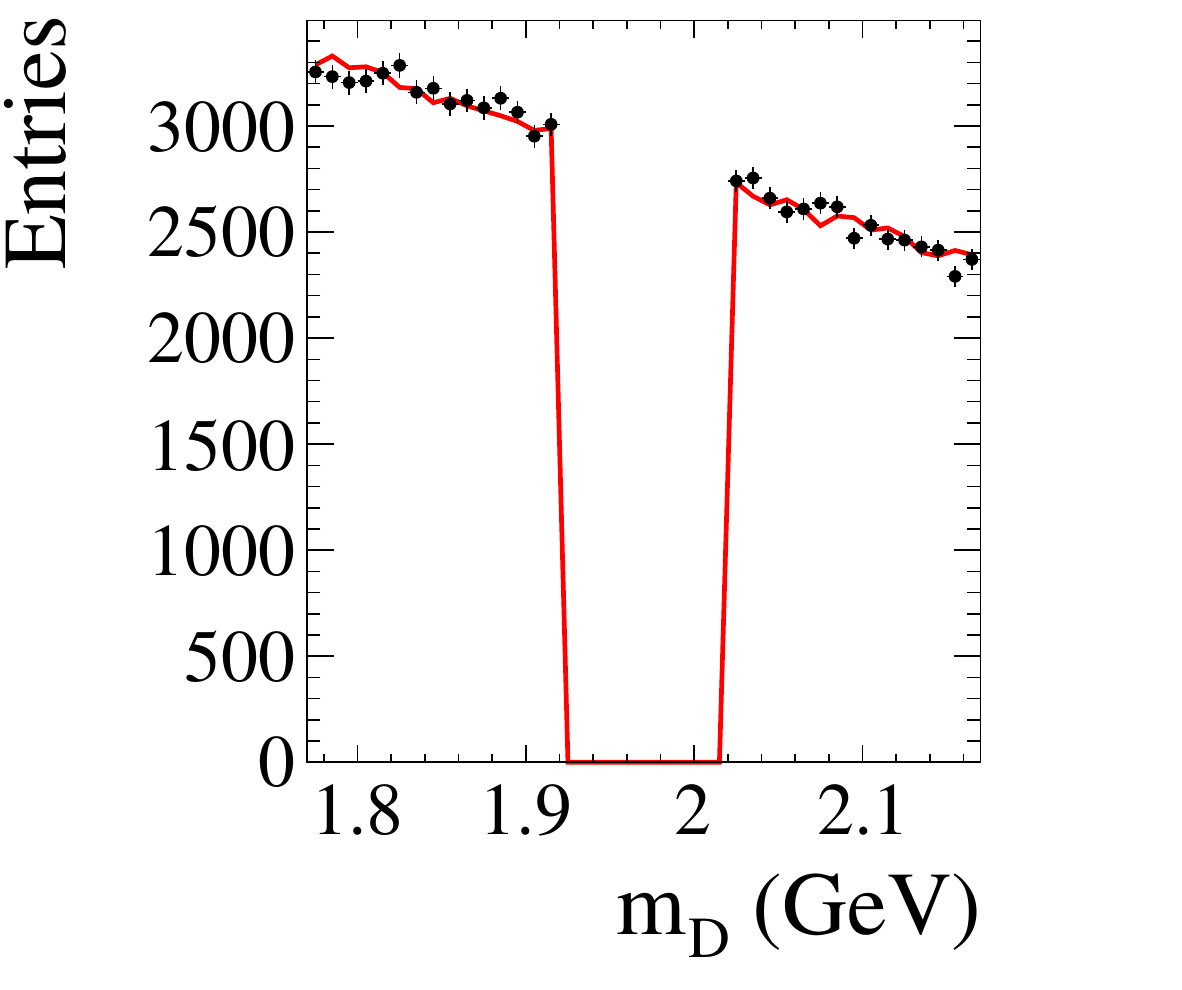}

  \caption{
  Results of the estimation of combinatorial background density in sideband regions using an ANN.
  Two-dimensional (first row) $m_D$ vs. $m'$, (second row) $\theta'$ vs. $m'$, and (third row) $m_D$ vs. $\theta'$ projections
  of the (left) simulated background sample and (right) density predictions from the fit model.
  (Bottom row) One-dimensional projections onto (from left to right) $m'$, $\theta'$ and $m_D$.
  }
  \label{fig:bkg_train_3d}
\end{figure}

\begin{figure}


  \includegraphics[width=0.33\textwidth]{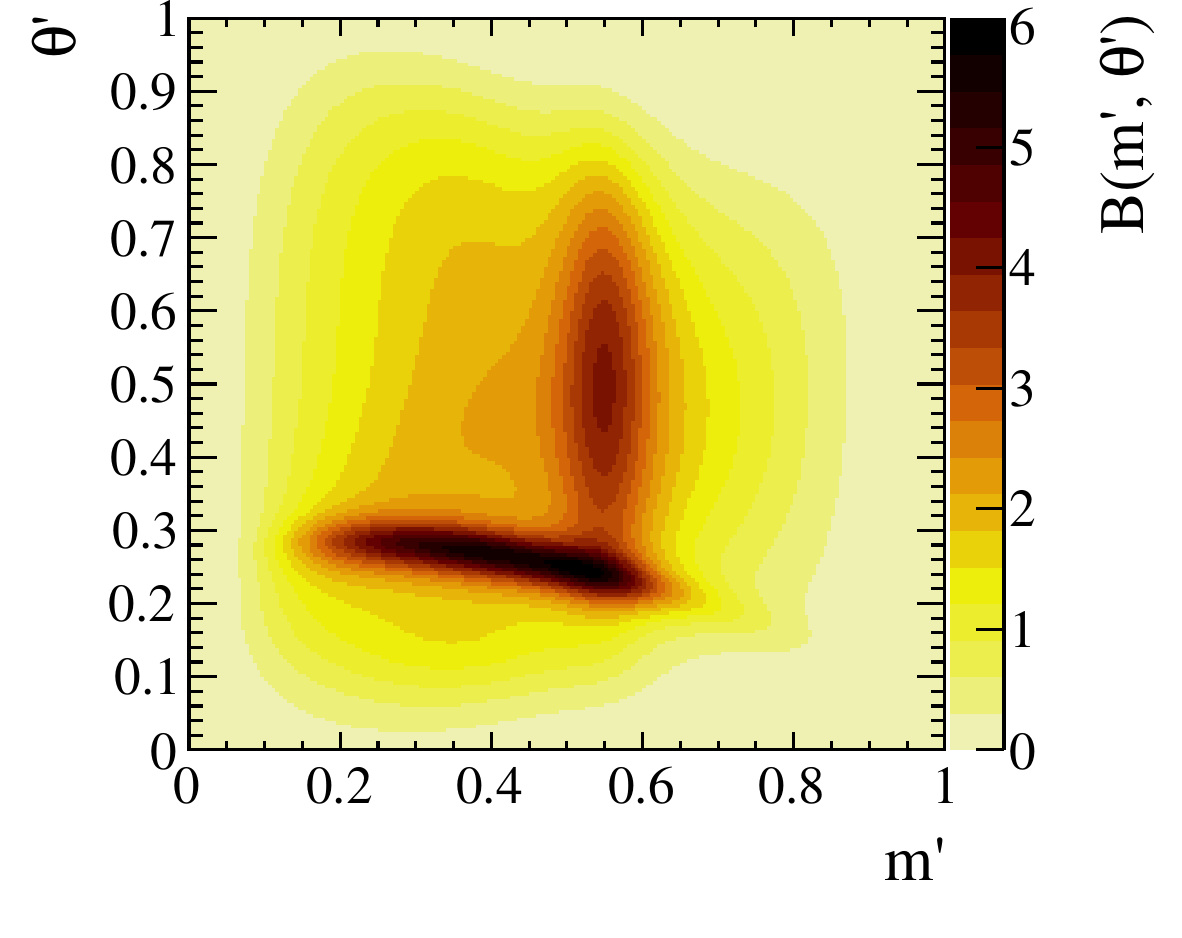}
  \put(-54, 100){\colorbox{white}{\small (a)}}
  \includegraphics[width=0.33\textwidth]{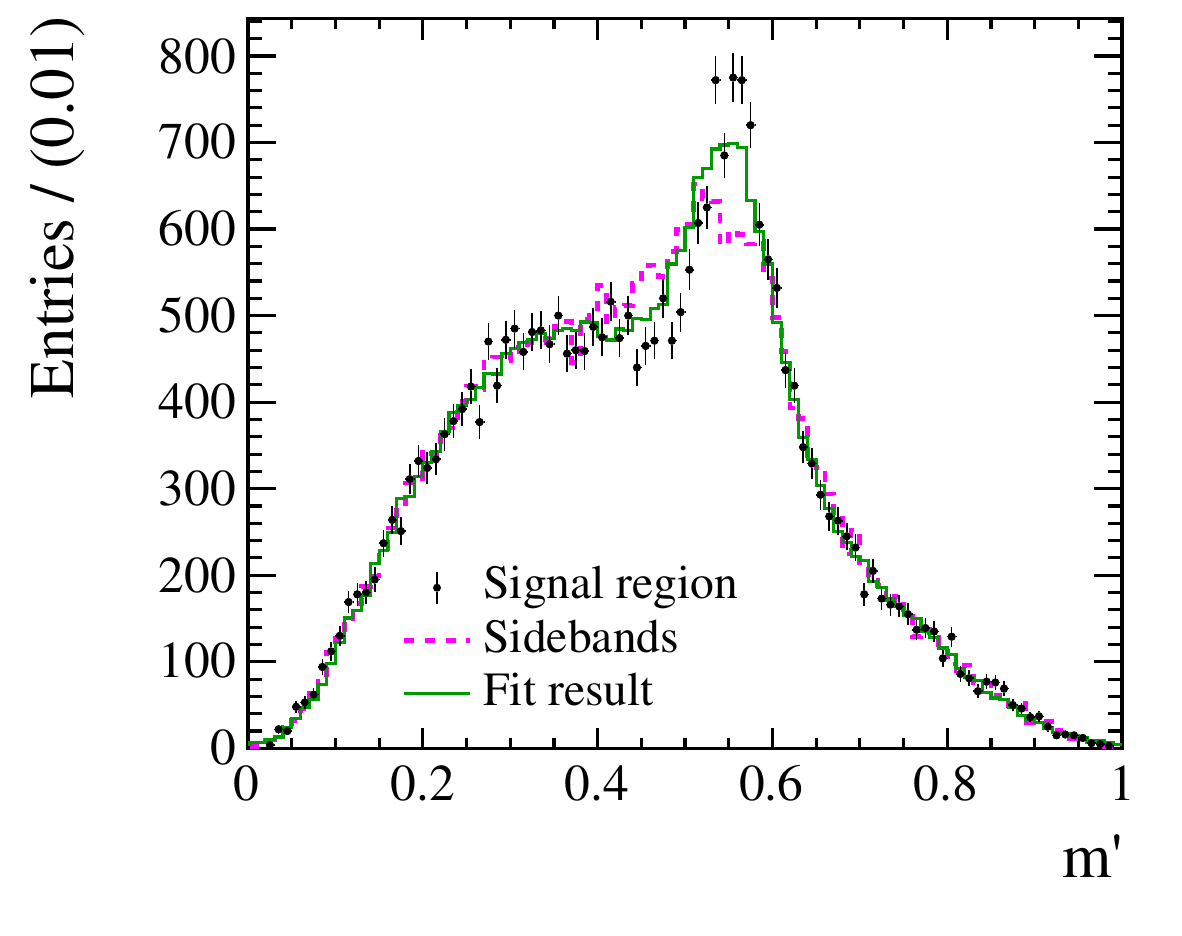}
  \put(-32, 100){\colorbox{white}{\small (b)}}
  \includegraphics[width=0.33\textwidth]{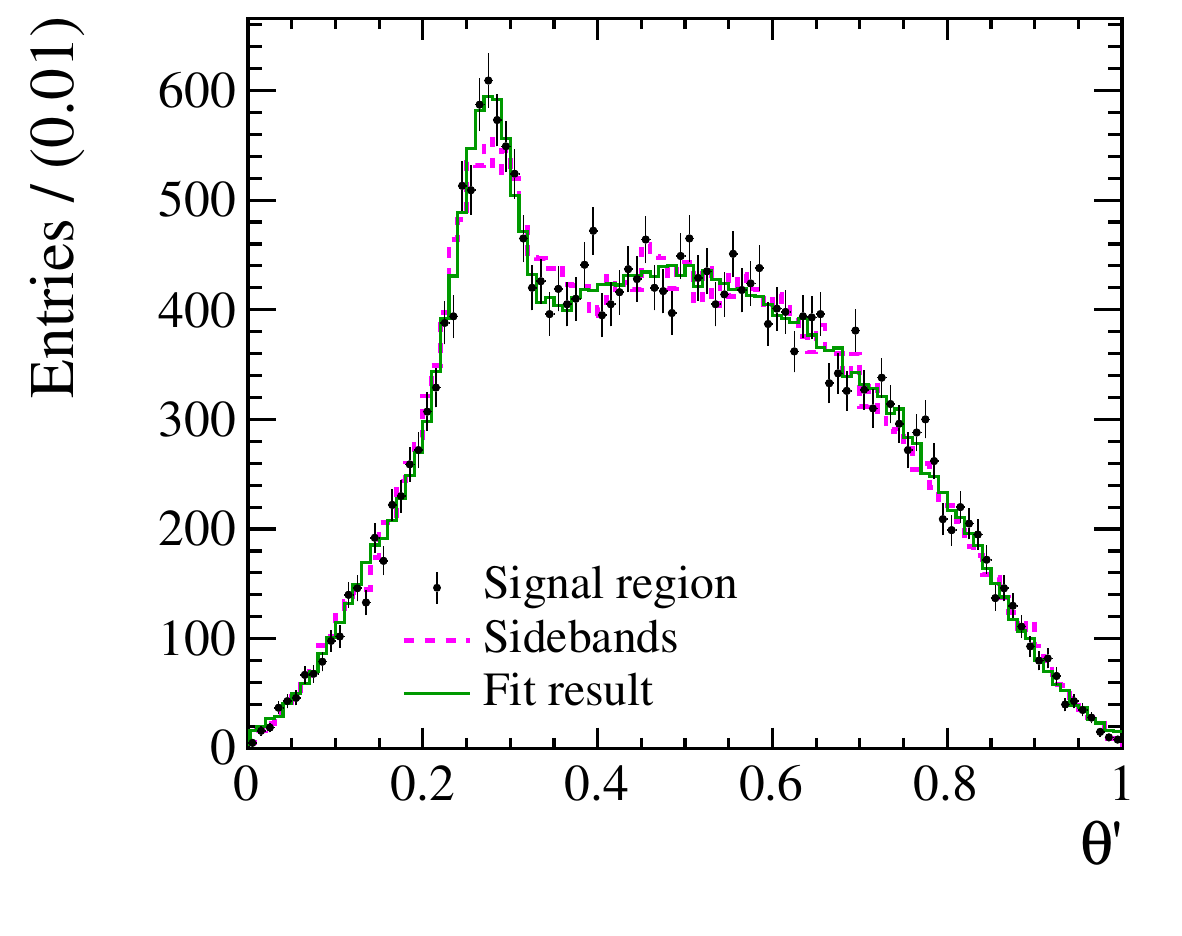}
  \put(-32, 100){\colorbox{white}{\small (c)}}

  \caption{Results of the interpolation of the combinatorial background density in the signal region using an ANN:
           (a) its two-dimensional density and (b,c) its projections. }
  \label{fig:bkg_fit_3d}
\end{figure}

The fit in the sideband regions to the simulated combinatorial background of the $\Ds\to\Kp\pim\pip$ decay using the ANN
is shown in Fig.~\ref{fig:bkg_train_3d}.
For the neurons in the first layer that take the $m_D$ dimension as input, the regularisation parameter $\lambda_2$ is set equal to $10$, while for the other neurons this is equal to $1$.
In Fig.~\ref{fig:bkg_fit_3d}, the predicted combinatorial background density in the signal region using ANN approach is shown, where the ANN is trained using Adam optimised for $50\,000$ epochs with the learning rate of 0.0002. The $\chi^2$ value for $50\times 50$ bins equals $2457.9$. 

\section{Model-assisted density parameterisation with neural networks}

\label{sec:madenn}

In the training of the ANN, it is often difficult to replicate specific features of the resulting density,
especially in the case of limited training data. The only handles on the generic ANN training are the
topology of the network and the generic regularisation terms, and careful tuning of these is needed to obtain a reasonable
description of the density.

The procedure of parameterising the background or acceptance using only the input data,
without any external knowledge of the processes that govern the features of the distributions, is
not the most optimal approach. In general, it is known, for instance, that the acceptance function should
be relatively smooth with a fall--off at the boundaries of the phase space due to kinematic selection requirements,
or that the combinatorial background should contain contributions from certain two-body resonances. The implication of this is that the behaviour is constrained much more than conventional parameterisation techniques assume, and an ideally efficient procedure should take this prior information into account. For example, one can introduce
a simplified model of these processes, and extract only the parameters of this simplified model from the training data samples.

In the case of background and efficiency distributions, even a simple analytical model for this would be difficult to express. Instead, here we propose a technique to perform nonparametric estimation of these distribution, using the formalism described in Section~\ref{sec:annde}, with the assumption that the complex observed behaviour explicitly depends only on a few underlying parameters that are sufficient to describe the efficiency or background behaviour in the region of interest. The values of these latent parameters can then be inferred from the observed data, or detailed simulation in the case of a description for the acceptance, in order to parameterise the distributions.

This way, the features of the resulting density are controlled by
the simplified model, which leverages prior information on the correlations between the model parameters and kinematic observables. This results in more stable training of the ANN, as only data for the simplified parameters are required, rather than resource-intensive detailed simulation, and therefore larger sample sizes can be generated. Crucially, reliance on a few latent parameters also results in a an \emph{ad hoc} regularisation effect, and as such the density obtained via this procedure is less sensitive to statistical fluctuations when obtaining the values of the simplified parameters. A similar technique has recently been independently proposed that utilises generative adversarial networks~\cite{Alonso-Monsalve:2018aqs}.

\subsection{Implementation}

In the initial stage, an estimate of the joint probability distribution $P \equiv P(\xvec, \thetavec)$, is constructed, in terms of the kinematic observables $\xvec$, in which the background or efficiency description is required, and the latent parameters on which the background or efficiency depend, $\thetavec$. The variables $\xvec$ could comprise the (square) Dalitz-plot variables, such as in the examples in this paper, but could also be anything else that is required to be parameterised in a physics analysis, such as the invariant mass of the reconstructed particle, or its decay time. The parameters $\thetavec$ are those that directly control physical constraints on the system, and influence $\xvec$ via their correlations. These parameters necessarily vary between analyses, but it is likely that these would include, \eg, effective threshold values on the final state particle momenta, parameters that describe the shape of these momentum distributions, or fractions of potential background contributions.

An estimate of the joint probability distribution is parameterised using an ANN, obtained via the probability density estimation technique described in Section~\ref{sec:annde}. The ANN is trained using a sample of simulated data, $\mathbf{S}_{\rm train} = \{\mathbf{X}_{\rm train}, \thetavec_{\rm train}\}$, that encapsulates dependencies between the kinematic observables, $\xvec$, and the latent parameters $\thetavec$. These data are required to span the space of possible model parameter values, however accurate description of any specific configuration of $\thetavec$ or $\xvec$ is not required (that is, there is no requirement for the set of input data points in this initial construction to overlap with the set of eventual evaluation points, due to the model smoothing). Effectively, the ANN parametrisation obtained at this stage will represent the functional representation of the model that is implemented using simulation. 

Secondly, an estimate of the specific values of the latent parameters, $\thetavec_{\rm pred}$, that correspond to the background data or detailed simulation, $\mathbf{X}_{\rm data}$, is obtained. This is done by fixing the weights of the ANN obtained at the first stage, and performing a maximum-likelihood fit for these values,
treating the ANN output as the probability of the latent parameters conditioned on the known vector of kinematic observables, $P(\thetavec | \Xvec_{\rm data})$,
such that $\thetavec_{\rm pred} = {\rm arg\,max}_{\thetavec} \ P(\thetavec | \mathbf{X}_{\rm data})$.

Lastly, again using the ANN as a joint probability function, the sample $\mathbf{X}$ is drawn from the distribution
$P(\xvec | \thetavec_{\rm pred})$, to obtain the probability density of kinematic observables. This approach is illustrated below for the estimation of the acceptance and combinatorial background distributions of the $\Ds\to\Kp\pim\pip$ decay, described in Section~\ref{sec:toymc}.

It is worth noting that, whilst these latent parameters should in principle comprise the set of features on which the parameters of interest depend strongly upon, this set need not be exhaustive. Providing that the included parameters are at least reasonably correlated with any additional parameters that are not considered, yet influence the kinematic distributions, an ``effective'' value of these parameters can be obtained in the maximum likelihood fit stage. As such, these latent parameters can differ from those that can be calculated directly from the dataset that the model is evaluated on, in such a way that the efficiency or background distribution can nevertheless be correctly parameterised using these.

\subsection{Acceptance parameterisation}

To demonstrate the feasibility of the model-assisted approach for the parameterisation of acceptances,
the same model as the one described in Section~\ref{sec:toymc_acc} is used, however the requirements on
the reconstructed \Ds mesons and their decay products are varied in the generation of the training sample
used to construct the ANN, as in step one above. The range of the variations for each of the
five parameters of the model is given in Table~\ref{tab:eff}, where the entire sample consists of $2\,000\,000$
events that satisfy the selection requirements. Since the efficiency model is relatively simple,
generation of the training sample takes only a few minutes, and therefore this can be arbitrarily large. Since the events that do not satisfy the selection requirements are rejected during
generation, the initial uniform distributions of the model parameters can become significantly non-uniform
for the accepted events in the training sample. To compensate for this effect, the model parameters
are sampled from non-uniform prior distributions (exponential in our case), where the parameters of the prior distributions
are tuned to ensure that the distribution of model parameters for the accepted events are roughly uniform.

The functional form of the efficiency model, a $7$-dimensional probability density
function $p(m', \theta' , \thetavec)$, where $\thetavec$ is the vector of parameters listed in Table~\ref{tab:eff},
is obtained by the ANN density estimation procedure described in Section~\ref{sec:annde}. The ANN topology
and training parameters are the same as those highlighted in that Section~\ref{sec:annde}, with the exception that here
the $L_2$ regularisation parameter is taken to be $\lambda_2 = 5\times 10^{-3}$.
The projections of the result of the ANN training after $50\,000$ epochs in the square Dalitz-plot variables, as well as the correlations between the Dalitz-plot variables and the latent model parameters, are given in Appendix~\ref{app:eff}.

\begin{figure}


  \includegraphics[width=0.34\textwidth]{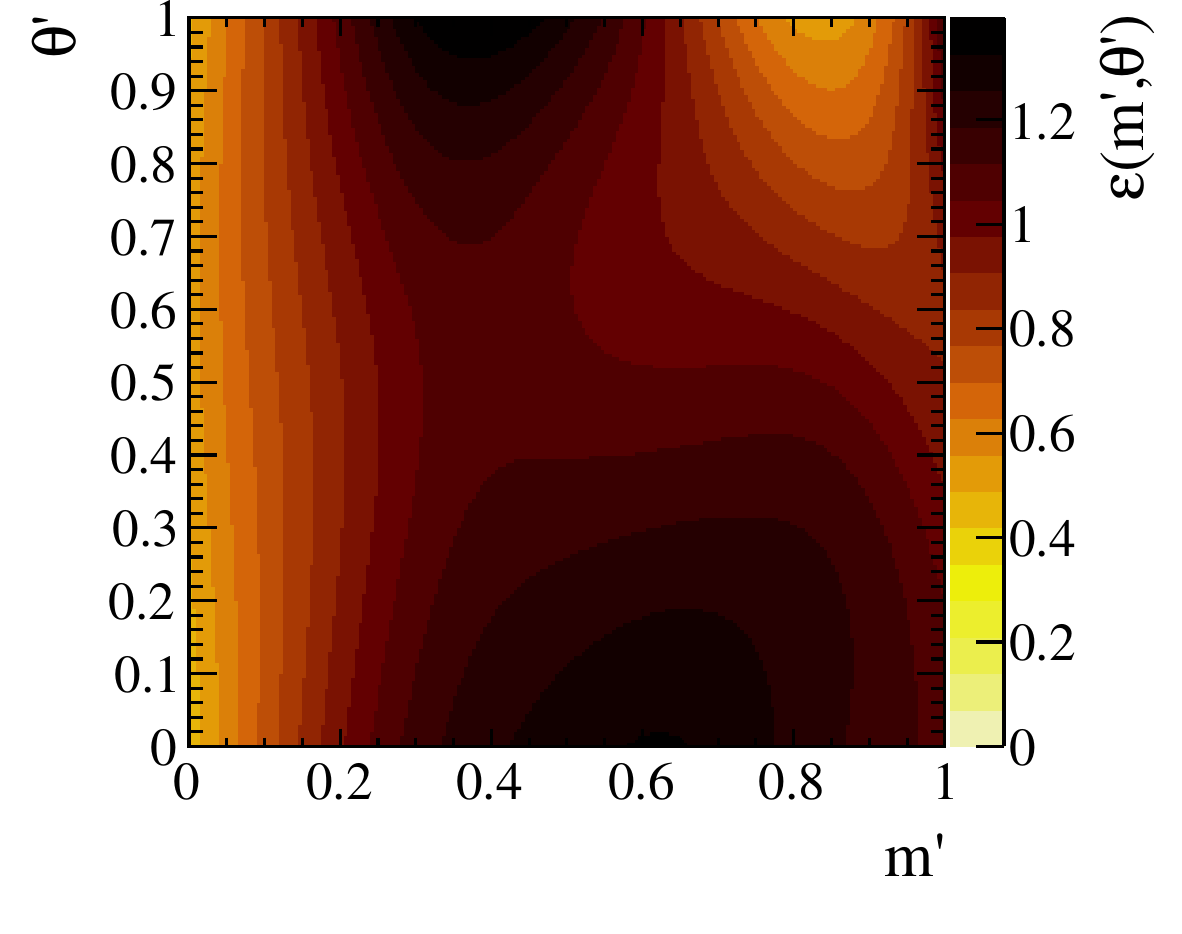}
  \put(-54, 35){\colorbox{white}{\small (a)}}
  \includegraphics[width=0.32\textwidth]{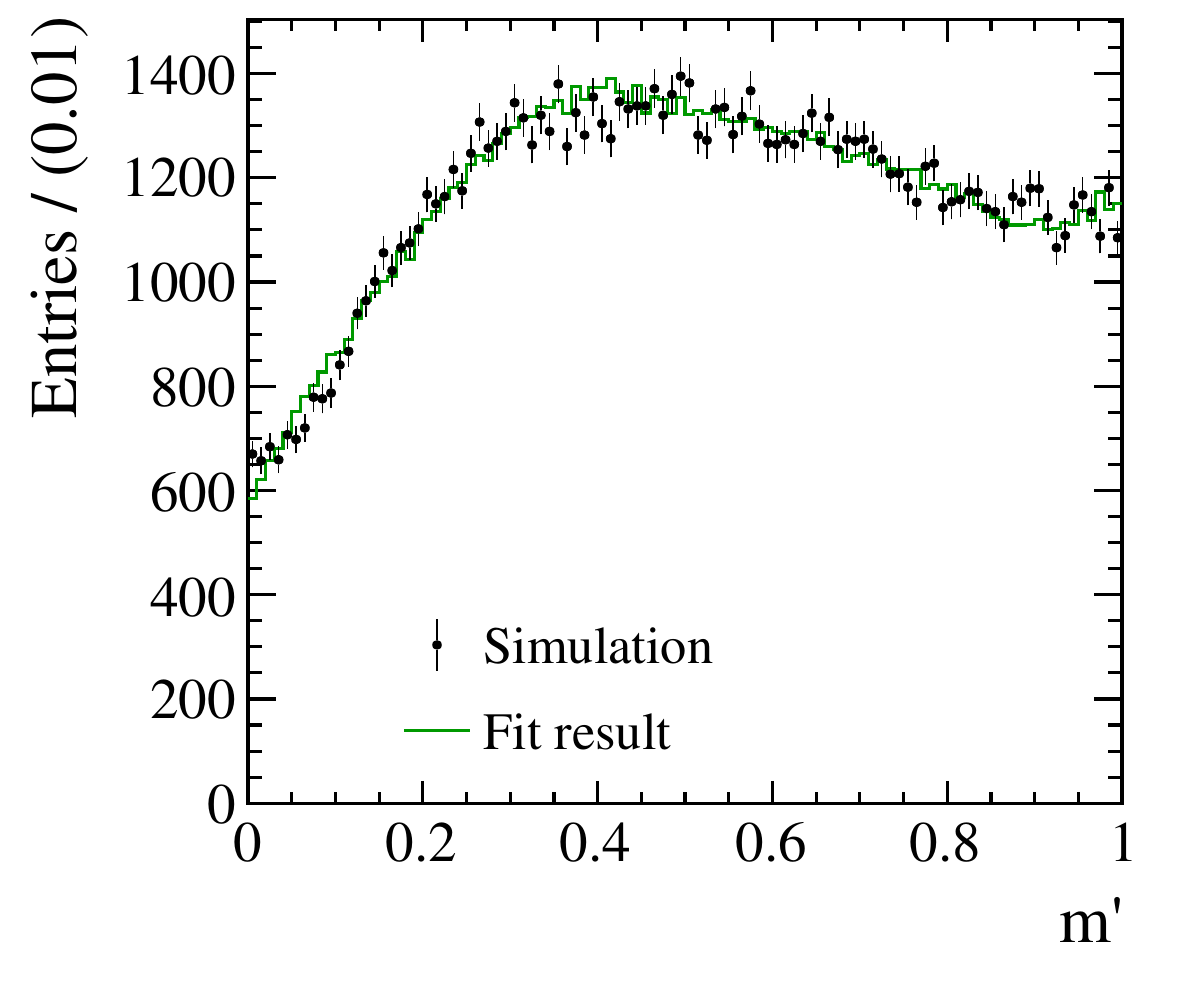}
  \put(-32, 35){\colorbox{white}{\small (b)}}
  \includegraphics[width=0.32\textwidth]{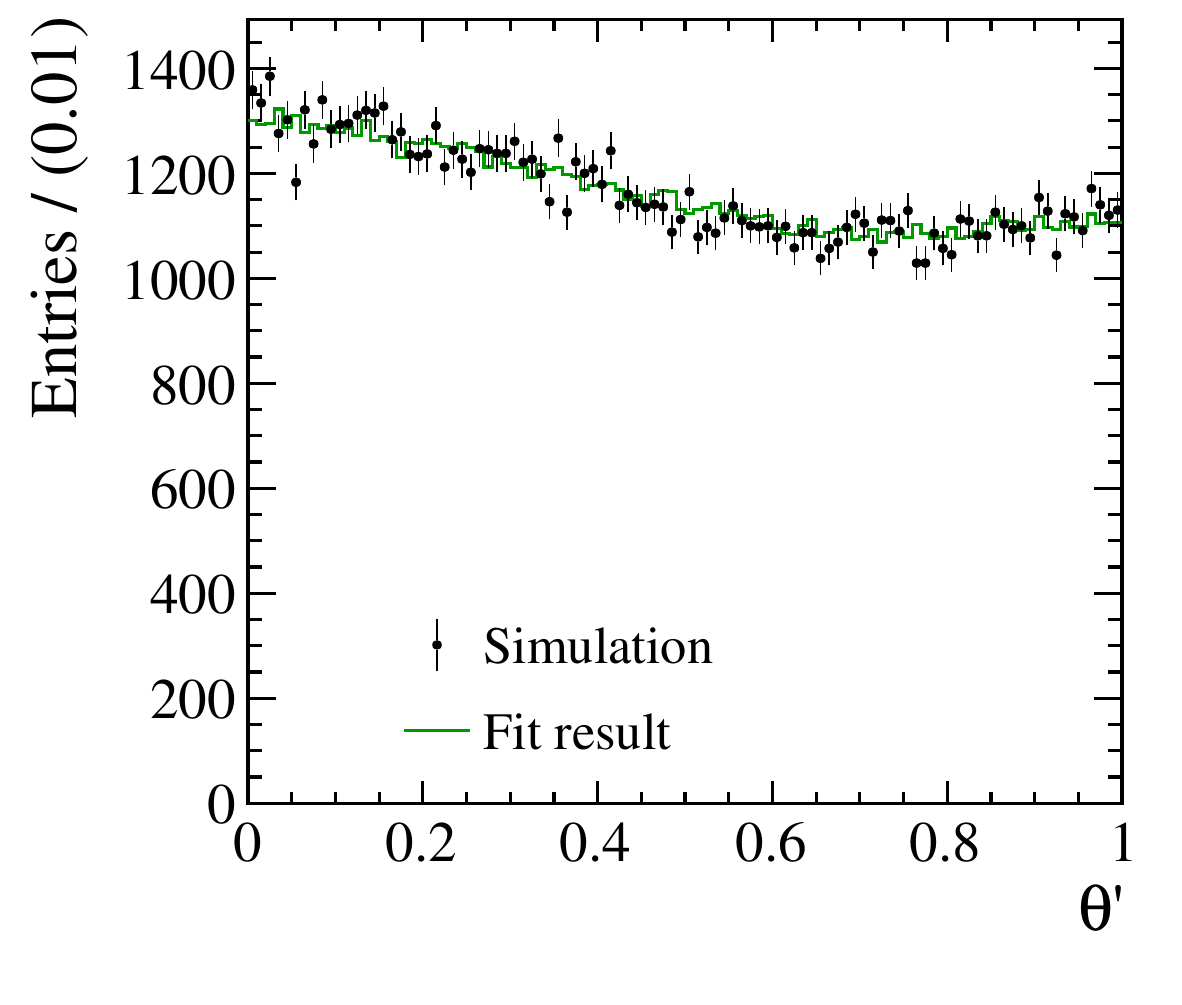}
  \put(-32, 35){\colorbox{white}{\small (c)}}

  \caption{Result of the density estimation of the simulated sample of $\Ds\to\Km\pip\pim$ decays using the model-assisted ANN
          (a) in the two square Dalitz plot variables $m'$ and $\theta'$, and the projections of the two-dimensional distribution
          on to the(b) $m'$ and (c) $\theta'$ variables. }
  \label{fig:eff_fit_model}
\end{figure}

An unbinned maximum likelihood fit is then performed to obtain the effective model parameters for a test sample corresponding to that described in Section~\ref{sec:toymc_acc}. The results of the fit are presented in Fig.~\ref{fig:eff_fit_model}, and the model parameters,
both the true generated values and the values obtained from the fit, are given in Table~\ref{tab:eff}.
The maximum likelihood fit is performed using {\tt iminuit}~\cite{hans_dembinski_2020_4283509}, and the interface between
the ANN implemented in TensorFlow and {\tt iminuit} is provided by TensorFlowAnalysis package~\cite{tfa}.

Since the same underlying model is used to generate both the test samples and the samples used for the ANN training,
one should expect the reconstructed parameters to be statistically consistent with the true generated ones.
Certain tension is observed between the parameters in Table~\ref{tab:eff}, which may indicate some deficiency
or ambiguities in the ANN model. Nevertheless, this gets corrected by the maximum likelihood fit to the test dataset,
and a good-quality parameterisation of the distribution results is obtained with the fit quality calculated with $50\times 50$ binning equal to $\chi^2/\mbox{nDoF}=2528.9/2494$. 
In the case of estimating the acceptance distribution from the detailed simulation,
the quality of the parameterisation will depend on how well the simplified efficiency model approximates the
experimental selection.

\begin{table}
  \caption{Parameters of the $\Ds\to\Kp\pim\pip$ efficiency model:
  the range of parameter variations used at the ANN training stage,
  the true values used in the generated test sample,
  and the reconstructed values extracted from the fit of the ANN model to the test sample.}
  \label{tab:eff}
  \begin{center}
  \begin{tabular}{|l|c|c|c|}
    \hline
    Model parameter        & Range & True value & Reconstructed value \\
    \hline
    Track $\pt$ cut (GeV) & $(0.1, 1)$ & 0.4 & $0.408\pm  0.010$ \\
    Track $p$ cut (GeV)   & $(1, 10)$  & 3.0 &  $4.5\pm  0.4$ \\
    \Ds $\pt$ cut (GeV)   & $(0, 5)$   & 2.0 &  $1.82\pm  0.19$ \\
    max $\pt$ cut (GeV)   & $(0.5, 3)$ & 1.0 &  $1.23\pm  0.05$ \\
    sum $\pt$ cut (GeV)   & $(2.5, 6)$ & 3.0 &  $3.19\pm  0.10$ \\
    \hline
  \end{tabular}
  \end{center}
\end{table}

\subsection{Background parameterisation}

The estimation of the combinatorial background density is performed in a similar way to that of the acceptance parameterisation,
except that the ANN training sample also includes the full range of the selection variable, $m_D$, and the test sample
only contains the events in the sidebands (defined in Section~\ref{sec:interpolation}) to reproduce a more realistic analysis scenario. As in the case of the acceptance parameterisation, the background model from Section~\ref{sec:toymc_bkg} is used to both generate samples for the initial joint density estimation by the ANN, and the test dataset for the subsequent maximum likelihood fit. The list of generated parameters of the model, with ranges of their variation, is given in Table~\ref{tab:bkg}, where the size of the training sample is $1\,000\,000$ events.
Here, the topology of the ANN is unchanged from that described in Section~\ref{sec:interpolation}, and the penalty factor in the $L_2$ regularisation term is taken to
be $\lambda_2 = 1$. The projections of ANN variables, after $50\,000$ training epochs, in
the $m'$, $\theta'$ and $m_D$ variables, as well as the correlations of these variables
with each other and with the model parameters, are shown in Appendix~\ref{app:bkg}.

\begin{figure}


  \includegraphics[width=0.34\textwidth]{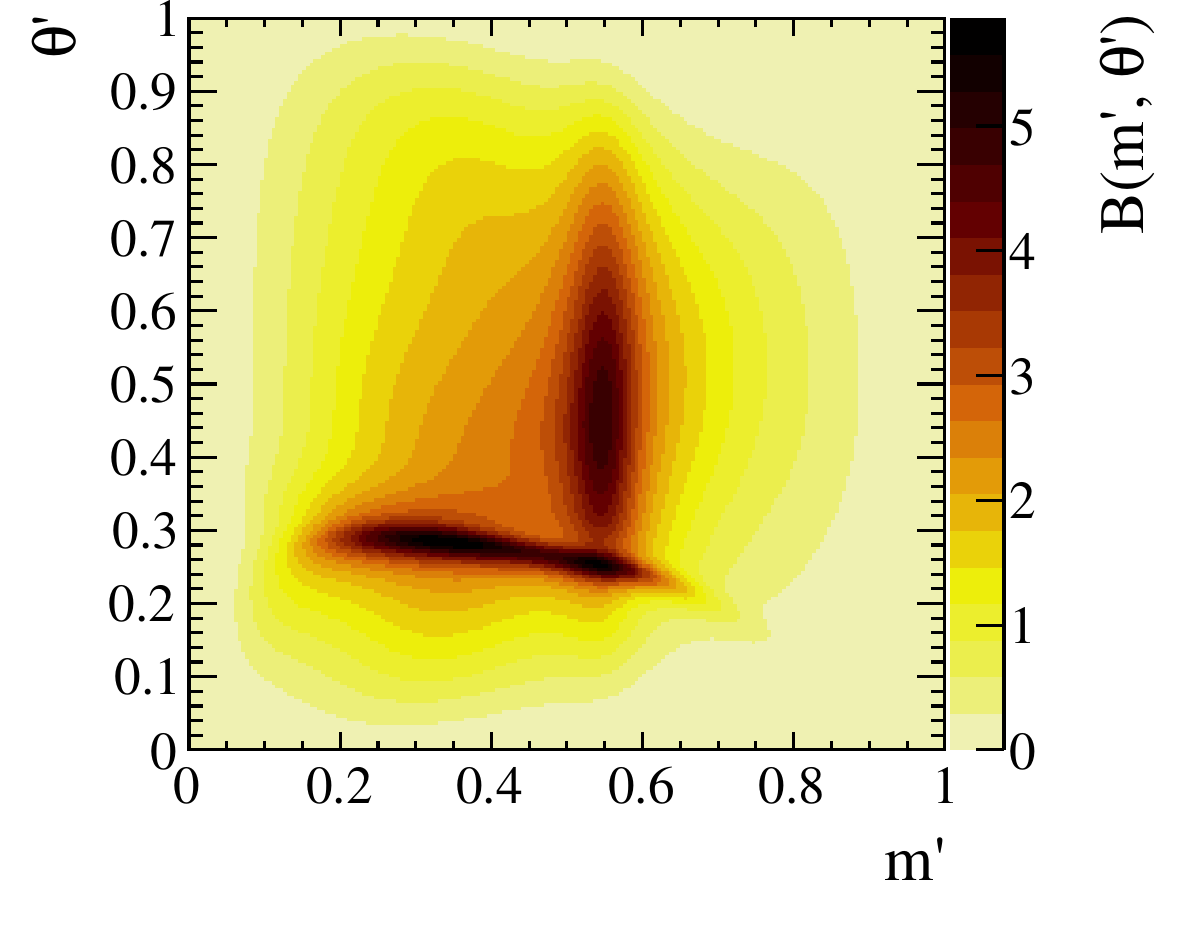}
  \put(-54, 100){\colorbox{white}{\small (a)}}
  \includegraphics[width=0.32\textwidth]{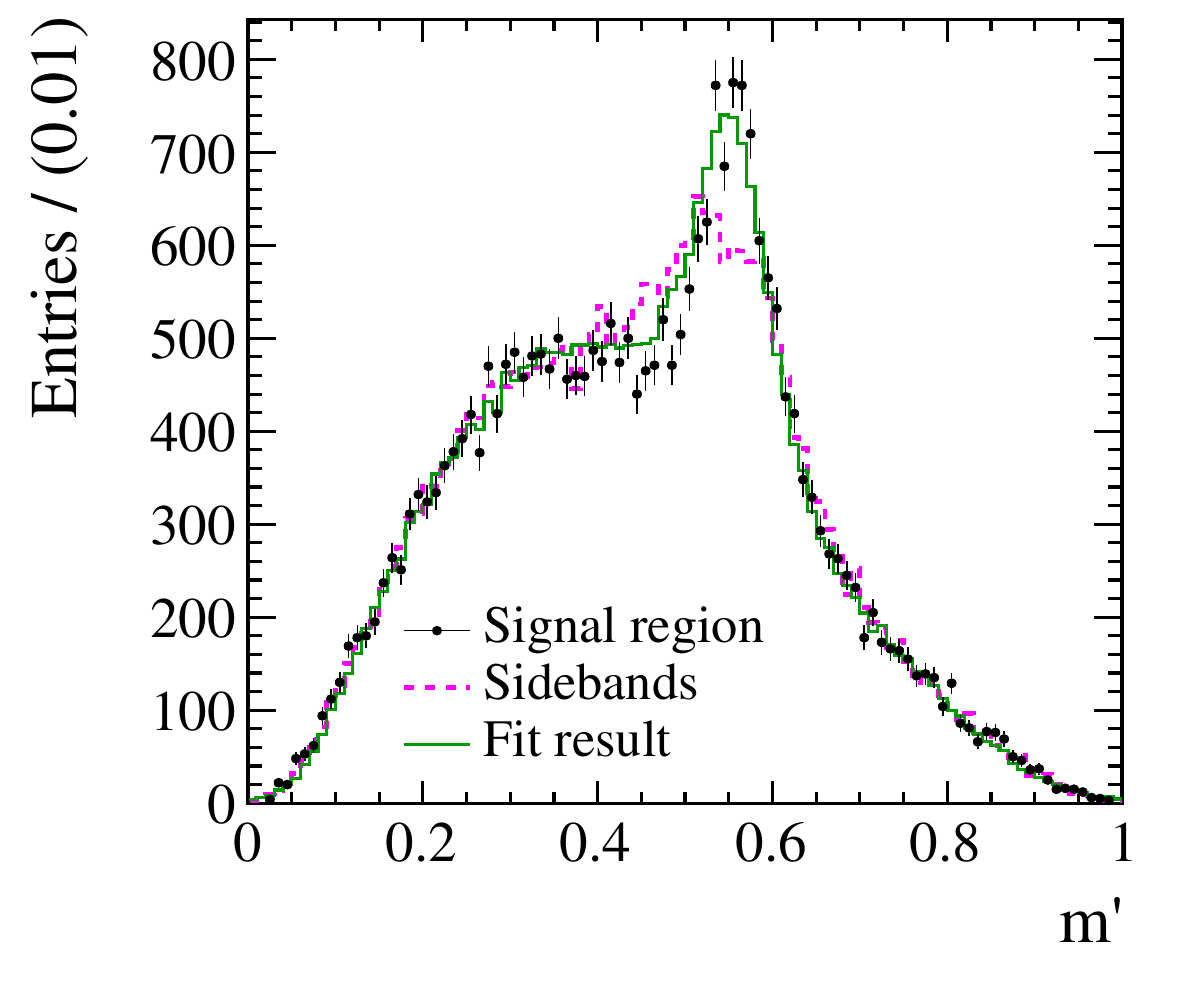}
  \put(-32, 100){\colorbox{white}{\small (b)}}
  \includegraphics[width=0.32\textwidth]{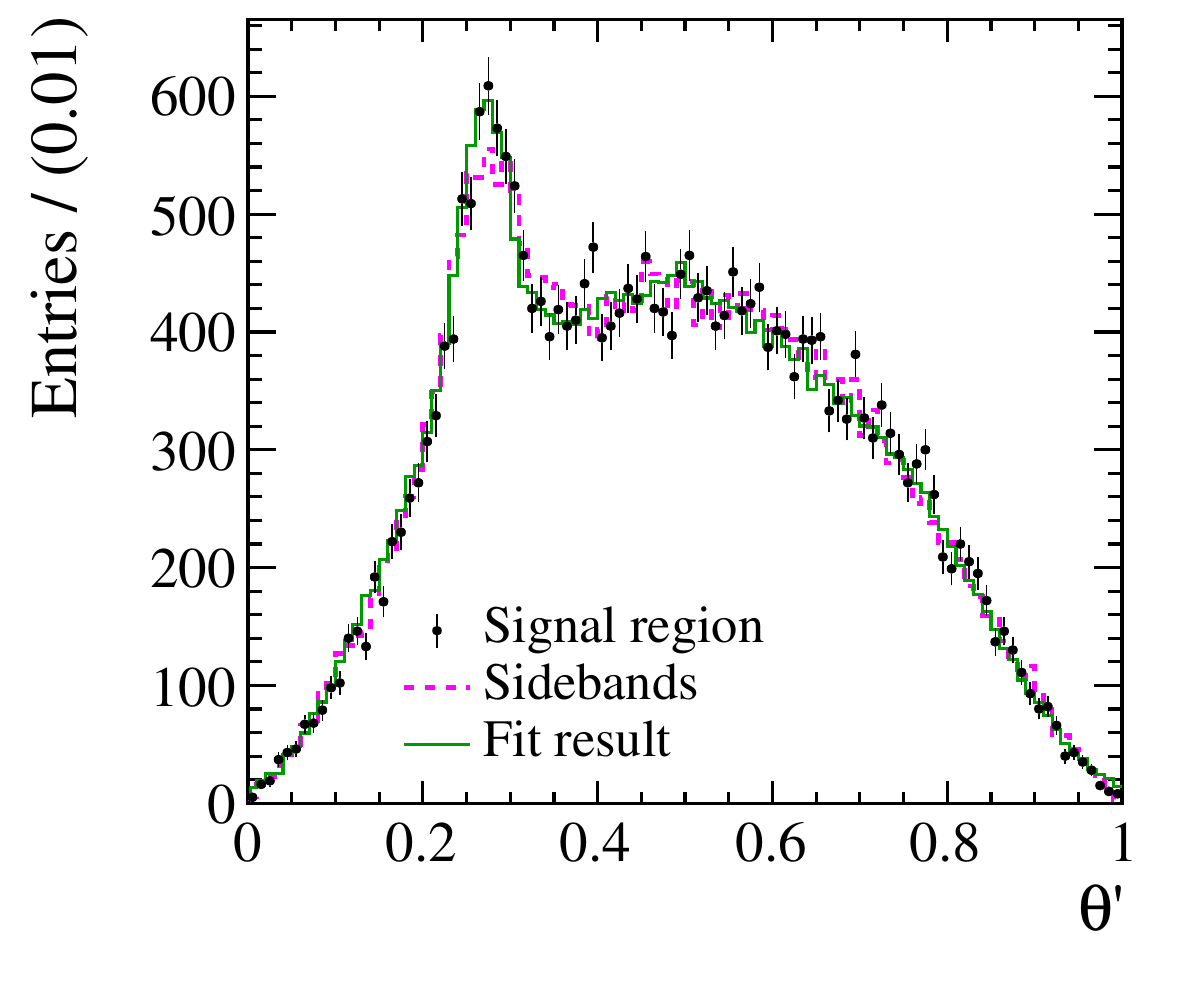}
  \put(-32, 100){\colorbox{white}{\small (c)}}

  \caption{Results of the interpolation of the combinatorial background density in the signal region using the model-assisted ANN:
           (a) the two-dimensional density and (b,c) its projections.}
  \label{fig:bkg_fit_model}
\end{figure}

The projections of the resulting background density estimation are shown in Fig.~\ref{fig:bkg_fit_model}, and the
true and reconstructed values of the model parameters are given in Table~\ref{tab:bkg}. The values
of the reconstructed parameters are consistent with the true values within their uncertainties,
which indicates a good quality of the $11$-dimensional ANN parameterisation of the background model. The fit quality calculated with $50\times 50$ binning is $\chi^2/\mbox{nDoF}=2472.6/2491$. 

\begin{table}
  \caption{Parameters of the $\Ds\to\Kp\pim\pip$ background model:
           the range of parameter variations used at the ANN training stage,
           the true values used in the generated test sample,
           and the reconstructed values extracted from the fit of the ANN model to the test sample.}
  \label{tab:bkg}
  \begin{center}
  \begin{tabular}{|l|c|c|c|}
    \hline
    Model parameter         & Range & True value & Reconstructed value \\
    \hline
    Mean $\pt(K)$ (GeV)    & $(0.2, 1)$   & 0.3 & $0.277\pm 0.010$  \\
    Mean $\pt(\pi)$ (GeV)  & $(0.2, 1)$   & 0.6 & $0.588\pm 0.014$  \\
    Mean $\pt(K^*)$ (GeV)  & $(0.5, 3)$   & 2.0 & $2.12 \pm 0.19$   \\
    Mean $\pt(\rho)$ (GeV) & $(0.5, 3)$   & 2.0 & $1.78 \pm 0.30$   \\
    $\Kstarz$  fraction     & $(0, 0.3)$   & 0.1 & $0.096\pm 0.003$ \\
    $\rhoz$ fraction        & $(0, 0.3)$   & 0.2 & $0.212\pm 0.005$  \\
    Track $\pt$ cut (GeV)  & $(0.1, 0.5)$ & 0.3 & $0.291\pm 0.003$  \\
    Track $p$ cut (GeV)    & $(1, 4)$     & 3.0 & $3.23 \pm 0.14$   \\
    \hline
  \end{tabular}
  \end{center}
\end{table}

\section{Conclusion}

Techniques have been proposed here to efficiently parameterise background and acceptance variations that are an essential component to multidimensional fits of hadronic decay amplitudes. Often, treatments of the acceptance variations are sub-optimal, as they either do not exploit rudimentary assumptions of local smoothness, and require large quantities of computationally expensive simulated data, or a sizeable systematic uncertainty results from the use of an inefficient parameterisation. For the background description, assumptions often have to be made on the functional form of specific backgrounds, or on the validity of an extrapolation into the signal region. These assumptions are difficult to validate, and therefore in these cases, the background can be a considerable source of systematic uncertainty.

Here, several new applications of Gaussian processes and neural networks are proposed that attempt to mitigate these issues, by utilising a more efficient parameterisation, or imposing regularisation constraints on a model with a large number of degrees of freedom. Additionally, a method is proposed that utilises a neural network to extract latent dependencies on physics parameters, which permits a more physically motivated way of imposing constraints on the resulting probability density.

The techniques proposed in this paper reduce the overall systematic uncertainty from the aforementioned effects by providing a more efficient, regularised description of the acceptance variations. Furthermore, by doing this they also reduce the computational burden of generating sufficient simulated data to obtain robust results.

These approaches also mitigate biases in the estimation of the background contributions due to correlations with selection variables; permit extrapolation into the signal region of backgrounds that, for example, are not constant throughout the control variables due to decay kinematics or kinematical constraints; and scale efficiently with increasing dimensionality.

\section*{Acknowledgements}

The authors would like to thank their colleagues from the LHCb collaboration
(Tim Gershon, Thomas Latham, Mark Whitehead, Sneha Malde, Maurizio Martinelli, and others) for fruitful discussions.
This work is supported by the Science and Technology Facilities Council (United Kingdom).

\section*{Appendices}

\appendix

\section{Neural network efficiency model}

\label{app:eff}

\begin{figure}
  \centering
  \includegraphics[width=0.245\textwidth]{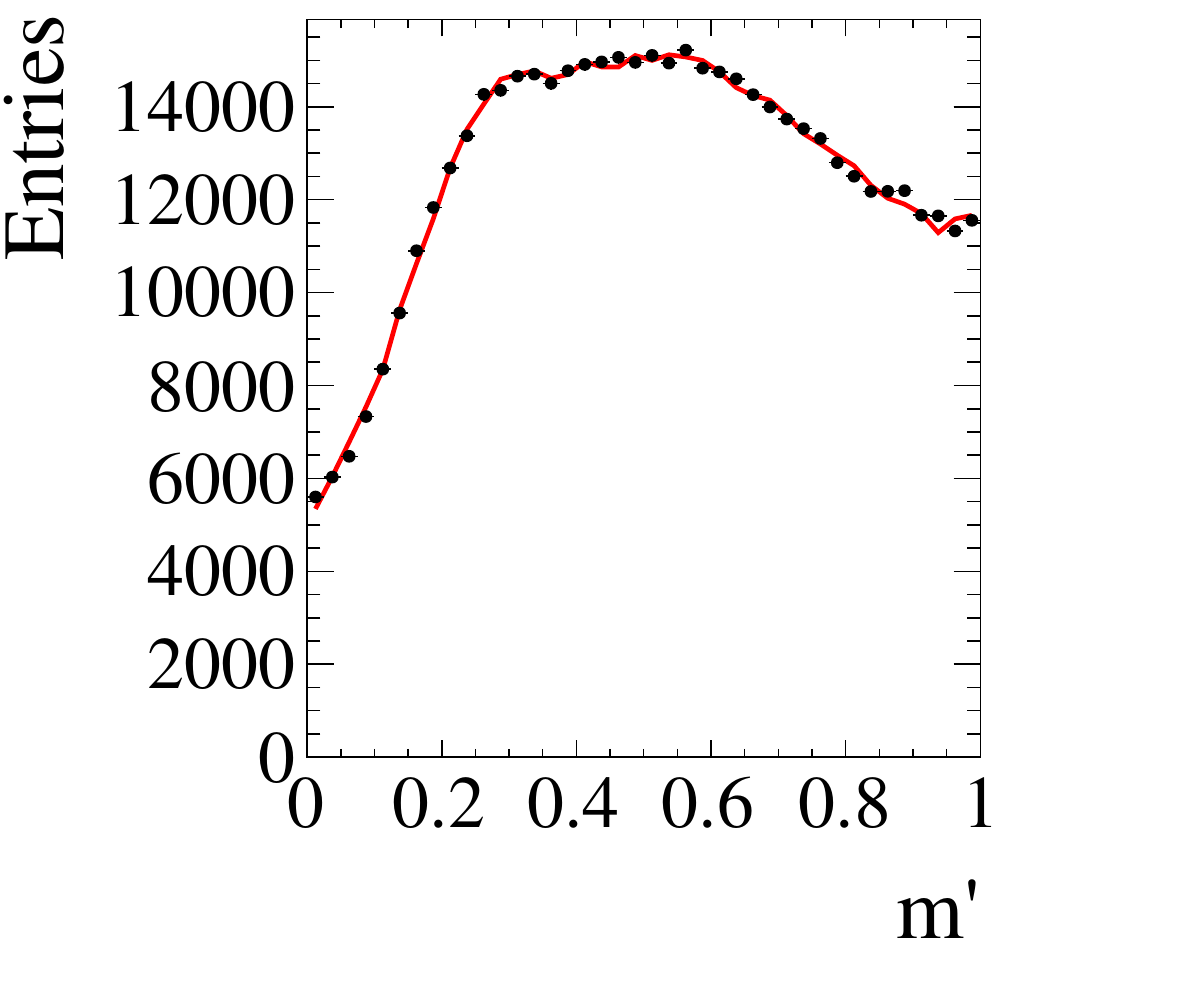}
  \includegraphics[width=0.245\textwidth]{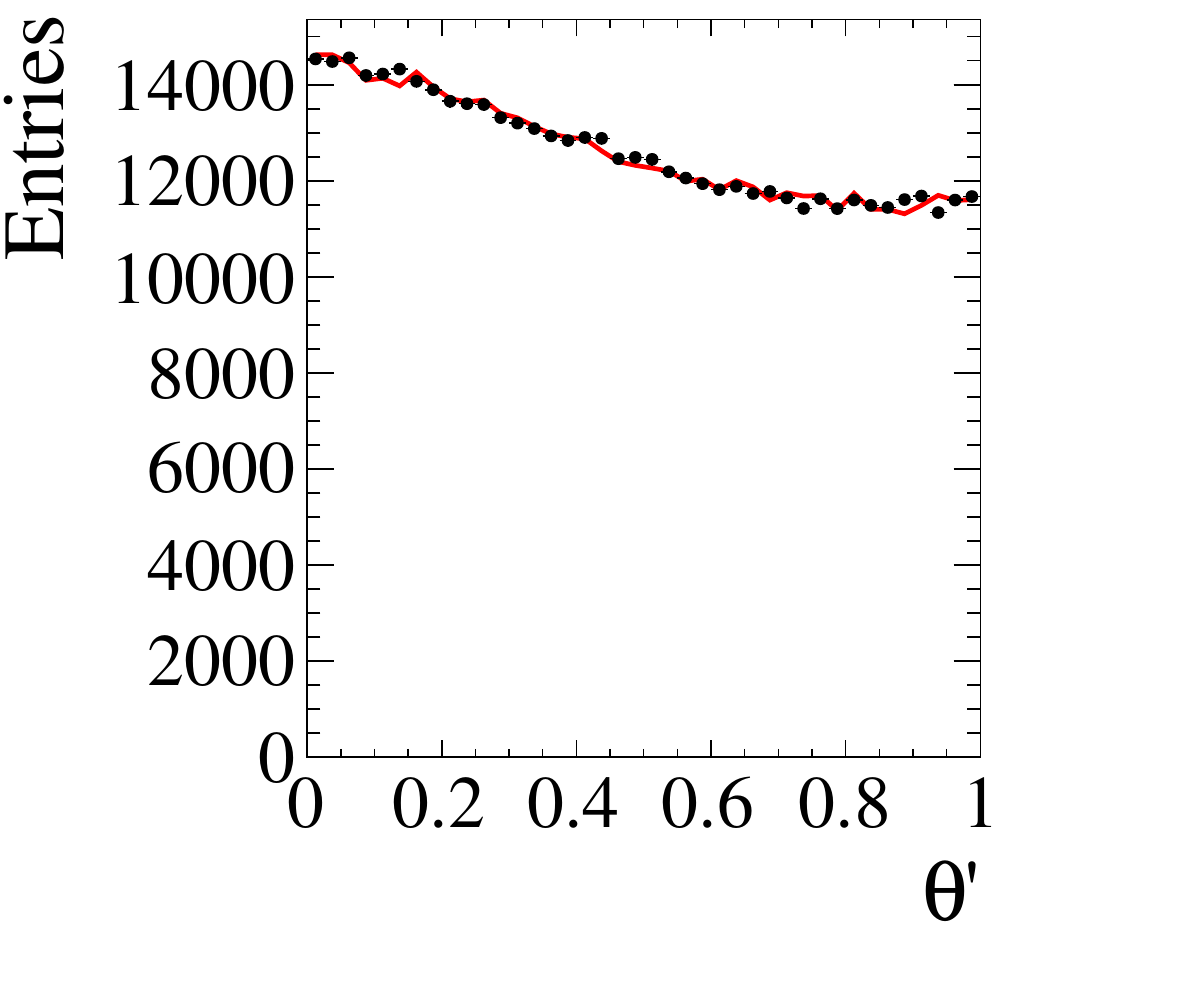}
  \includegraphics[width=0.494\textwidth]{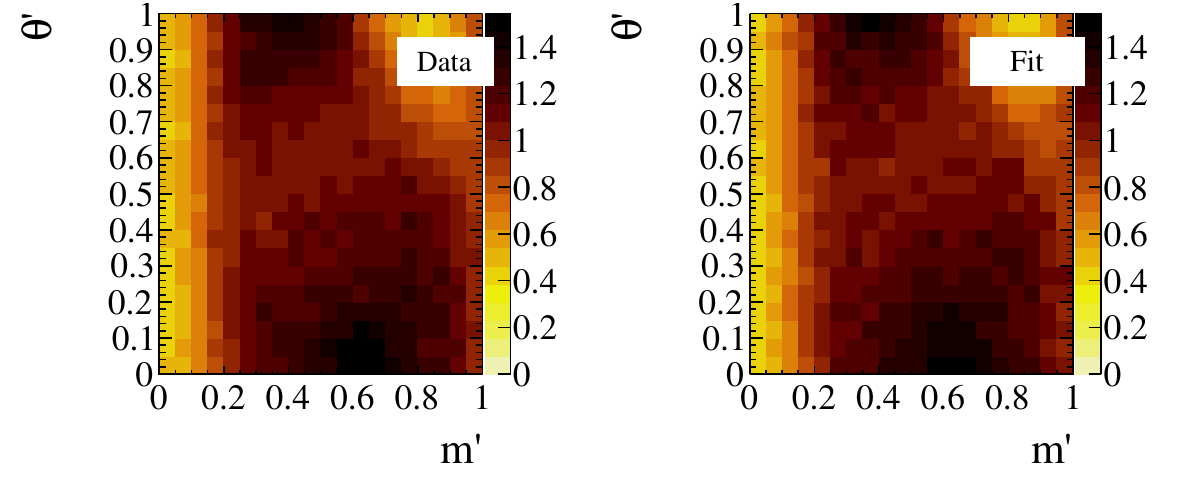}

  \includegraphics[width=0.494\textwidth]{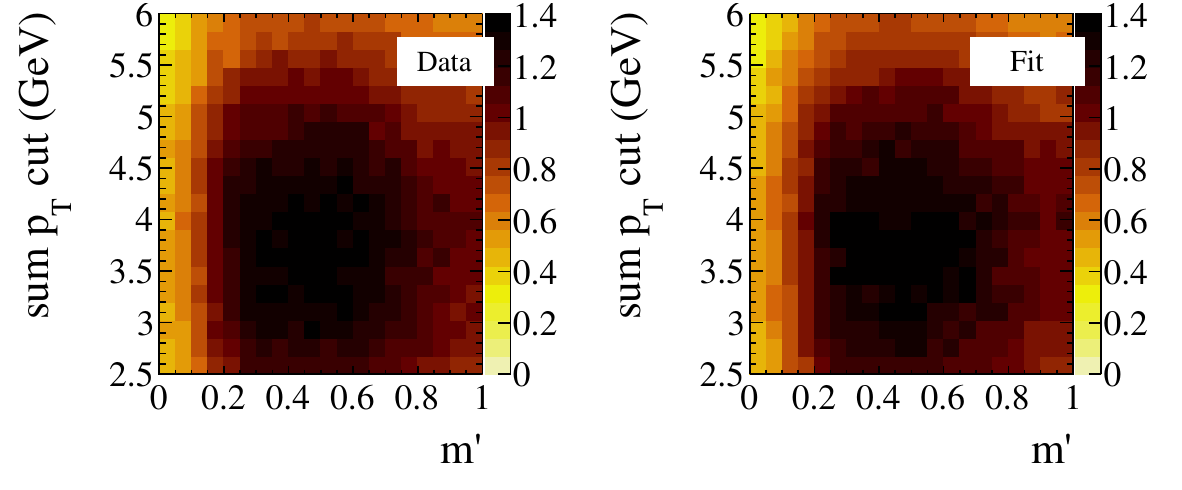}
  \includegraphics[width=0.494\textwidth]{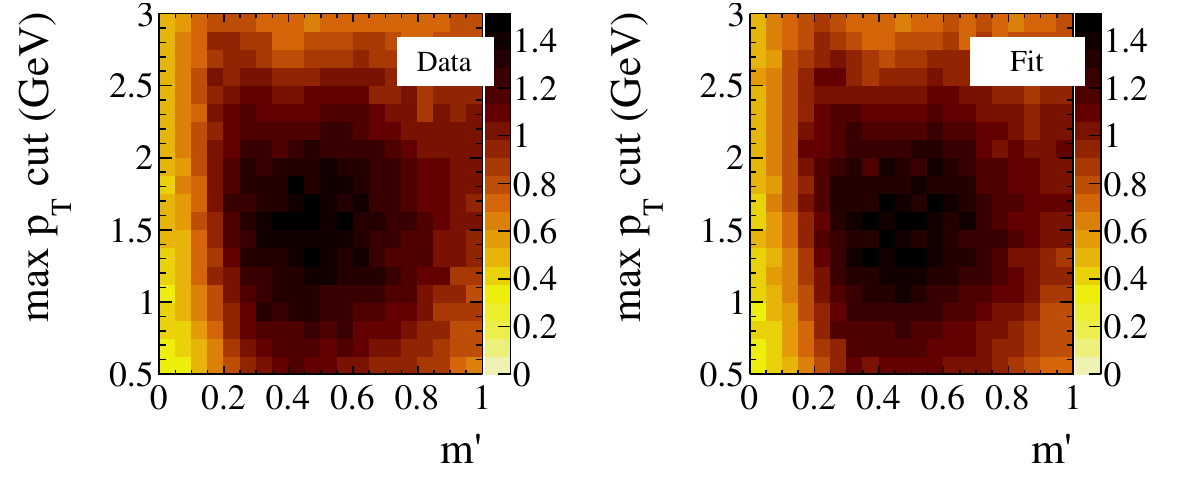}

  \includegraphics[width=0.494\textwidth]{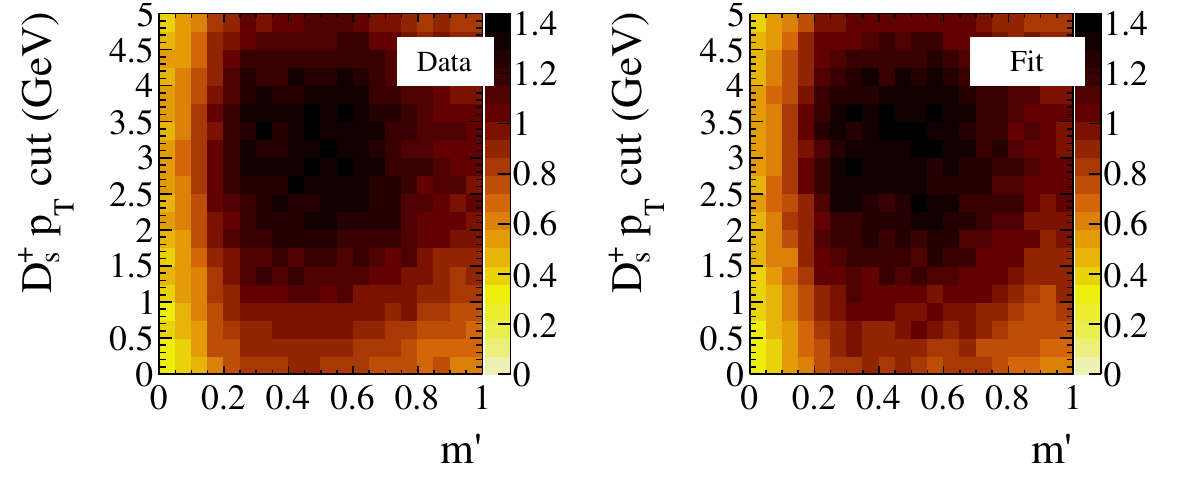}
  \includegraphics[width=0.494\textwidth]{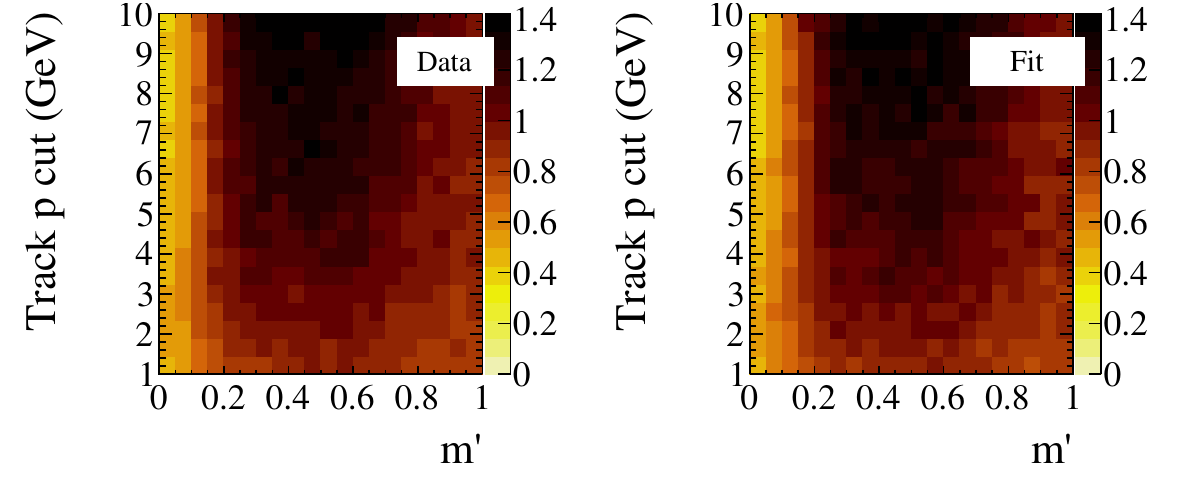}

  \includegraphics[width=0.494\textwidth]{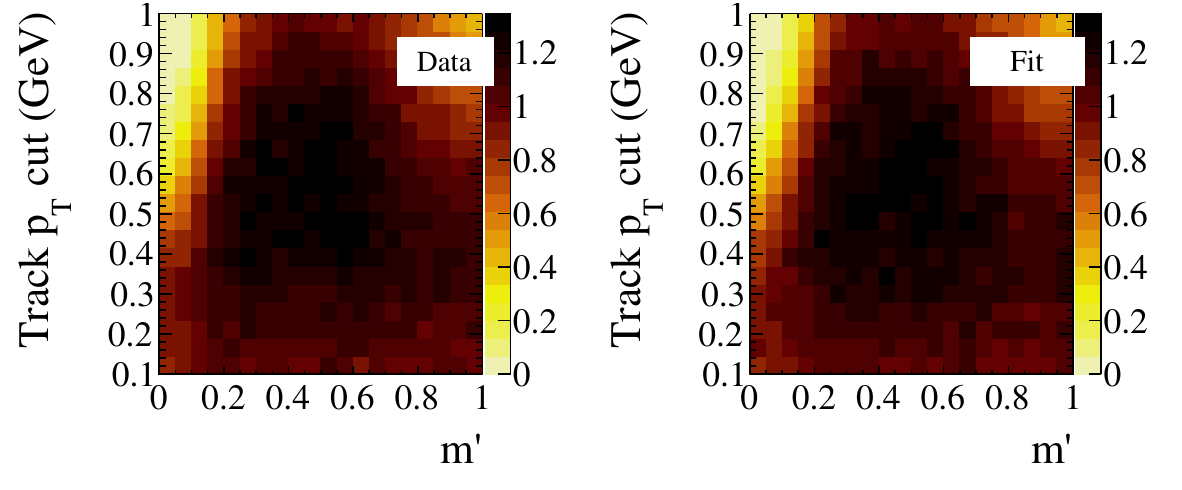}
  \includegraphics[width=0.494\textwidth]{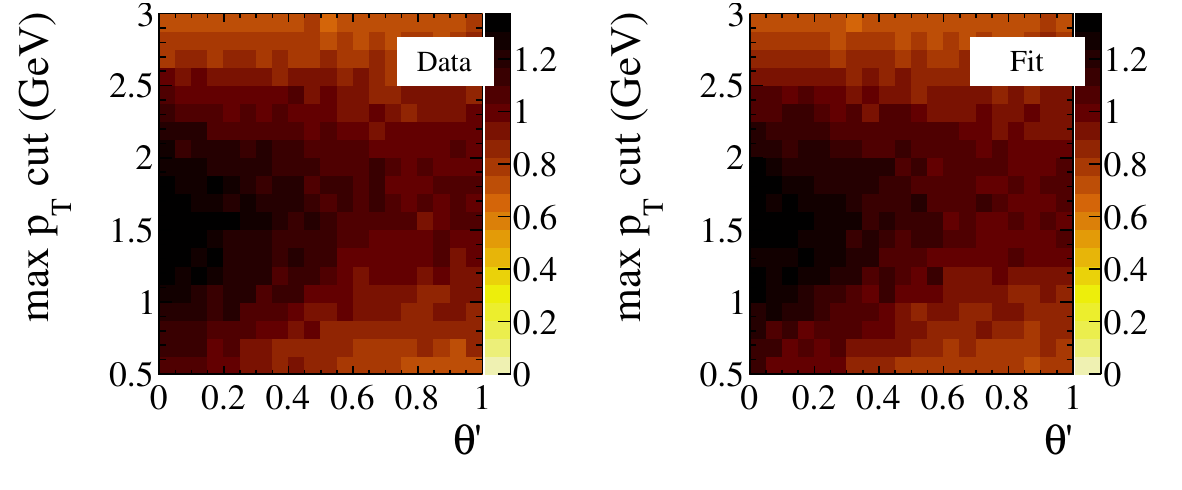}

  \includegraphics[width=0.494\textwidth]{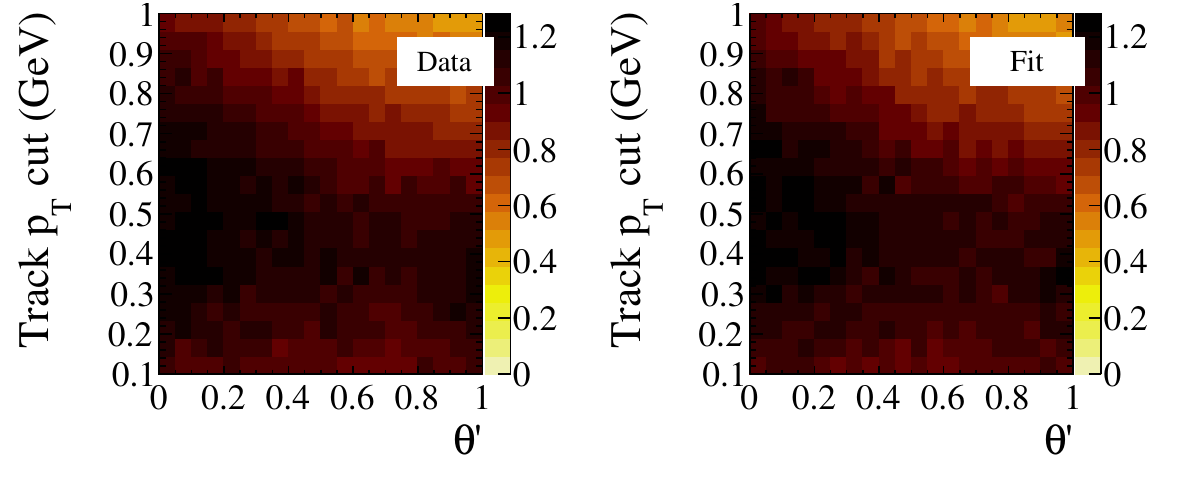}
  \includegraphics[width=0.494\textwidth]{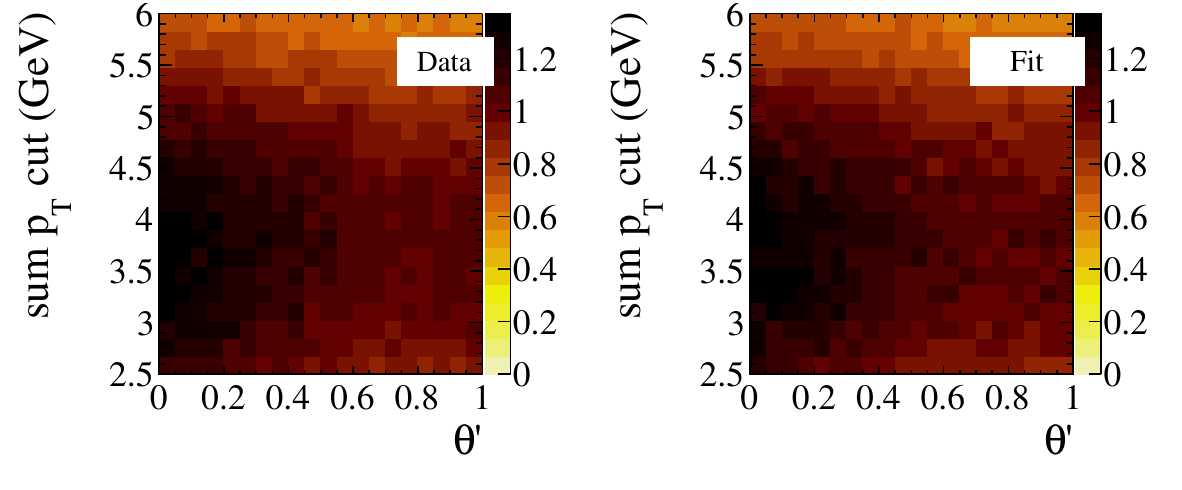}

  \includegraphics[width=0.494\textwidth]{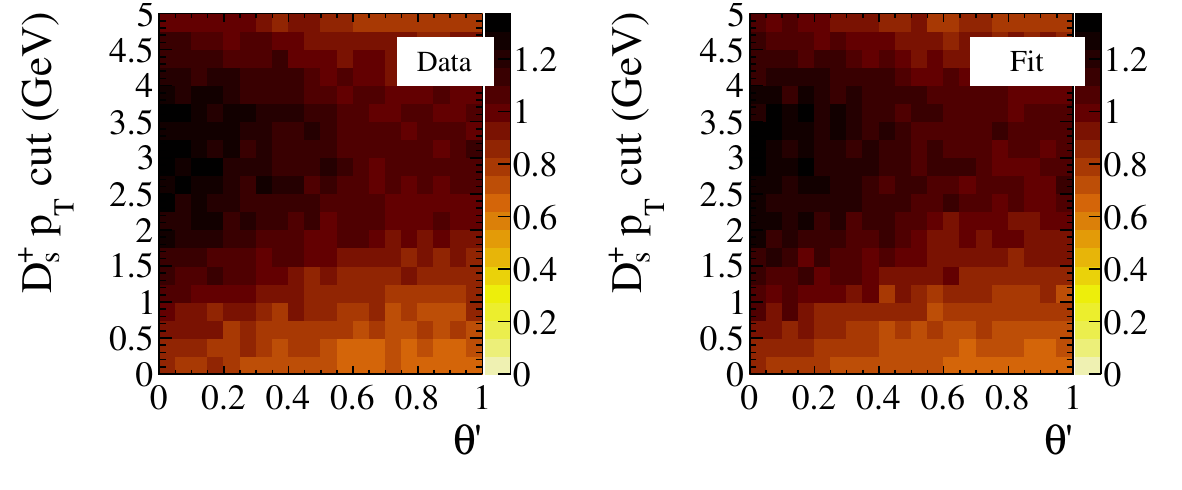}
  \includegraphics[width=0.494\textwidth]{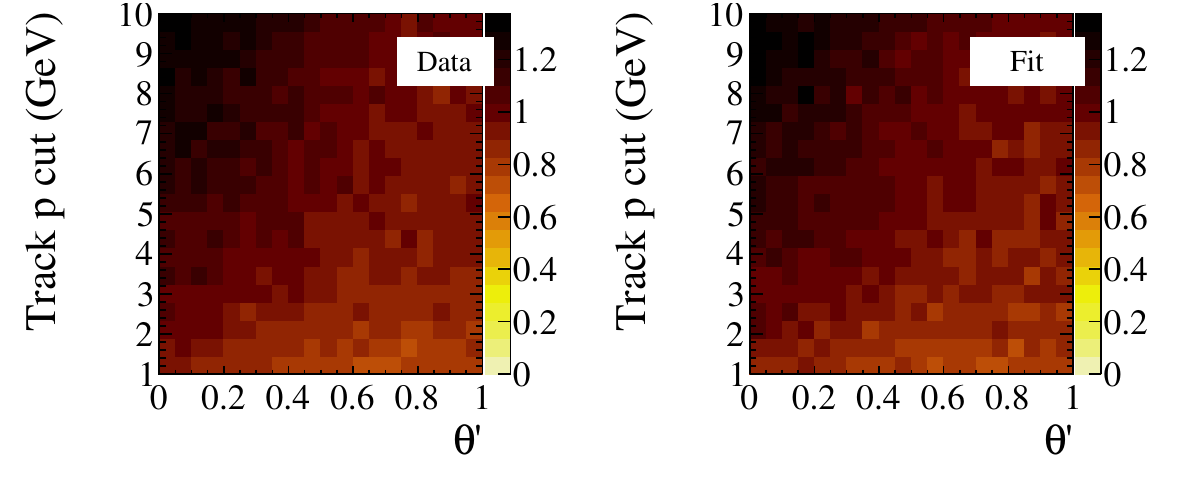}

  \caption{Results of the estimation of the simulated acceptance distribution, as a function of the effective model parameters, using an ANN.}
  \label{fig:eff_train_model}
\end{figure}

The results of the ANN training to estimate the $7$-dimensional density of a generated efficiency model, as a function of a set of effective model parameters, is shown in Fig.~\ref{fig:eff_train_model}.
These plots show the projections of the $m'$ and $\theta'$ variables (top row, two leftmost plots), as well as the
normalised two-dimensional distributions for the correlations between $m'$ and $\theta'$
(top row, two rightmost plots) and correlations between $m'$ or $\theta'$ and the model parameters.
The plots marked as ``Data" show the training data distributions, those marked with ``Fit" show the high-statistics
distributions generated from the result of the ANN training. As in the rest of the paper, the two-dimensional distributions
are normalised in such a way that the average of the projected distribution equals $1$.

\section{Neural network background model}

\label{app:bkg}

\begin{figure}
  \centering
  \includegraphics[width=0.245\textwidth]{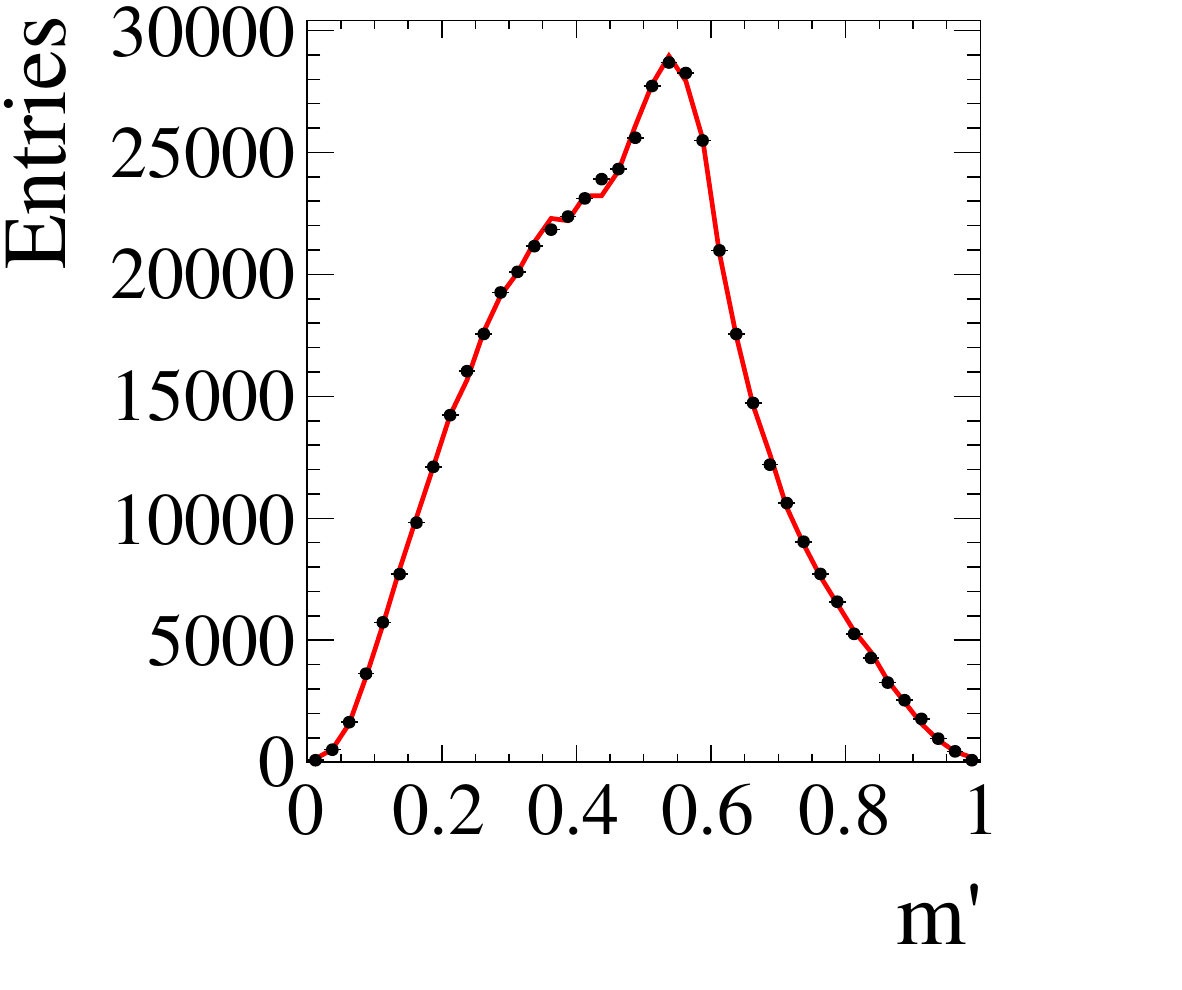}
  \includegraphics[width=0.245\textwidth]{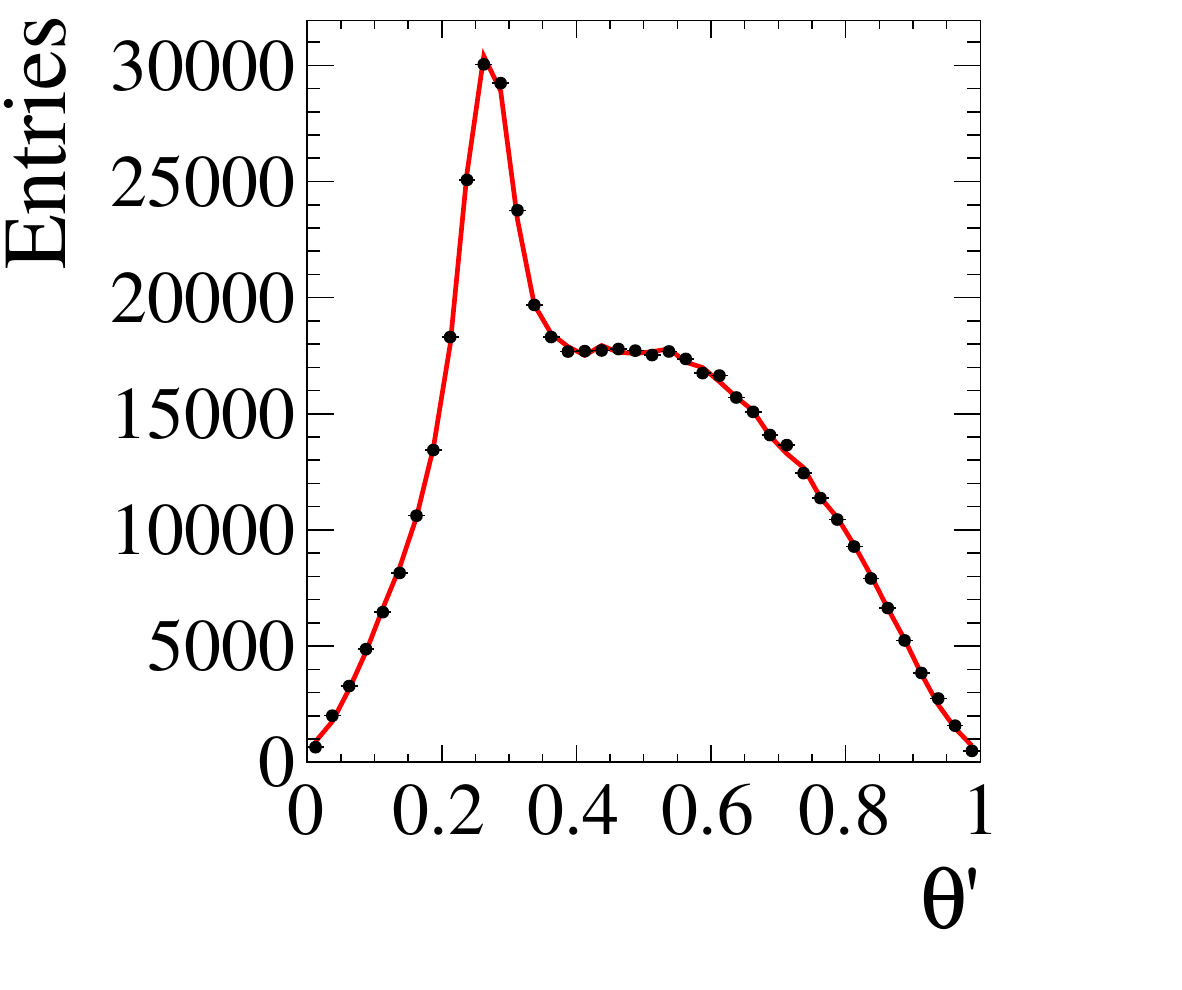}
  \includegraphics[width=0.245\textwidth]{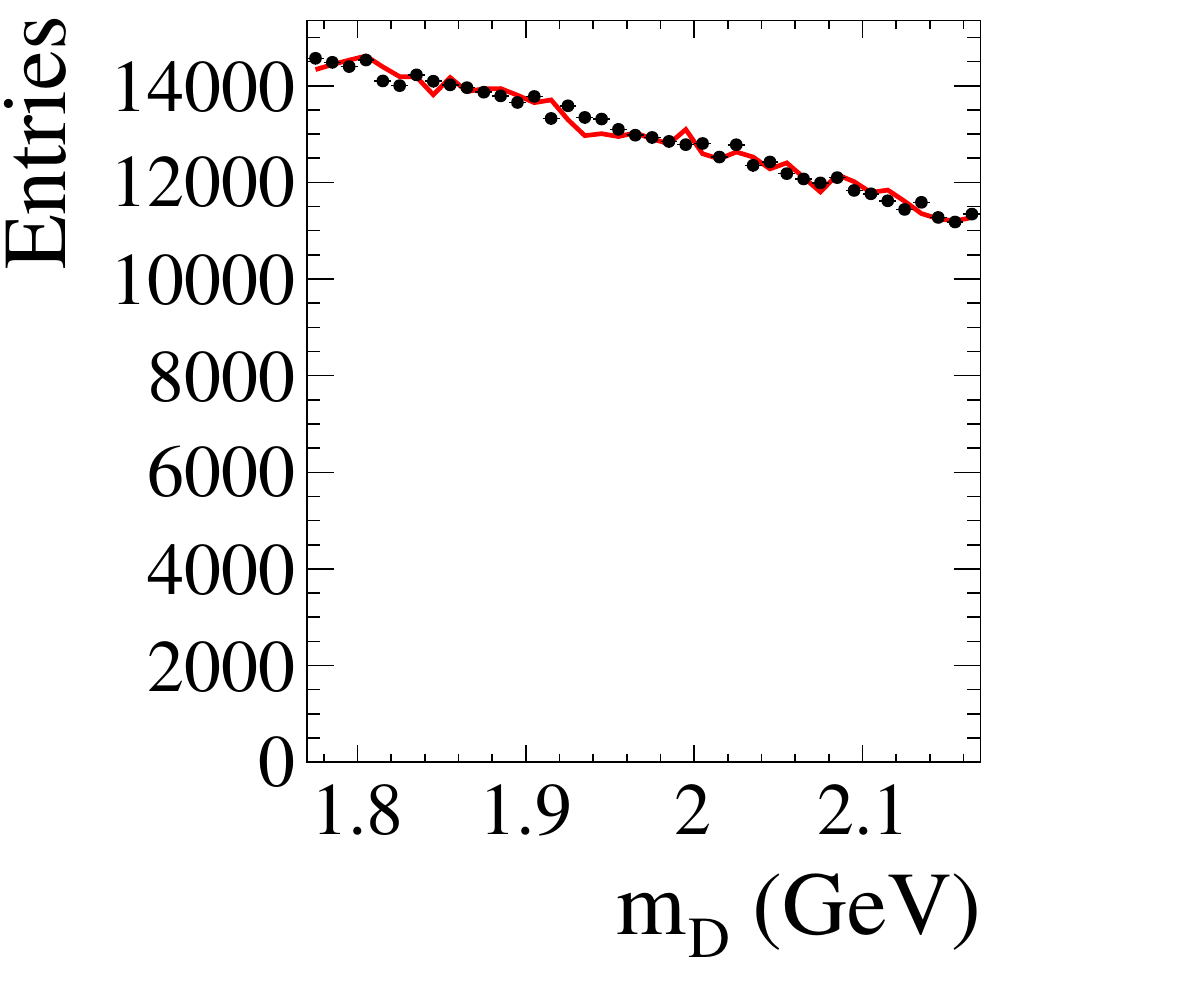}

  \includegraphics[width=0.494\textwidth]{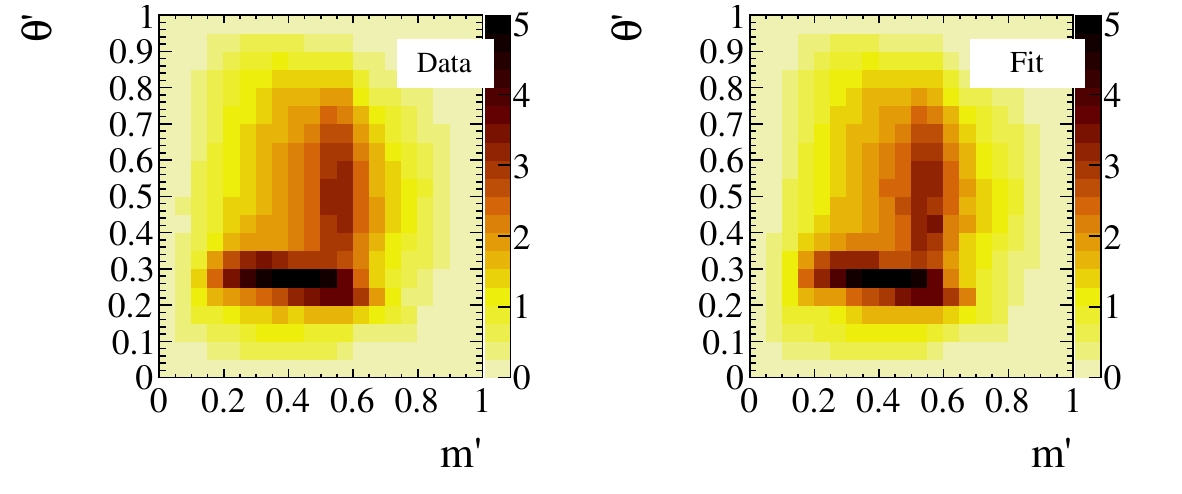}
  \includegraphics[width=0.494\textwidth]{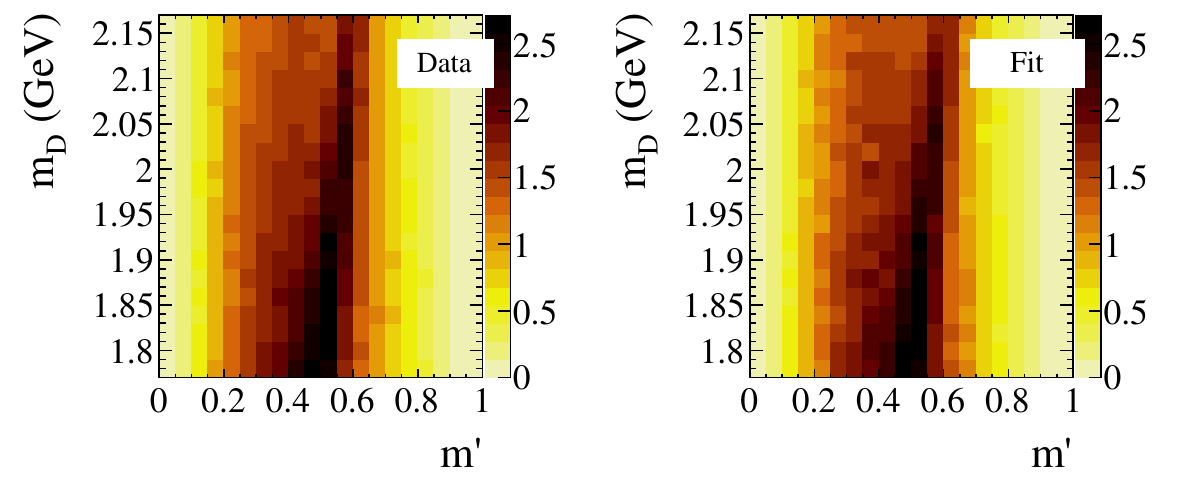}

  \includegraphics[width=0.494\textwidth]{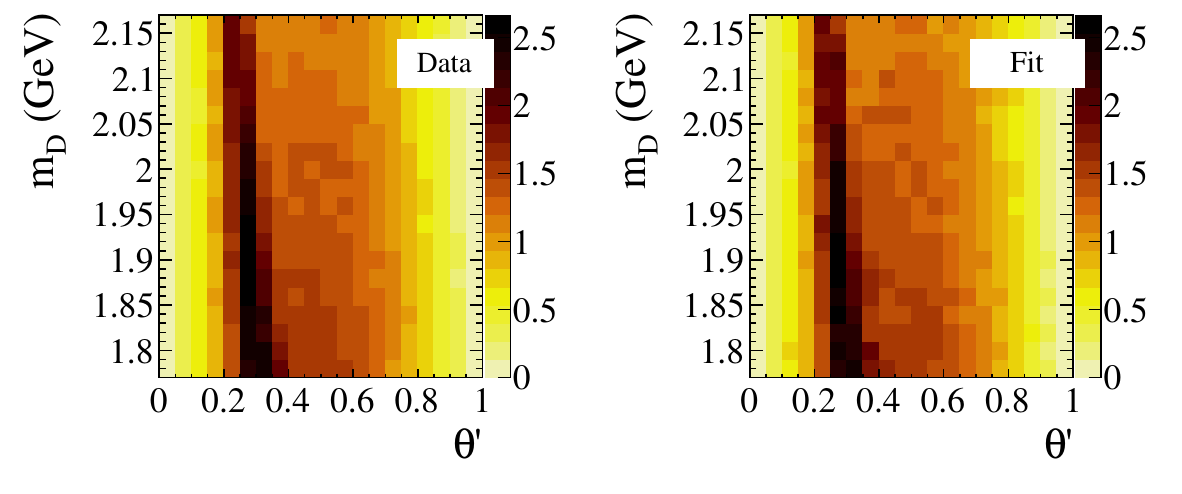}
  \includegraphics[width=0.494\textwidth]{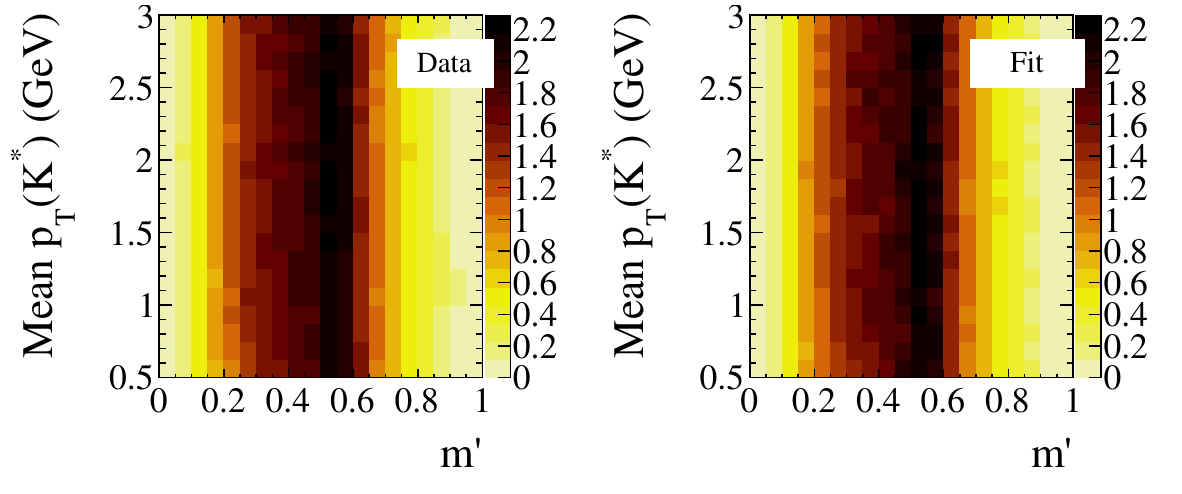}

  \includegraphics[width=0.494\textwidth]{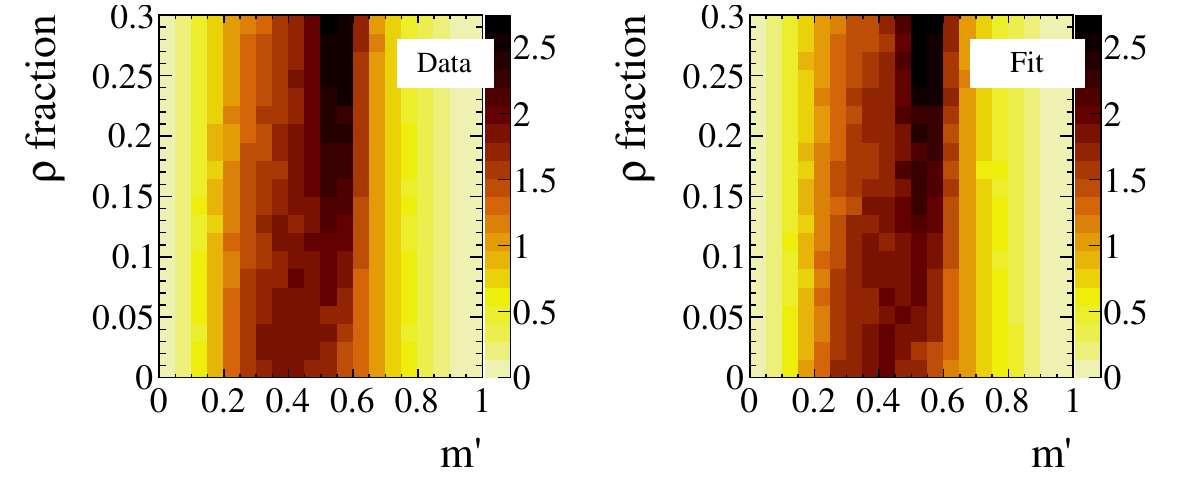}
  \includegraphics[width=0.494\textwidth]{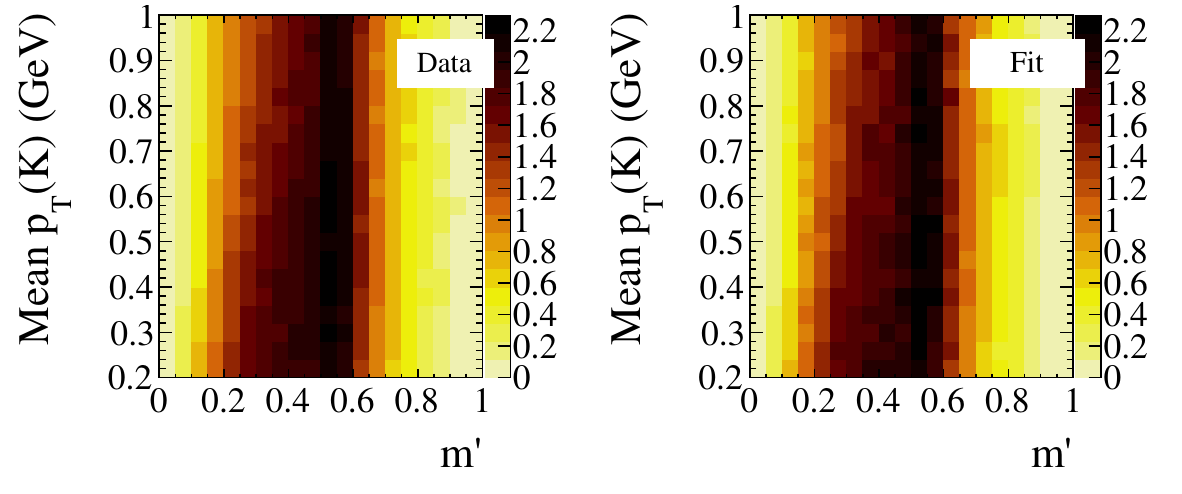}

  \includegraphics[width=0.494\textwidth]{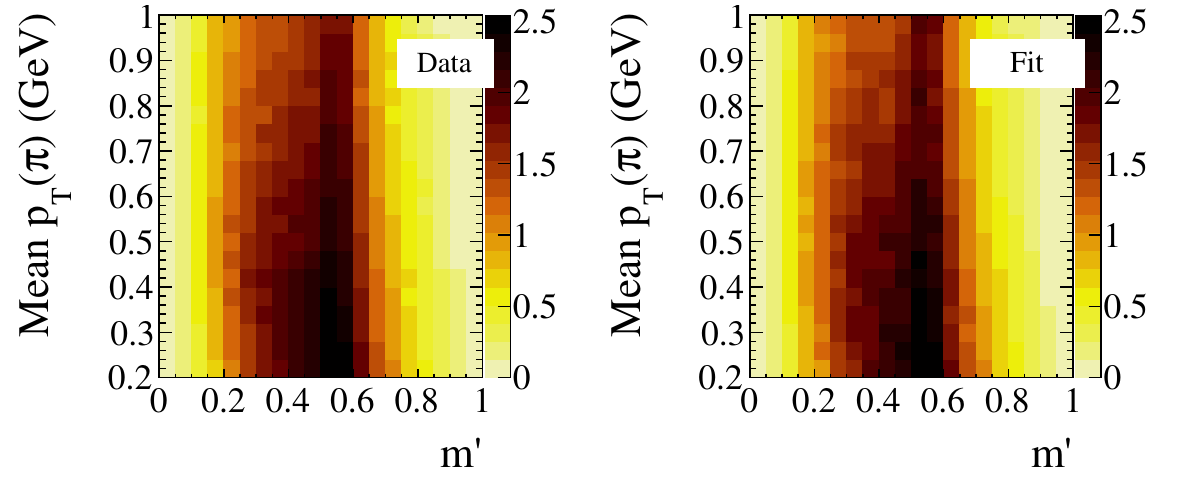}
  \includegraphics[width=0.494\textwidth]{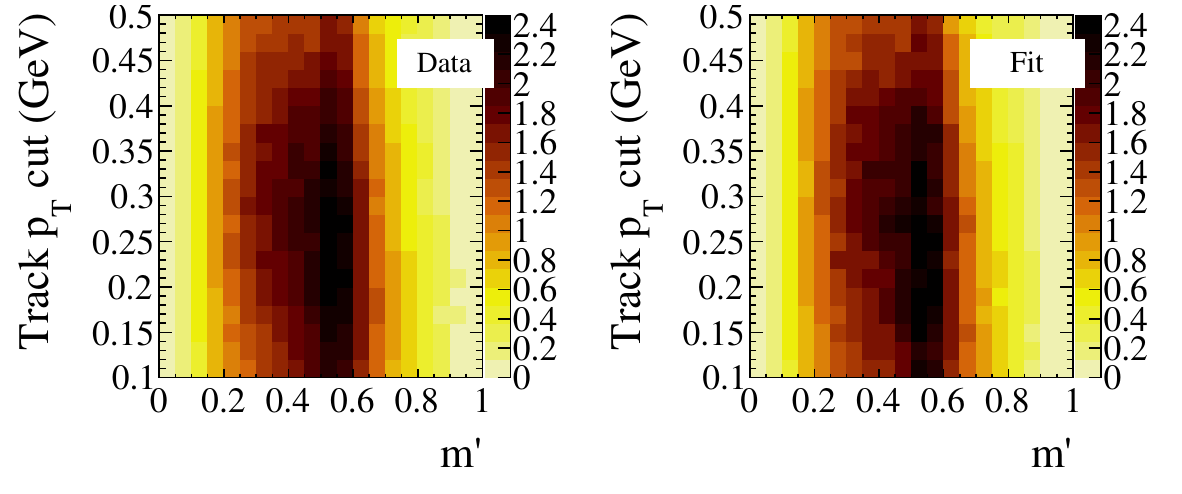}

  \includegraphics[width=0.494\textwidth]{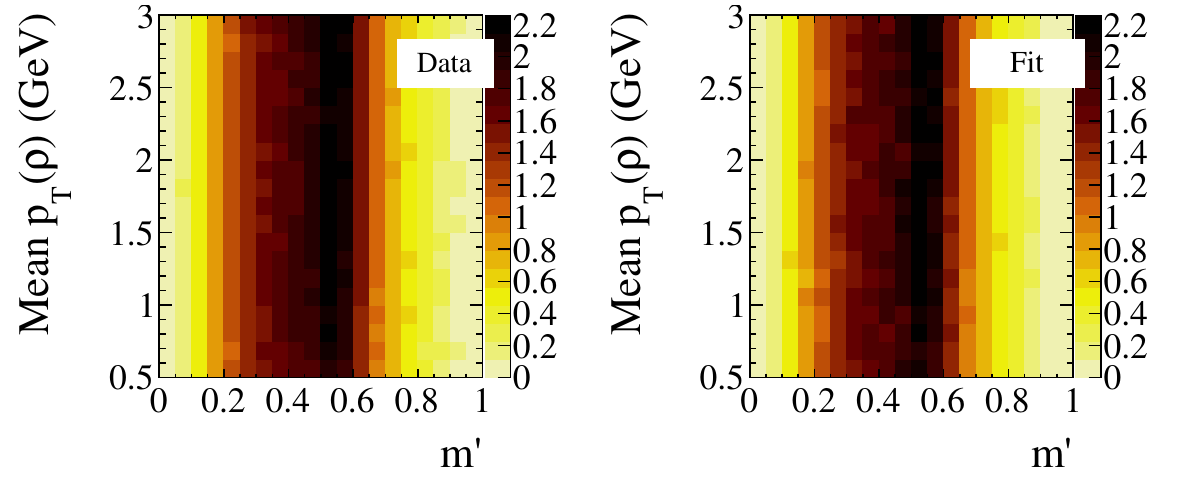}
  \includegraphics[width=0.494\textwidth]{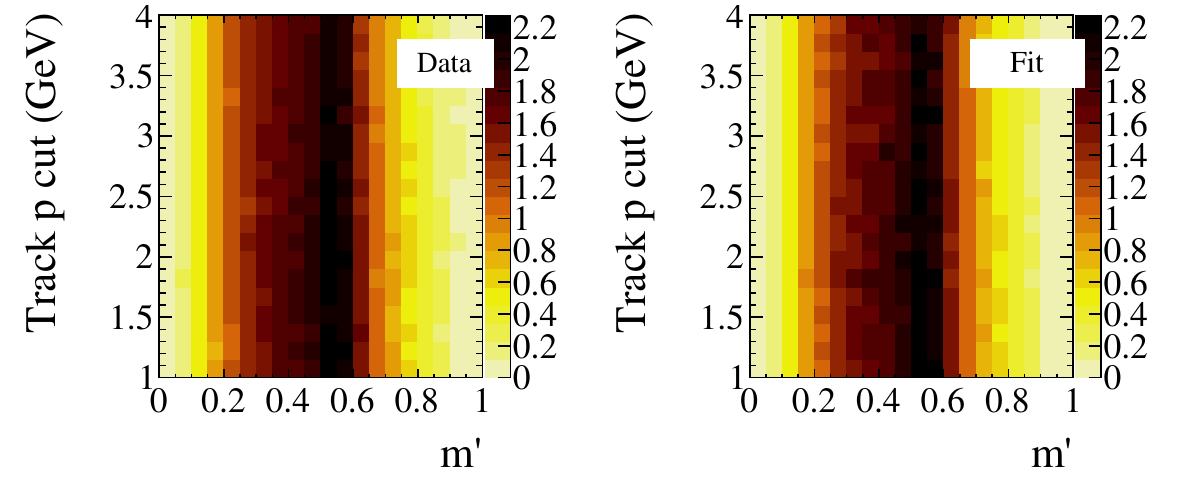}

  \caption{Results of the estimation of the simulated combinatorial background density, as a function of effective model parameters, using an ANN. }
  \label{fig:bkg_train_model}
\end{figure}

\begin{figure}
  \centering
  \includegraphics[width=0.494\textwidth]{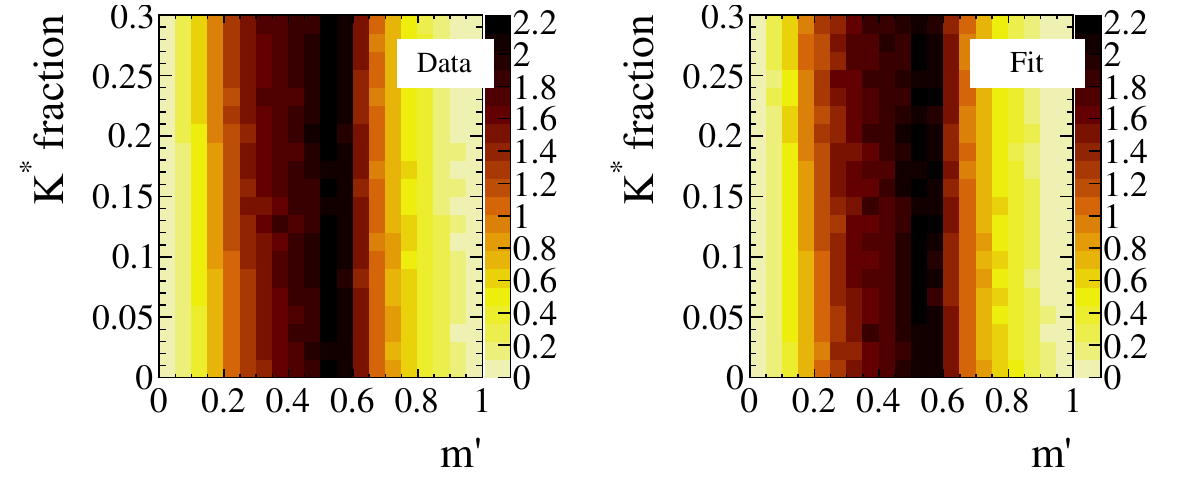}
  \includegraphics[width=0.494\textwidth]{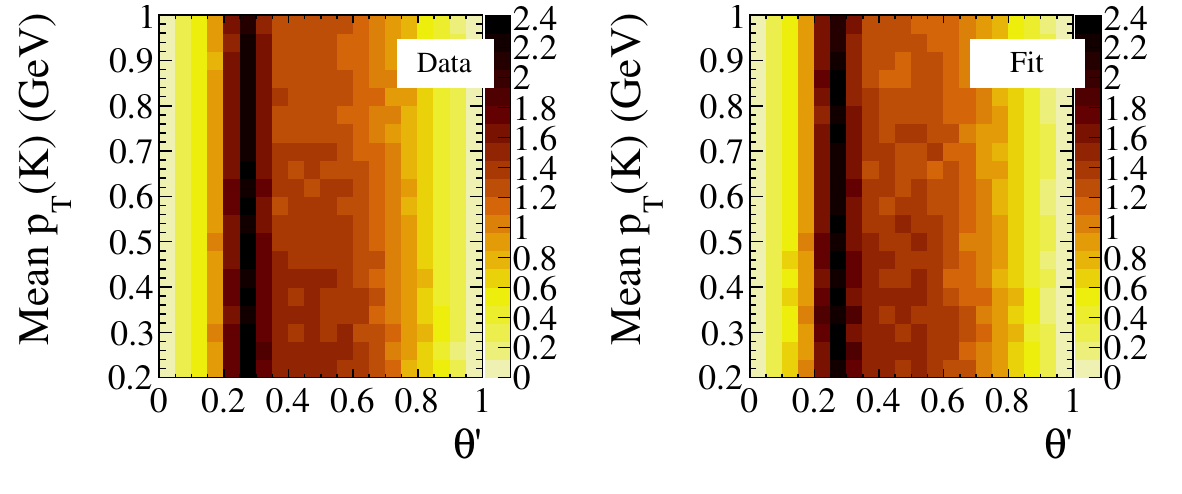}

  \includegraphics[width=0.494\textwidth]{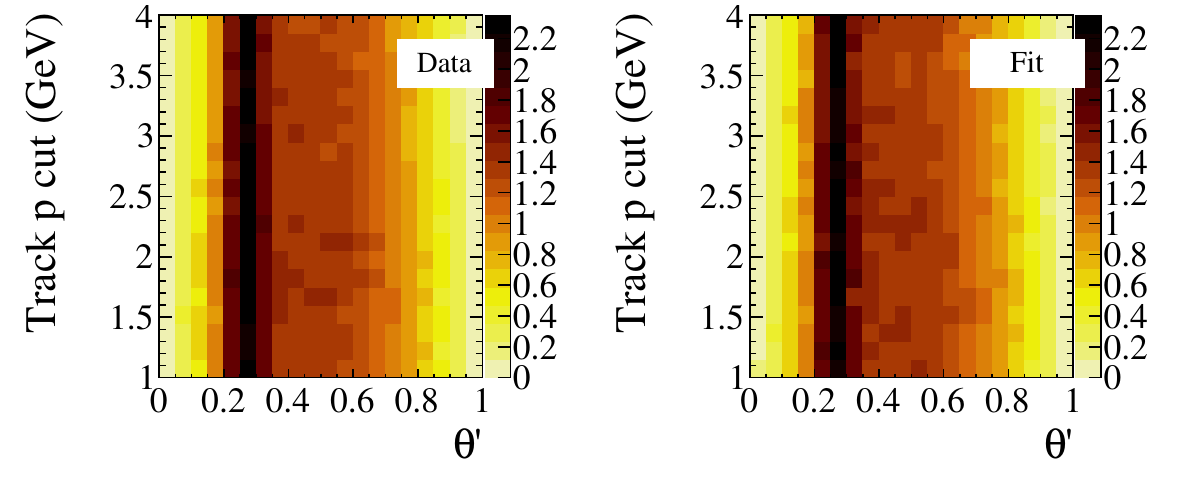}
  \includegraphics[width=0.494\textwidth]{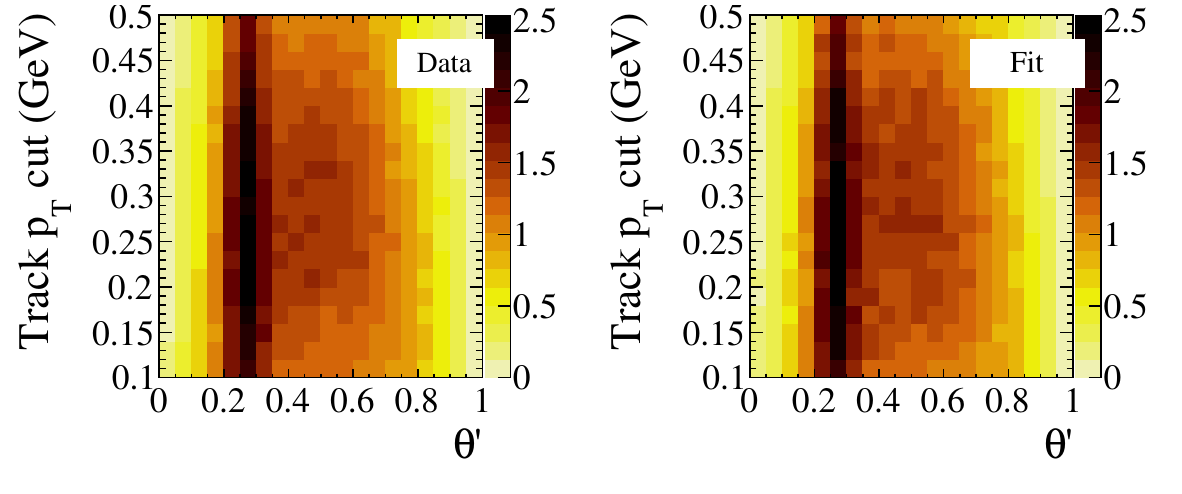}

  \includegraphics[width=0.494\textwidth]{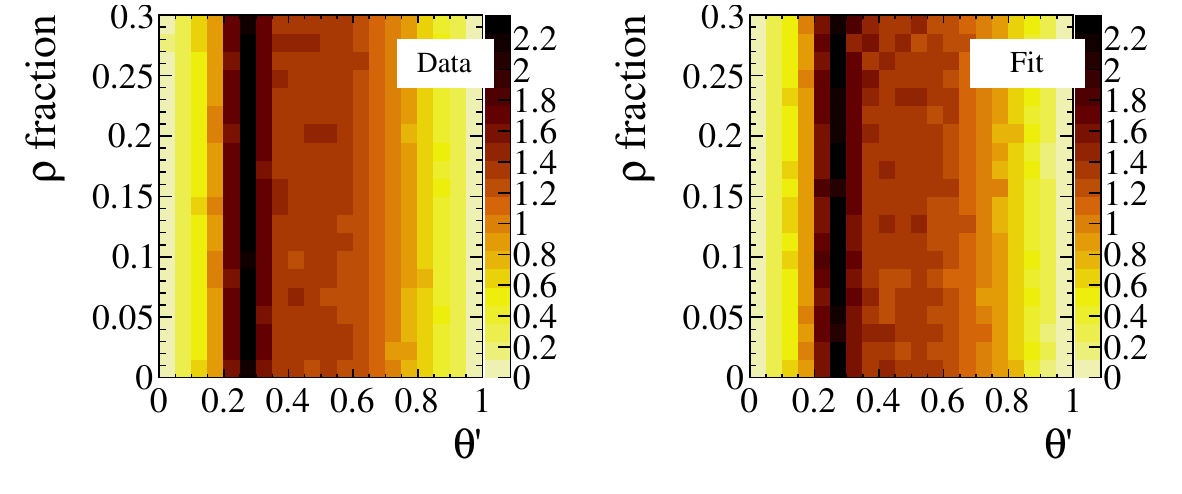}
  \includegraphics[width=0.494\textwidth]{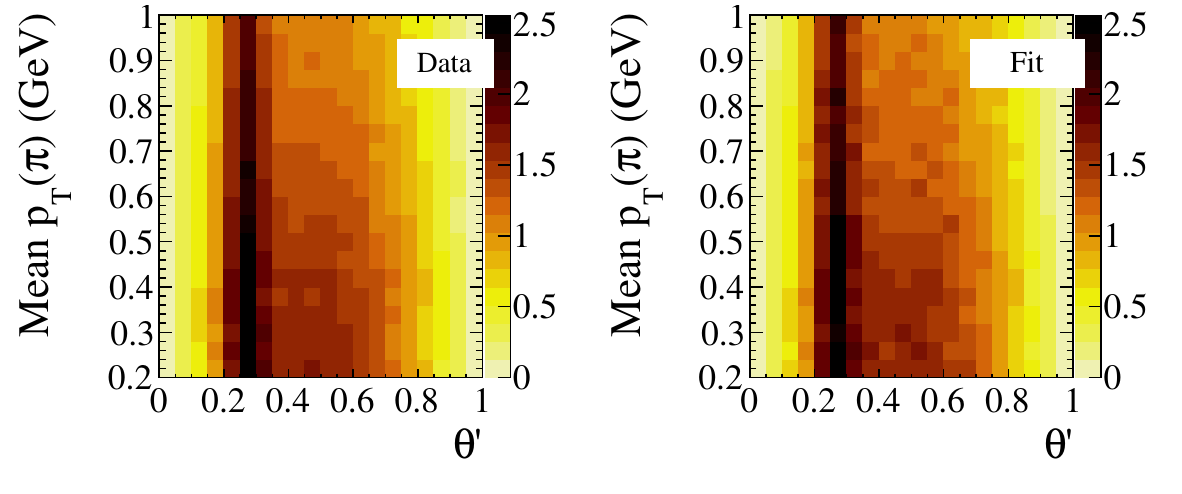}

  \includegraphics[width=0.494\textwidth]{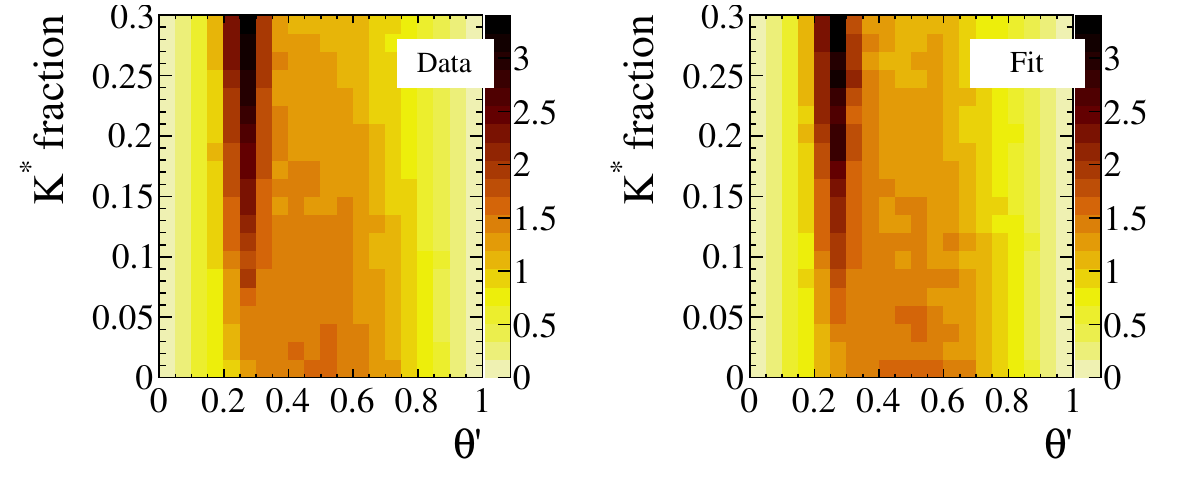}
  \includegraphics[width=0.494\textwidth]{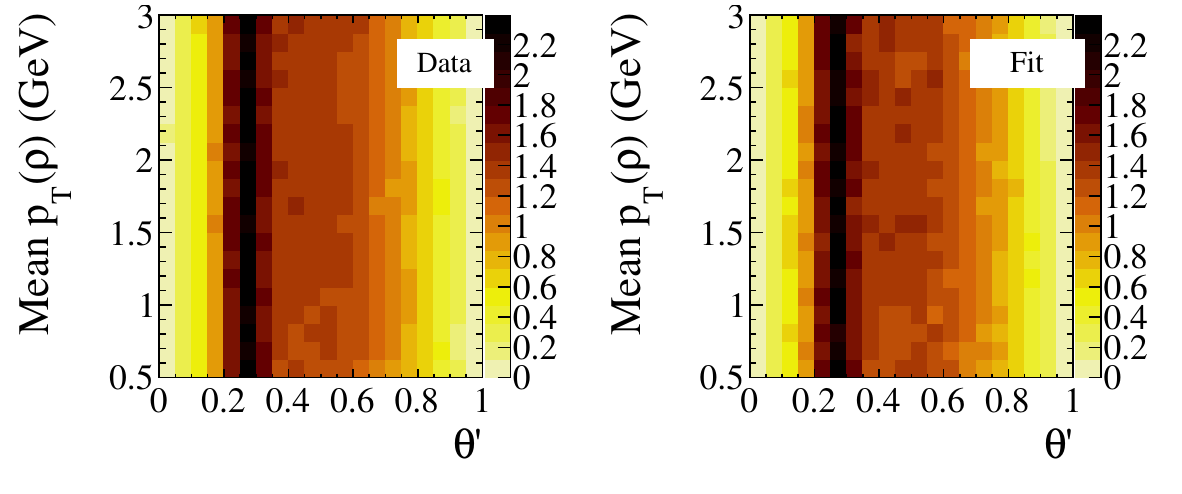}

  \includegraphics[width=0.494\textwidth]{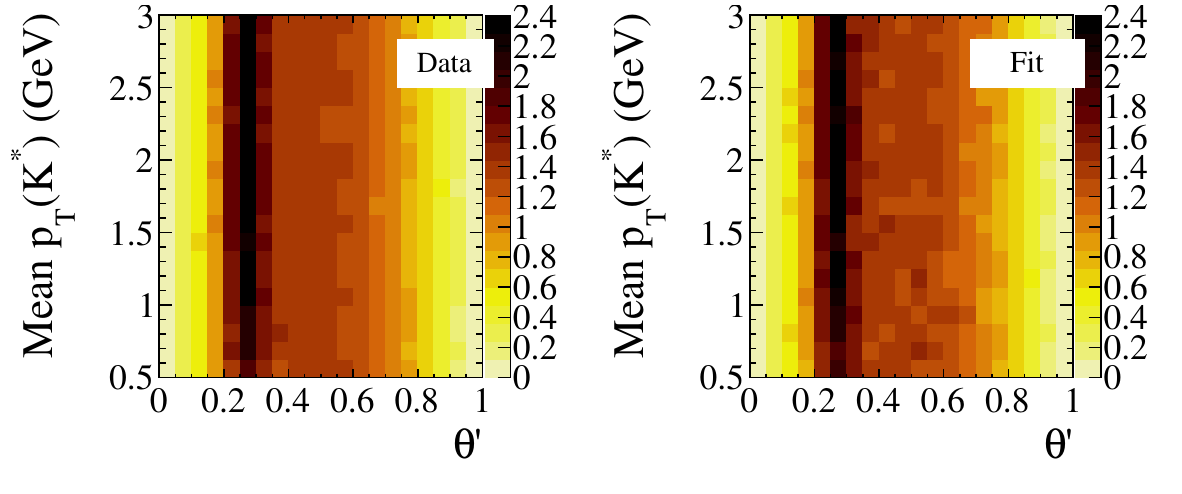}
  \includegraphics[width=0.494\textwidth]{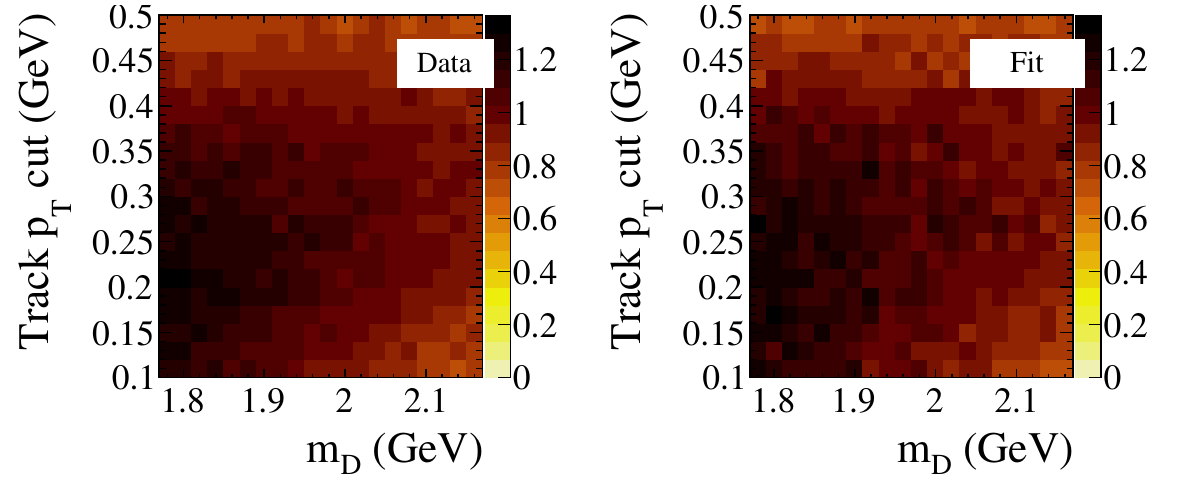}

  \includegraphics[width=0.494\textwidth]{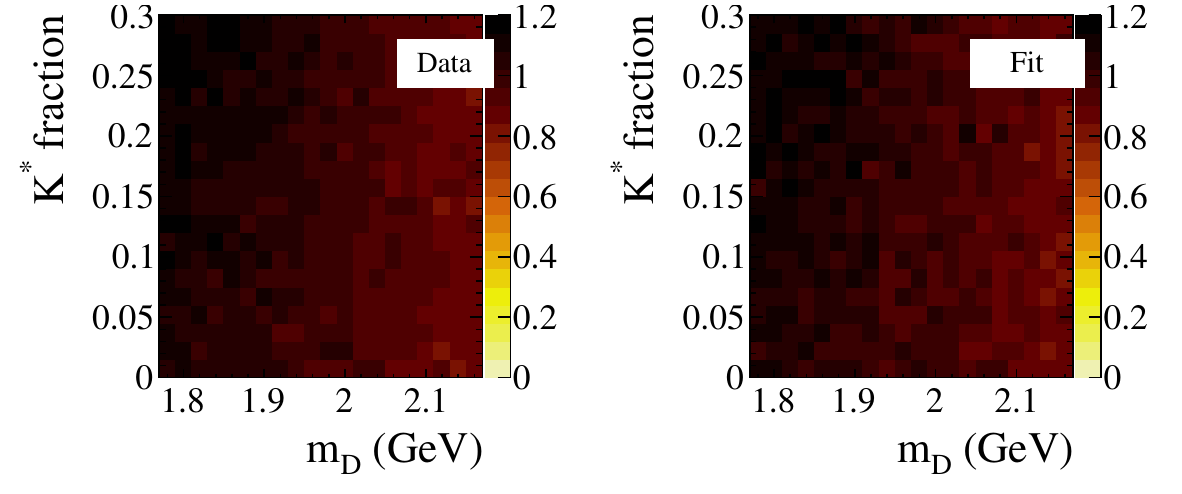}
  \includegraphics[width=0.494\textwidth]{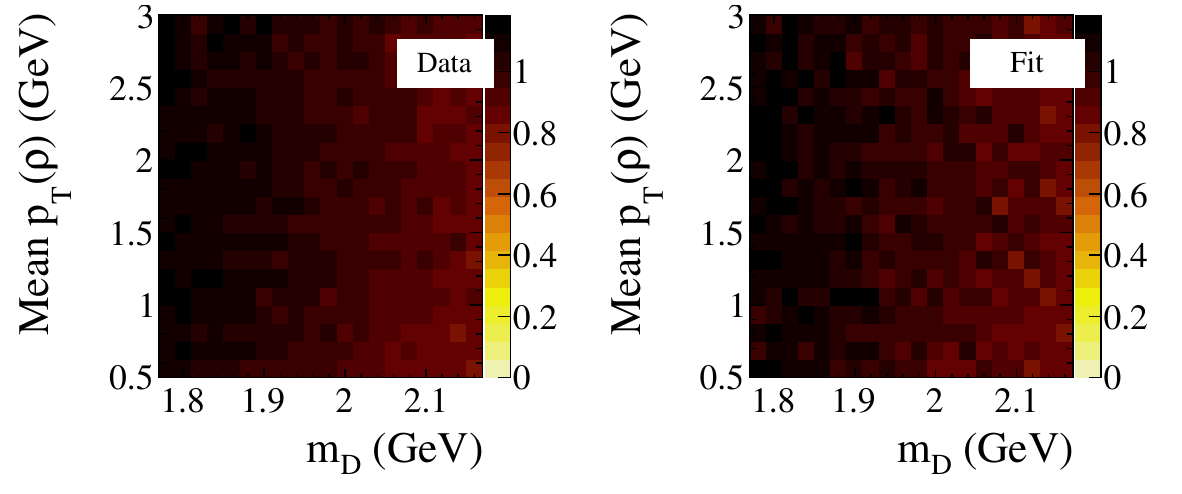}

  \caption{Results of the estimation of the simulated combinatorial background density, as a function of effective model parameters, using an ANN. }
  \label{fig:bkg_train_model2}
\end{figure}

\begin{figure}
  \centering
  \includegraphics[width=0.494\textwidth]{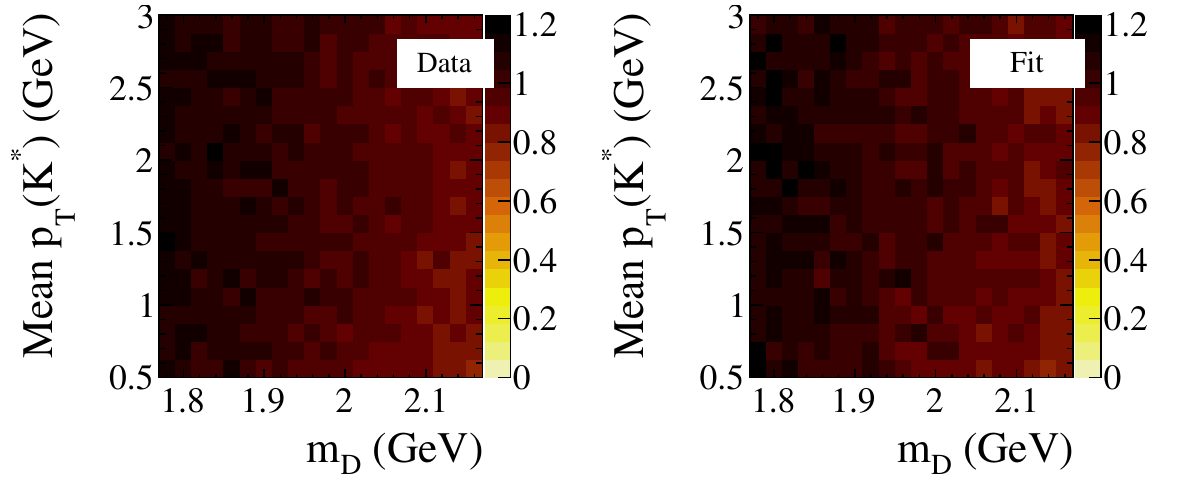}
  \includegraphics[width=0.494\textwidth]{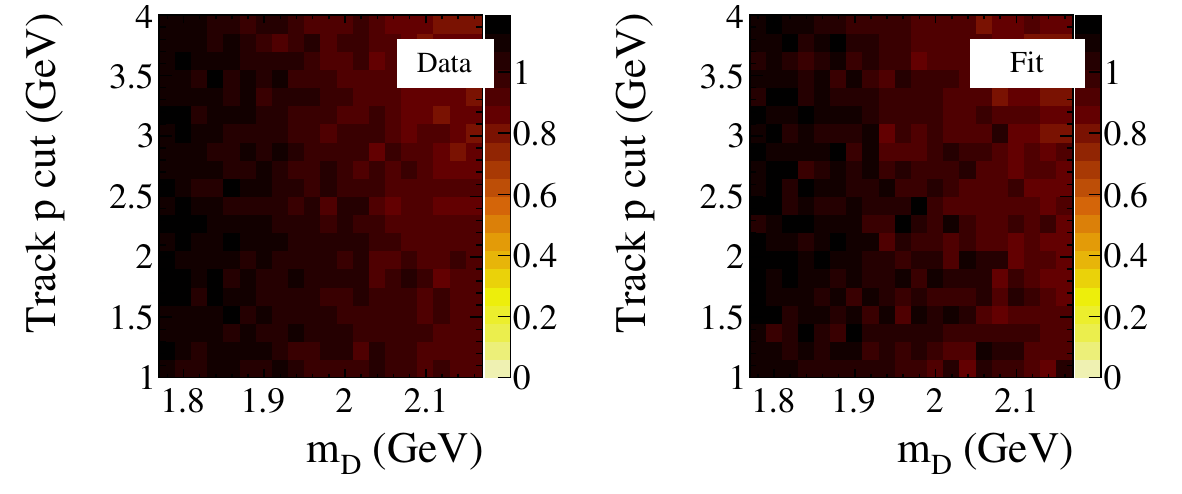}

  \includegraphics[width=0.494\textwidth]{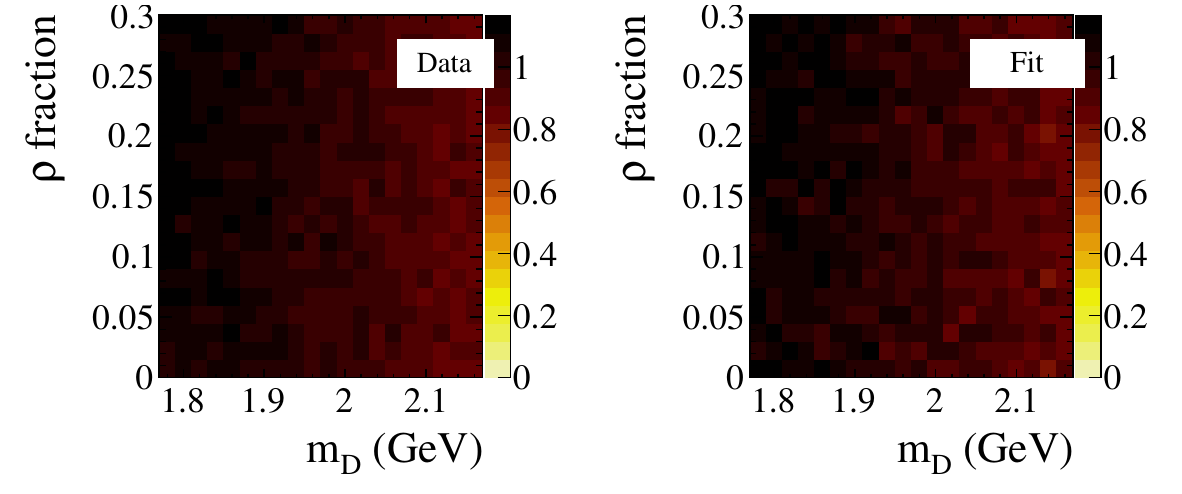}
  \includegraphics[width=0.494\textwidth]{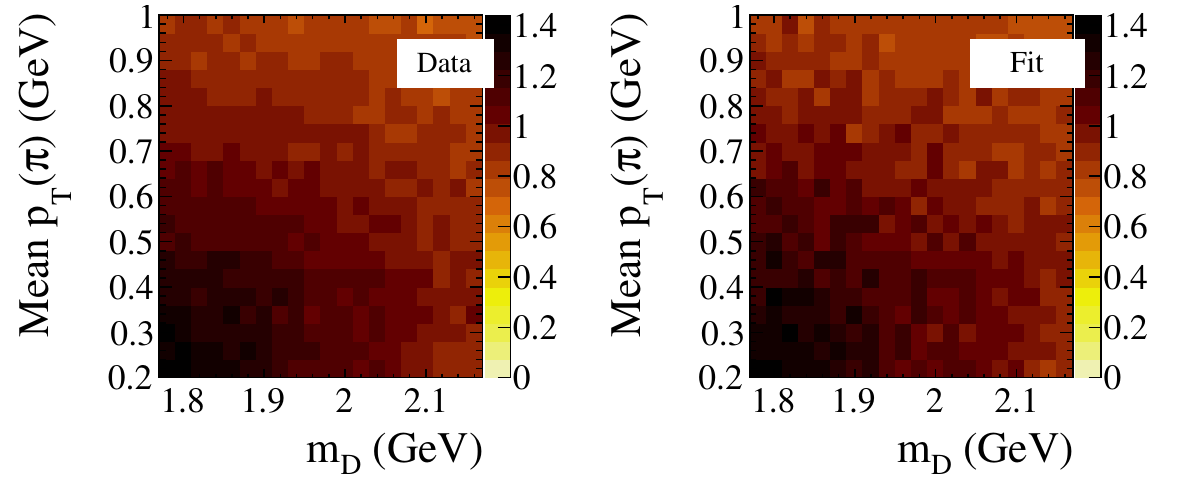}

  \includegraphics[width=0.494\textwidth]{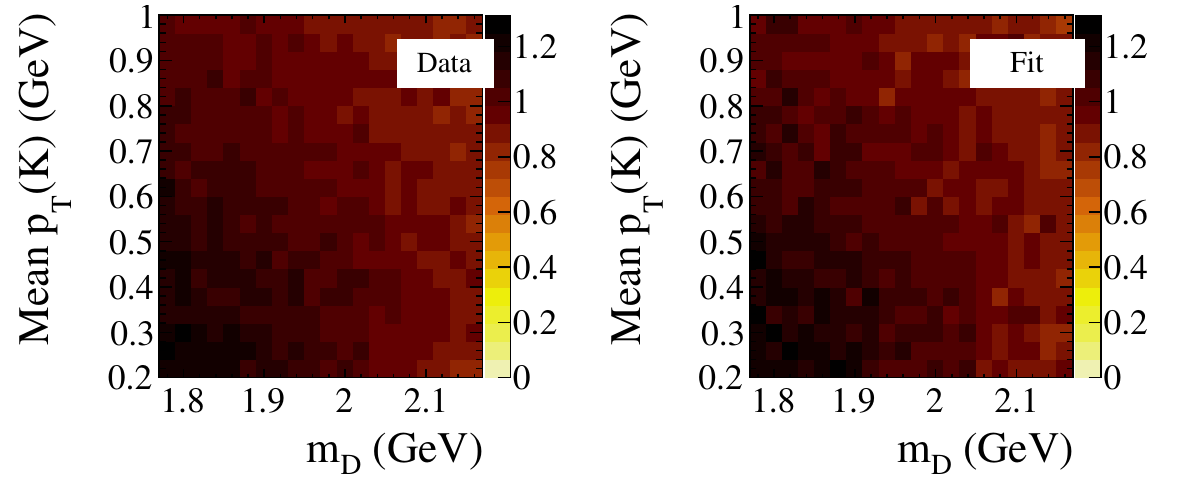}

  \caption{Results of the estimation of the simulated combinatorial background density, as a function of effective model parameters, using an ANN. }
  \label{fig:bkg_train_model3}
\end{figure}

The results of the ANN training to estimate the $11$-dimensional density of generated combinatorial background
model as a function of model parameters are shown in Fig.~\ref{fig:bkg_train_model}--\ref{fig:bkg_train_model3}.
These plots show the projections of the $m'$, $\theta'$ and $m_D$ variables (Fig.~\ref{fig:bkg_train_model}, top row),
as well as the normalised two-dimensional distributions for the correlations between each pair of $m'$, $\theta'$ and $m_D$
variables (Fig.~\ref{fig:bkg_train_model}, second row and the two leftmost plots in the third row) and correlations between
$m'$, $\theta'$, or $m_D$, and the model parameters.
The plots marked as ``Data" show the training data distributions, those marked with ``Fit" show the high-statistics
distributions generated from the result of the ANN training. As in the rest of the paper, the two-dimensional distributions
are normalised in such a way that the average of the projected distribution equals $1$.

\bibliographystyle{LHCb}
\bibliography{main,LHCb-PAPER}

\end{document}